\pdfoutput=1
\newcommand*{\ATLASLATEXPATH}{latex/}

\newcommand{\AtlasCoordFootnote}{%
ATLAS uses a right-handed coordinate system with its origin at the nominal interaction point (IP)
in the centre of the detector and the $z$-axis along the beam pipe.
The $x$-axis points from the IP to the centre of the LHC ring,
and the $y$-axis points upwards.
Cylindrical coordinates $(r,\phi)$ are used in the transverse plane, 
$\phi$ being the azimuthal angle around the $z$-axis.
The pseudorapidity is defined in terms of the polar angle $\theta$ as $\eta = -\ln \tan(\theta/2)$.
Angular distance is measured in units of $\Delta R \equiv
\sqrt{(\Delta\eta)^{2} + (\Delta\phi)^{2}}$. The transverse momentum
is the momentum component in the transverse plane.}

\documentclass[cernpreprint,texlive=2015,txfonts,UKenglish]{\ATLASLATEXPATH atlasdoc}

\usepackage{\ATLASLATEXPATH atlaspackage}
\usepackage{\ATLASLATEXPATH atlasbiblatex}

\usepackage{\ATLASLATEXPATH atlascontribute}

\usepackage{\ATLASLATEXPATH atlasphysics}

\usepackage{enumerate}
\usepackage{rotating} 
\usepackage{mdwlist}
\usepackage{multirow}
\usepackage{multicol}
\usepackage{pdflscape}
\usepackage{longtable}
\usepackage{slashed}
\usepackage{verbatim}
\usepackage{hyperref} 
\usepackage{booktabs} 
\usepackage{pifont}
\usepackage{tabu}
\usepackage{amsmath}
\usepackage{slashed}
\usepackage{color}
\usepackage{placeins}
\usepackage{xspace}
\usepackage{adjustbox}
\usepackage{dcolumn}
\usepackage[flushleft]{threeparttable}

\graphicspath{{logos/}{figures/}}

\AtlasTitle{\boldmath Search for a scalar partner of the top quark in
the jets plus missing transverse momentum final state at $\rts=13\tev$ with the ATLAS detector}

\author{The ATLAS Collaboration}

\PreprintIdNumber{CERN-PH-2017-162}

\AtlasJournalRef{JHEP 12 (2017) 085}
\AtlasDOI{10.1007/JHEP12(2017)085}

\AtlasAbstract{%
  A search for pair production of a scalar partner of the top
  quark in events with four or more jets plus missing transverse
  momentum is presented. An analysis of 36.1 \ifb\ of ${\rts=13\tev}$
  proton--proton collisions collected using the ATLAS detector at the
  LHC yields no significant excess over the expected Standard Model background. 
  To interpret the results a simplified supersymmetric model is used
  where the top squark is assumed to decay via $\stop \to t^{(*)} \ninoone$ and $\stop\to b\chinoonepm \to b  W^{\left(\ast\right)} \ninoone$, where $\ninoone$ ($\chinoonepm$) denotes the lightest neutralino (chargino). Exclusion limits are placed in terms of the top-squark and
neutralino masses. Assuming a branching ratio of $100\%$ to $t \ninoone$, top-squark masses in the range
\stopLimLowLSP~GeV are excluded for $\ninoone$ masses below
  $160\GeV$. In the case where $\mstop\sim m_t+\mLSP$, top-squark
  masses in the range \stopLimDiag~GeV are excluded.}

\hypersetup{pdftitle={ATLAS document},pdfauthor={The ATLAS Collaboration}}

\begin{document}

\maketitle




\section{Introduction}
\label{sec:intro}

Supersymmetry (SUSY)~\cite{Golfand:1971iw,Volkov:1973ix,Wess:1974tw,Wess:1974jb,Ferrara:1974pu,Salam:1974ig} is an extension of the Standard Model (SM) that
can resolve, for example, the gauge hierarchy
problem~\cite{Dimopoulos:1981zb,Sakai:1981gr,PhysRevD.24.1681,IBANEZ1981439}
by introducing supersymmetric partners of the known bosons and fermions. The SUSY partner to the top quark, the top squark ($\tilde{t}$), plays an important role in cancelling potentially large top-quark loop corrections in the Higgs boson mass. The superpartners of the left- and right-handed top quarks, $\stopL$ and $\stopR$, mix to form the two mass eigenstates $\stopone$ and $\stoptwo$, where $\stopone$ is the lighter one. Throughout this paper it is assumed that the analysis is only sensitive to $\stopone$.

\begin{figure}[htb]
\begin{center}
\subfloat[$\stop\ra t^{(*)}\ninoone$]{\includegraphics[width=0.25\textwidth]{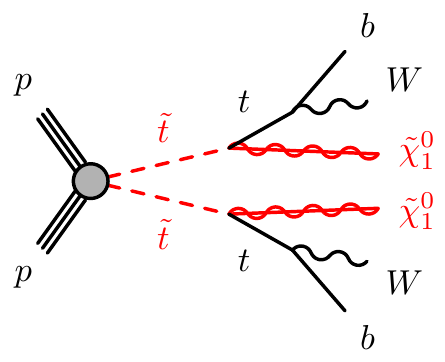}}\hspace{0.05\textwidth}
\subfloat[$\stop\ra b\chinoonepm\ra b \Wboson^{(*)}\ninoone$]{\includegraphics[width=0.25\textwidth]{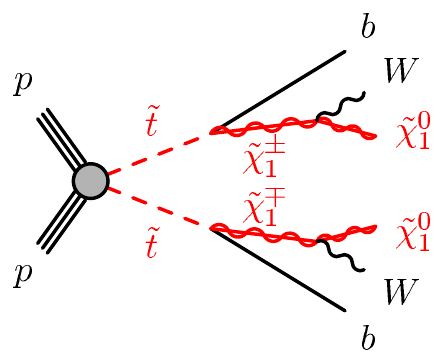}}\hspace{0.05\textwidth}
\subfloat[$\stop\ra t\ninotwo\to h/Z\ninoone$]{\includegraphics[width=0.25\textwidth]{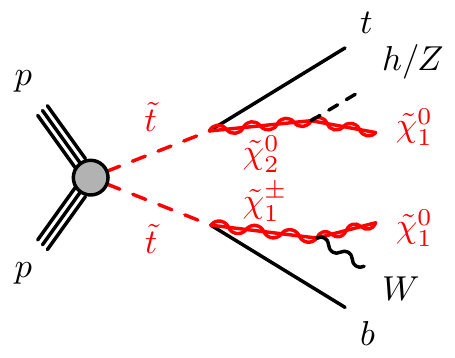}}\hspace{0.05\textwidth}
\subfloat[$\gluino \ra t\stop\ra t\ninoone +$soft]{\includegraphics[width=0.25\textwidth]{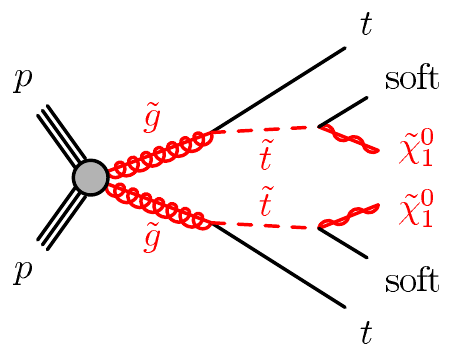}}\hspace{0.05\textwidth}
\end{center}
\caption{The decay topologies of the signal models considered with experimental signatures of four or more jets plus missing transverse momentum. Decay products that have transverse momenta below detector thresholds are designated by the term ``soft".}
\label{fig:feynDiagModels}
\end{figure}

In $R$-parity-conserving SUSY models~\cite{Farrar:1978xj}, the supersymmetric partners are produced in pairs. Top squarks are produced by strong interactions through quark--antiquark ($\qqbar$) annihilation or gluon--gluon fusion, and the cross section of direct top-squark pair production is largely decoupled from the specific choice of SUSY model parameters~\cite{Beenakker:1997ut,Beenakker:2010nq,Beenakker:2011fu,Borschensky:2014cia}. The decay of the top squark depends on the mixing of the superpartners of left- and right-handed top quarks, the masses of the top superpartner, and the mixing parameters of the fermionic partners of the electroweak and Higgs bosons. The mass eigenstates of the partners of electroweak gauge and Higgs bosons (binos, winos, higgsinos) are collectively known as charginos,
$\tilde{\chi}_{i}^{\pm}$, $i=1,2$, and neutralinos, $\tilde{\chi}_{i}^{0}$, $i=1, ..., 4$, where \ninoone\ is assumed to be the lightest supersymmetric particle (LSP) which is stable and a dark-matter candidate~\cite{Goldberg:1983nd,Ellis:1983ew}. For the models considered, either \ninotwo\ or \chinoonepm\ is assumed to be the next lightest supersymmetric particle (NLSP). 
Three different decay scenarios are considered in this search: (a) both top squarks decay via $\stop\rightarrow t^{(*)} \ninoone$, (b) at least one of the top squarks decays via $\stop\rightarrow b \chinoonepm \rightarrow b W^{(*)}\ninoone$, with various hypotheses for $\mLSP$ and $m_{\chinoonepm}$, and (c) where $m_{\ninotwo}$ is small enough for at least one top squark to decay via $\stop\to t \ninotwo \to h/Z \ninoone$, where $h$ is the SM-like Higgs boson with a mass of 125 GeV, as illustrated in 
Figure~\ref{fig:feynDiagModels}(a)$-$(c), respectively. The interpretation of the results uses simplified models
~\cite{Alwall:2008ve,Alwall:2008ag,Alves:2011wf} where only one or two decay steps are allowed. In the case with two allowed decays, referred to later in this paper as a natural SUSY-inspired mixed grid, the mass splitting between the \chinoonepm\ and the \ninoone, $\Delta m(\chinoonepm,\ninoone)$, is assumed to be 1~\gev. A grid of signal samples is generated across the plane of the top-squark and \ninoone\ masses with a grid spacing of $50\gev$ across most of the plane, 
assuming 
maximal mixing between the partners of the left- and right-handed top quarks. 
In both the one- and two-step decay scenarios the LSP is considered to be a pure bino state. Additionally, results are interpreted in two slices of phenomenological MSSM (pMSSM)~\cite{Djouadi:1998di,Berger:2008cq} models, referred to as wino-NLSP and well-tempered neutralino pMSSM models in the remainder of this paper. The pMSSM models are based on the more general MSSM~\cite{Fayet:1976et,Fayet:1977yc} but with the additional requirements of no new sources of CP violation and flavour-changing neutral currents, as well as first- and second-generation sfermion mass and trilinear coupling degeneracy. Finally, results are also interpreted in a simplified model which is inspired by the pMSSM and is referred to as non-asymptotic higgsino. Details of the models that are used in the various interpretations are given in Section~\ref{sec:result}.

In addition to direct pair production, top squarks can be produced indirectly through gluino decays, as shown in Figure~\ref{fig:feynDiagModels}(d). This search considers models where the mass difference between the top squark and the neutralino is small, i.e. $\Delta m(\stop,\ninoone)=5\GeV$. In this scenario, the jets originating from the \stop\ decays have momenta below the experimental acceptance, resulting in a signature nearly identical to that of $\stop\to t \ninoone$ signal models (Figure~\ref{fig:feynDiagModels}(a)).

This paper presents the search for top-squark pair production using a time-integrated luminosity of \lumi\ of proton--proton ($pp$) collisions data provided by the Large Hadron Collider (LHC) at a centre-of-mass energy of $\rts = 13\tev$. The data were collected by the ATLAS detector in 2015 and 2016. All-hadronic final states with at least four jets and large missing transverse momentum\footnote{\AtlasCoordFootnote} ($\ptmiss$, whose magnitude is referred to as \met) are considered, and the results are interpreted according to a variety of signal models as described above. Signal regions are defined to maximize the experimental sensitivity over a large region of kinematic phase space. Sensitivity to high top-squark masses $\sim1000\GeV$ (as in Figure~\ref{fig:feynDiagModels}(a)) and top squarks produced through gluino decays (as in Figure~\ref{fig:feynDiagModels}(d)) is achieved by exploiting techniques designed to reconstruct top quarks that are Lorentz-boosted in the lab frame. The dominant SM background process for this kinematic region is $Z\rightarrow \nu \bar{\nu}$ produced in association with jets initiated by heavy-flavour quarks (heavy-flavour jets). The sensitivity to the decay into $b\chinoonepm$ is enhanced by vetoing events containing hadronically decaying top-quark candidates to reduce the \ttbar\ background, leaving $Z\rightarrow \nu \bar{\nu}$ as the largest SM background. Sensitivity to the region where $m_{\stop}-m_{\ninoone} \sim m_t$, which typically has relatively low-\pT\ final-state jets and low \MET, is achieved by exploiting events in which high-\pT\ jets from initial-state radiation (ISR) boosts the di-top-squark system in the transverse plane. For this regime, \ttbar\ production gives the dominant background contribution. Similar searches based on $\rts = 8\tev$ and $\rts = 13\tev$ data collected at the LHC have been performed by both the ATLAS~\cite{Atlas8TeV,Atlas8TeVSummary,GtcStop1L,stop2L8TeV} and CMS~\cite{Khachatryan:2015pwa,Khachatryan:2016pxa,Sirunyan:2016jpr,Khachatryan:2017rhw,Khachatryan:2016xvy} collaborations.


\section{ATLAS detector}
\label{sec:detector}

The ATLAS experiment~\cite{DetectorPaper:2008} at the LHC is a multi-purpose particle detector
with a cylindrical forward-backward and $\phi$-symmetric
geometry
and an approximate $4\pi$ coverage in 
solid angle.
It consists of an inner tracking detector surrounded by a thin superconducting solenoid
providing a \SI{2}{\tesla} axial magnetic field, electromagnetic and hadron calorimeters, and a muon spectrometer.
The inner tracking detector covers the pseudorapidity range $|\eta| < 2.5$.
It consists of silicon pixel, silicon microstrip, and transition
radiation tracking detectors. The newly installed innermost layer of pixel sensors~\cite{IBL} was operational for the first time during the 2015 data-taking.
Lead/liquid-argon (LAr) sampling calorimeters provide electromagnetic (EM) energy measurements
with high granularity.
A hadron (steel/scintillator-tile) calorimeter covers the central pseudorapidity range ($|\eta| < 1.7$).
The end-cap and forward regions are instrumented with LAr calorimeters
for both the EM and hadronic energy measurements up to $|\eta| = 4.9$.
The muon spectrometer surrounds the calorimeters and features
three large air-core toroidal superconducting magnets with eight coils
each, providing coverage up to $|\eta| = 2.7$.
The field integral of the toroids ranges between 2.0 and \SI{6.0}{\tesla m} across most of the detector.
It includes a system of precision tracking chambers and fast detectors for triggering.


\section{Trigger and data collection}
\label{sec:trigger}

The data were collected from August to November 2015 and April to October 2016 at a $pp$ centre-of-mass energy of $13\tev$ with $25\ns$ bunch spacing. A two-level trigger system~\cite{trigger} is used to select events.
The first-level trigger is implemented in hardware and uses a subset of the detector information
to reduce the event rate to at most \SI{100}{\kilo\hertz}.
This is followed by a software-based trigger that reduces the accepted event rate to \SI{1}{\kilo\hertz} for offline storage.  

In all search regions, a missing transverse momentum trigger, which is fully efficient for offline calibrated $\met > 250$~\GeV\ in signal events, was used to collect data events.

Data samples enriched in the major sources of background were
collected with electron or muon triggers. The electron trigger
selects events based on the presence of clusters of energy in the electromagnetic calorimeter, with a
shower shape consistent with that expected for an electron, and a matching track
in the tracking system. The muon trigger selects events containing one or more muon candidates based on tracks
identified in the muon spectrometer and inner detector. 
The electron and muon triggers used are more than $99\%$ efficient for isolated electrons and muons with \pt\ above 28~\GeV.

Triggers based on the presence of high-\pt\ jets were used to collect data samples for the estimation of
the multijet and all-hadronic \ttbar~background.
The jet \pt~thresholds ranged from $20$ to $400\,\GeV$. In order
to stay within the bandwidth limits of the trigger system, only  a
fraction of the events passing these triggers was recorded to permanent
storage.



\section{Simulated event samples and signal modelling}
\label{sec:simulation}

Simulated events are used to model the SUSY signal and to aid in the description of the background processes. Signal models were all generated with {\scshape MG5\_aMC\/@NLO} 2.2-2.4~\cite{madgraph} interfaced to \pythiaeight~\cite{pythia8} for the parton showering (PS) and hadronization and with {\scshape EvtGen} 1.2.0~\cite{evtGen} for the $b$- and $c$-hadron decays. The matrix element (ME) calculation was performed at tree level and includes the emission of up to two additional partons for all signal samples. The parton distribution function (PDF) set used for the generation of the signal samples is NNPDF2.3LO~\cite{PDFs} with the A14~\cite{UEtune} set of tuned underlying-event and shower parameters (UE tune). The ME--PS matching was performed with the CKKW-L~\cite{CKKW} prescription, with a matching scale set to one quarter of the mass of the \stop, or \gluino\ for the gluino pair production model. All signal cross sections were calculated to next-to-leading order in the strong coupling constant, adding the resummation of soft-gluon emission at next-to-leading-logarithm accuracy (NLO+NLL)~\cite{Beenakker:1997ut,Beenakker:2010nq,Beenakker:2011fu}. The nominal cross section and the uncertainty were taken from an envelope of cross-section predictions using different PDF sets and factorization and renormalization scales, as described in Ref.~\cite{Borschensky:2014cia}. For pMSSM models, the sparticle mass spectra were calculated with Softsusy 3.7.3~\cite{Allanach:2001kg,Allanach:2013kza} while the decays of each sparticle were performed by HDECAY 3.4~\cite{hdecay} and SDECAY 1.5/1.5a~\cite{sdecay}.

SM background samples were generated with different MC event generators depending on the process. The background sources of \Zjets\ and \Wjets\ events were generated with \sherpa\ 2.2.1~\cite{sherpa} using the NNPDF3.0NNLO~\cite{PDFs} PDF set and the UE tune provided by \sherpa. Top-quark pair production where at least one of the top quarks decays semileptonically and single-top production were simulated with \textsc{Powheg-Box}~2~\cite{powheg-box} and interfaced to \pythia 6~\cite{pythia6} for PS and hadronization, with the CT10~\cite{CT10} PDF set and using the {\scshape Perugia2012}~\cite{Perugia2012} set of tuned shower and underlying-event parameters. {\scshape MG5\_aMC\/@NLO} interfaced to \pythiaeight\ for PS and hadronization was used to generate the \ttbar+$V$ (where $V$ is a  \Wboson\ or \Zboson\ boson) and \ttbar+$\gamma$ samples at NLO with the NNPDF3.0NLO PDF set. The underlying-event tune used is A14 with the NNPDF2.3LO PDF set. Diboson production was generated with \sherpa~2.2.1 using the CT10 PDF set. Finally, $V\gamma$ processes were generated with \sherpa~2.1 using the CT10 PDF set. Additional information can be found in Refs.~\cite{Vjets,multibosons,top,ttV,MC15val}.

The detector simulation~\cite{ATLdetSim} was performed using either \geant 4~\cite{GEANT} or a fast simulation framework where the showers in the electromagnetic and hadronic calorimeters are simulated with a parameterized description~\cite{fastSim} and the rest of the detector is simulated with \geant 4. The fast simulation was validated against full \geant 4 simulation for several selected signal samples and subsequently used for all signal samples because of the large number of signal grid points needed for interpretation. All SM background samples used the GEANT4 set-up. All MC samples were produced with a varying number of simulated minimum-bias interactions overlaid on the hard-scattering event to account for multiple $pp$ interactions in the same or nearby bunch crossing (pile-up). These events were produced using \pythiaeight\ with the A2 tune~\cite{A2} and MSTW 2008 PDF set~\cite{MSTW}. The simulated events were reweighted to match the distribution of the number of $pp$ interactions per bunch crossing in data. Corrections were applied to the simulated events to correct for differences between data and simulation for the lepton-trigger and reconstruction efficiencies, momentum scale, energy resolution, isolation, and for the efficiency of identifying jets containing $b$-hadrons, together with the probability for mis-tagging jets containing only light-flavour and charm hadrons.




\section{Event reconstruction}
\label{sec:reconstruction}

Events are required to have a primary vertex~\cite{vertexReco} reconstructed from at
least two tracks with $\pT>400\mev$. 
Among the vertices found, the vertex with the largest summed $\pT^2$ of the associated tracks is chosen.
\begin{sloppypar}
Jets are reconstructed from three-dimensional topological clusters of noise-suppressed calorimeter cells~\cite{caloClusters} using the anti-$k_t$ jet algorithm~\cite{antiKt,Cacciari:2011ma} with a radius parameter
$R=0.4$. An area-based correction is applied to account for energy
from additional $pp$ collisions based on an estimate of the pile-up activity in a given event~\cite{pileupSub}. Calibrated~\cite{jetsCalibrated} jet candidates are required to have
$\pT>20 \gev$ and $|\eta|<2.8$. Events containing jets arising from
non-collision sources or detector noise~\cite{jetCleaning} are removed (``no bad jets'' requirement). 
Additional selections based on track information are applied to
jets with $\pT<60 \gev$ and $|\eta|<2.4$ to reject jets that originate from
pile-up interactions~\cite{jetTagger}.
\end{sloppypar}
Jets containing $b$-hadrons and which are within the inner detector
acceptance ($|\eta|<2.5$) are identified (as $b$-tagged jets) with a multivariate algorithm that exploits
the impact parameters of the charged-particle tracks, the presence of secondary
vertices, and the reconstructed flight paths of $b$- and $c$-hadrons inside
the jet~\cite{btagging,MV10c,ATL-PHYS-PUB-2014-014}. The output of the multivariate algorithm is a single $b$-tagging weight which signifies the likelihood of a jet containing $b$-hadrons. The average identification efficiency of jets containing $b$-hadrons is $77\%$ as determined in simulated \ttbar\ events. A rejection factor of approximately 130 is reached for jets initiated by light quarks and
gluons and 6 for jets initiated by charm quarks.

Electron candidates are reconstructed from clusters of energy deposits in the
electromagnetic calorimeter that are matched to a track in the inner
detector. They are required to have $|\eta|<2.47$, $\pT>7\gev$ and
must pass a variant of the ``very loose'' likelihood-based selection~\cite{egamma,egamma2}. The electromagnetic shower of an electron can also form a jet such that a procedure is required to resolve this ambiguity. In the case where the separation between an electron candidate and a non-$b$-tagged ($b$-tagged) jet is $\Delta R < 0.2$,\footnote{For the overlap removal, rapidity, defined as $\frac{1}{2} \ln \frac{E+p_z}{E-p_z}$, is used instead of pseudorapidity in the $\Delta R$ definition.} the candidate is considered to be an electron ($b$-tagged jet). If the
separation between an electron candidate and any jet satisfies $0.2<
\Delta R < 0.4$, the candidate is considered to be a jet, and the electron candidate is removed.

Muons are reconstructed by matching tracks in the inner detector to tracks 
in the muon spectrometer and are required to have $|\eta|<2.7$ and
$\pT>6\gev$. If the separation between a muon and any jet is $\Delta R
< 0.4$, the muon is omitted. Events containing muons identified as originating from cosmic rays ($|d_0| > 0.2$ mm and $|z_0| > 1$ mm) or as poorly reconstructed ($\sigma(q/p)/|(q/p)| > 0.2$) are removed (``cosmic and bad muon'' requirement). 
Here, $d_0$ is the transverse impact parameter of a track with respect to the primary vertex, $z_0$ is the distance of this point from the primary vertex projected onto the $z$-axis, and $\sigma(q/p)/|(q/p)|$ provides a measure of the momentum uncertainty for a particle with charge $q$.

The $\ptmiss$ vector is the negative vector sum 
of the \pT\ of all selected and calibrated electrons, muons, and jets in the
event. An extra term is added to account for small energy depositions in the event that are not associated with any of the selected objects. This ``soft'' term is calculated from inner detector tracks with $\pT > 400 \MeV$ matched to the primary vertex, to make it resilient to pile-up contamination, not associated with physics objects~\cite{met}. The missing transverse momentum from the tracking system (denoted by \ptmisstrk, with magnitude \mettrk)
is computed from the vector sum of the reconstructed inner detector tracks with $\pt > 400\MeV$, $|\eta|<2.5$, that are associated with the primary vertex in the event. The \ptmisstrk\ and \mettrk\ are used to reject events with large calorimeter-based \met\ due to pile-up contamination or jet energy mismeasurements. These events, where the \ptmisstrk\ tends to not be aligned with the \ptmiss\ and the \met\ tends to be much larger than the \mettrk, are rejected by requiring that the $\Delta\phi$ between the \ptmiss\ and \ptmisstrk\ is less than $\pi$/3 and that the $\mettrk>30$ GeV.

The requirements on electrons and muons are tightened for the
selection of events in background control regions (described in
Section~\ref{sec:background}) containing leptons. Electron and muon candidates are required to have $\pT>20\GeV$ ($\pT>28\GeV$) for regions using the \met\ (lepton) triggers and to satisfy $\pt$-dependent track- and calorimeter-based isolation criteria. 
The calorimeter-based isolation is determined by taking the ratio of the sum of energy deposits in a cone of $R=0.2$ around the electron or muon candidate and the energy deposits associated with the electron and muon. The track-based isolation is estimated in a similar way but using a variable cone size with a maximum value of $R=0.2$ for electrons and $R=0.3$ for muons. An isolation requirement is made that is 95\% efficient for electron or muon candidates with $\pt=25$ GeV and 99\% for candidates with $\pt=60$ GeV.

Electron candidates are required to pass a ``tight'' likelihood-based selection~\cite{egamma}.
The impact parameter of the electron in the transverse plane with respect to the reconstructed event primary
vertex is required to be less than five times the impact
parameter uncertainty ($\sigma_{d_0}$). The impact parameter along the
beam direction, $\left|z_0 \times \sin\theta\right|$, is required to be less than $0.5$~mm. Further
selection criteria are also imposed on reconstructed muons: muon candidates are required to pass a ``medium" quality selection~\cite{PERF-2015-10}. In addition, the requirements $|d_0| < 3\sigma_{d_0}$ and
$\left|z_0 \times \sin\theta\right| <0.5$~mm are imposed for muon candidates.



\section{Signal region definitions}
\label{sec:signalregions}

The main experimental signature for all signal topologies is the presence of multiple jets (two of which are $b$-tagged), no muons or electrons, and significant missing transverse momentum. 

Five sets of signal regions (SRA--E) are defined to target each topology and kinematic regime. SRA (SRB) is sensitive to production of high-mass \stop\ pairs with large (intermediate) $\Delta m(\stop,\LSP)$. Both SRA and SRB employ top-mass reconstruction techniques to reject background. SRC is designed for the highly compressed region with $\Delta m(\stop,\LSP)\sim m_{t}$. In this signal region, initial-state radiation (ISR) is used to improve sensitivity to these decays. SRD is targeted at $\stop\to b\chinoonepm$ decays, where no top-quark candidates are reconstructed. SRE is optimized for scenarios with highly boosted top quarks that can occur in gluino-mediated top-squark production. 

A common preselection is defined for all signal regions. At least four jets are required, of which at least one must be $b$-tagged. The four leading jets (ordered in \pt) must satisfy $\ptzerotothree>80,80,40,40 \gev$ due to the tendency for signal events to have more energetic jets than background. Events containing reconstructed electrons or muons are vetoed. The $\met$ trigger threshold motivates the requirement $\met>250\gev$ and rejects most of the background from multijet and all-hadronic $\ttbar$
events. In order to reject events with mismeasured \met\
originating from multijet and hadronic \ttbar\ decays, an angular
separation between the azimuthal angle of the two highest-\pT\ jets
and the \ptmiss\ is required: $\dphijettwomet 
> 0.4$. Further rejection of such events is achieved by requiring the \ptmisstrk\ to be aligned in $\phi$
with respect to the \ptmiss\ calculated from the calorimeter system:
$\mettrk > 30 \gev$ and $\dphimettrk <\pi/3$.

\subsubsection*{Signal Regions A and B}

\begin{figure}[t]
  \begin{center}
    \subfloat[]{\includegraphics[width=0.5\textwidth]{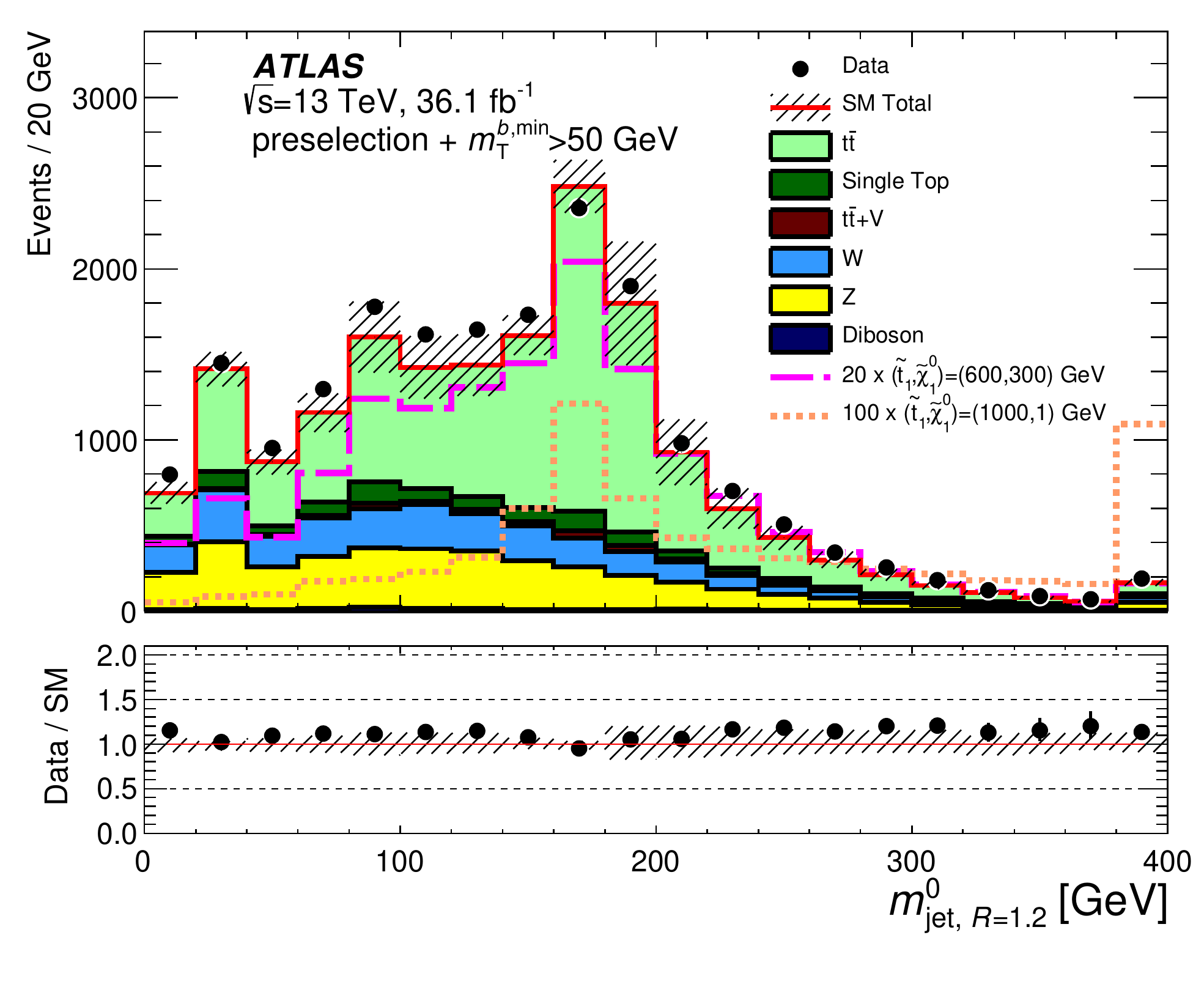}}
    \subfloat[]{\includegraphics[width=0.5\textwidth]{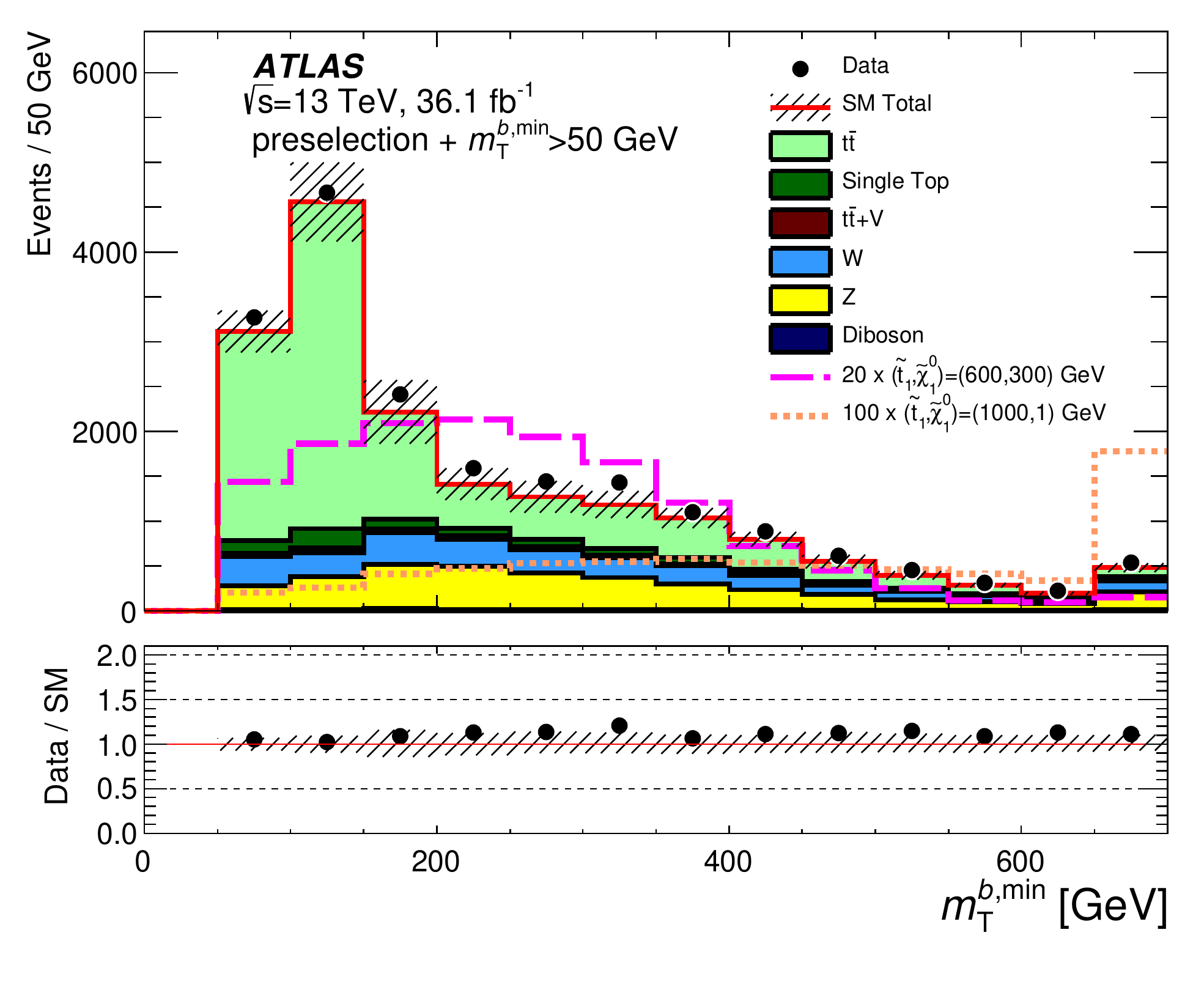}}
    \caption{Distributions of the discriminating variables (a)~\mantikttwelvezero\ and (b)~\mtbmin\ after the common preselection and an additional $\mtbmin>50\gev$ requirement. The stacked histograms show the SM prediction before being normalized using scale factors derived from the simultaneous fit (detailed in Section~\ref{sec:simultaneousfit}) to all dominant backgrounds. The ``Data/SM" plots show the ratio of data events to the total SM prediction. The hatched uncertainty band around the SM prediction and in the ratio plots illustrates the combination of statistical and detector-related systematic uncertainties. The rightmost bin includes overflow events.}
    \label{fig:preselection}
  \end{center}
\end{figure}

SRA and SRB are targeted at direct top-squark pair production where the top squarks decay via $\stop
\rightarrow t \ninoone$ with $\Delta m(\stop,\ninoone) > m_t$. SRA is optimized for $\mstop=1000\gev$ and $\mLSP=1\gev$, while SRB is optimized for $\mstop=600\gev,\mLSP=300\gev$. At least two $b$-tagged jets ($\nBJet\ge2$) are required and an additional requirement on the $\Delta\phi$ of the three leading jets and \ptmiss\ of $\dphijetthreemet > 0.4$ is made.

\begin{figure}[t]
  \begin{center}
    \includegraphics[width=0.7\textwidth]{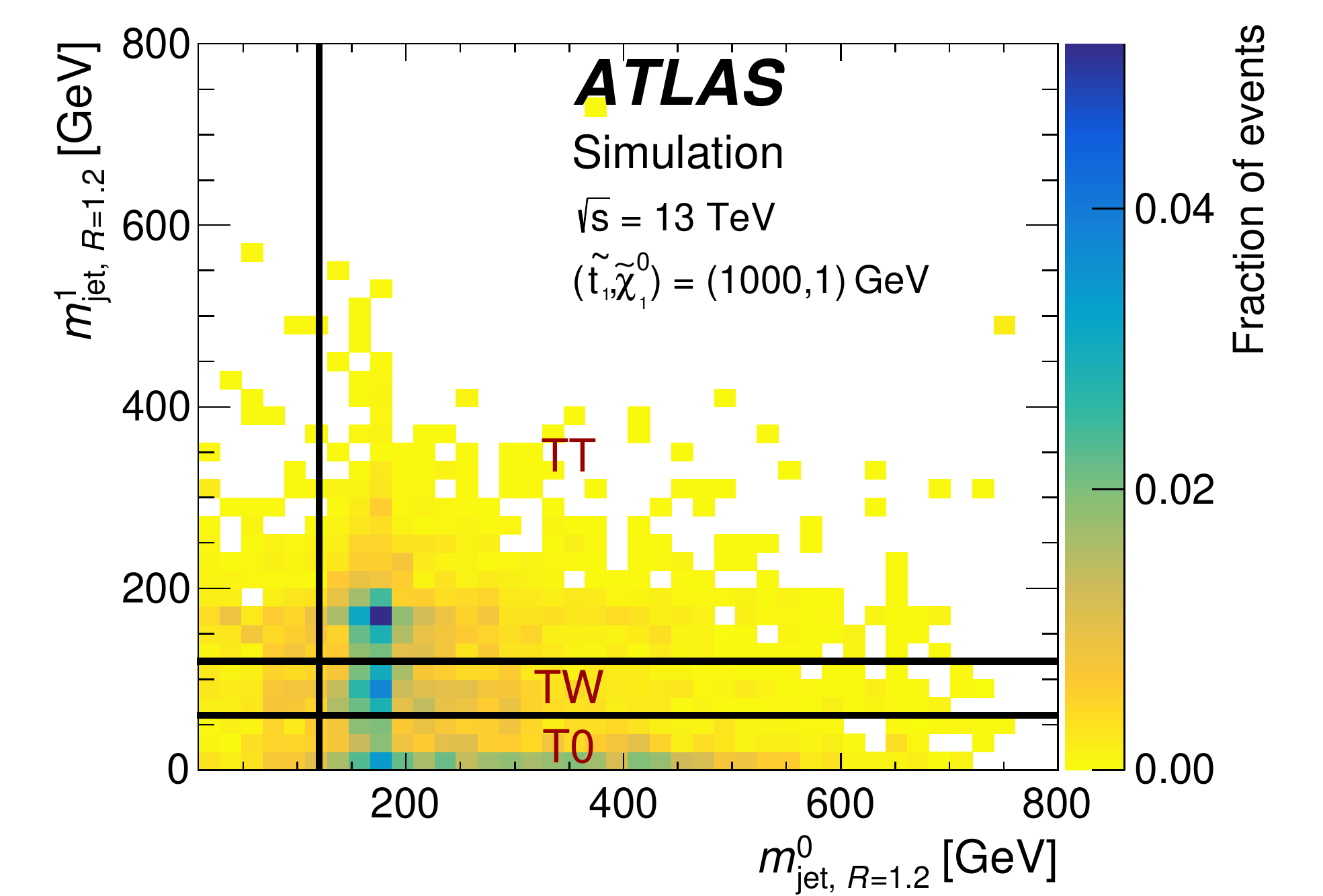}
    \caption{Illustration of signal-region categories (TT, TW, and T0) based on the $R=1.2$ reclustered top-candidate masses for simulated direct top-squark pair production with $(m_{\stop},m_{\ninoone})=(1000,1) \GeV$ after the loose preselection requirement described in the text. The black lines represent the requirements on the reclustered jet masses.}
    \label{fig:categories}
  \end{center}
\end{figure}

The decay products of the $\ttbar$ system in the all-hadronic decay mode can often be reconstructed as six distinct $R=0.4$ jets. The transverse shape of these jets is typically circular with a radius equal to this radius parameter, but when two of the jets are less than $2R$ apart in $\eta$--$\phi$ space, the one-to-one correspondence of a jet with a top-quark daughter may no longer hold. Thus, the two hadronic top candidates are reconstructed by applying the \antikt\ clustering algorithm~\cite{antiKt} to the $R=0.4$ jets, using reclustered radius parameters of $R=0.8$ and $R=1.2$. Two $R=1.2$ reclustered jets are required; the mass of the highest-\pT\ $R=1.2$ reclustered jet is shown in Figure~\ref{fig:preselection}(a). The events are divided into three categories based on the resulting $R=1.2$ reclustered jet masses ordered in \pt, as illustrated in Figure~\ref{fig:categories}: the ``TT'' category includes events with two top candidates, i.e.\ with masses $\mantikttwelvezero>120\gev$ and $\mantikttwelveone>120\gev$; the ``TW'' category contains events with one top candidate and a $\Wboson$ candidate, i.e.\ where $\mantikttwelvezero>120\gev$ and $60<\mantikttwelveone<120\gev$; and the ``T0" category represents events with only one top candidate, i.e.\ where $\mantikttwelvezero>120\gev$ and $\mantikttwelveone<60\gev$. Since the signal-to-background ratio is different in each of these categories, they are optimized individually for SRA and SRB.

The most powerful discriminating variable against SM \ttbar\ production is the $\met$ value, which for the signal results from the undetected $\ninoone$ neutralinos. Substantial $\ttbar$ background rejection is provided by additional requirements that reject events in which one $\Wboson$ boson decays via a charged lepton plus neutrino. The first requirement is that the transverse mass (\mt) calculated from the \met\ and
the $b$-tagged jet with minimum distance in $\phi$ to the $\ptmiss$ direction is above $200\GeV$:

\begin{equation*}
\mtbmin\ = \sqrt{2\,\ptb\,\met \left[1-\cos{\Delta\phi\left(\vecptb,\ptmiss\right)}\right]} > 200\,\GeV,
\end{equation*}

since its upper bound (ideally, without consideration of resolution effects) is below the top-quark mass for the \ttbar\ background, as illustrated in Figure~\ref{fig:preselection}(b). An additional requirement is made on the mass of the leading (in \pt) $R=0.8$ reclustered jet to be consistent with a \Wboson\ candidate: $\mantikteightzero>60\gev$. Additionally, requirements on the stransverse mass (\mttwo)~\cite{Lester:1999tx,Barr:2003rg}
are made which are especially powerful in the T0 category where a $\chi^2$ method is applied to reconstruct top quarks with lower momenta where reclustering was suboptimal. 
The \mttwo\ variable is constructed from the direction and magnitude of the \ptmiss\ vector in the transverse plane as well as the direction of two top-quark candidates reconstructed using a $\chi^2$ method. 
The minimization in this method is done in terms of a $\chi^2$-like penalty function, $\chi^2 = (m_{\mathrm{cand}}-m_{\mathrm{true}})^2/m_{\mathrm{true}}$, where $m_{\mathrm{cand}}$ is the candidate mass and $m_{\mathrm{true}}$ is set to 80.4 GeV for \Wboson\ candidates and 173.2 GeV for top candidates. 
Initially, single or pairs of $R=0.4$ jets form \Wboson\ candidates which are then combined with additional $b$-tagged jets in the event to construct top candidates. The top candidates selected by the $\chi^2$ method are only used for the momenta in \mttwo\ while the mass hypotheses for the top quarks and the invisible particles are set to 173.2 GeV and 0 GeV, respectively. Finally, a ``$\tau$-veto'' requirement is applied to reject semi-hadronically decaying $\tau$-lepton candidates
likely to have originated from a $W \rightarrow \tau \nu$ decay. Here, events that contain
a non-$b$-tagged jet within $|\eta|<2.5$ with fewer than four associated charged-particle tracks
with $\pT > 500 \mev$, and where the $\Delta \phi$ between the jet and the
$\ptmiss$ is less than $\pi / 5$, are vetoed. The systematic uncertainties for this requirement are found to be negligible~\cite{Atlas8TeV}. In SRB, additional discrimination is provided by $\mtbmax$ and $\Delta R(b,b)$. The former quantity is analogous to \mtbmin\ except that the transverse mass is computed with the $b$-tagged jet that has the largest $\Delta\phi$ with respect to the $\ptmiss$ direction. 
The latter quantity provides additional discrimination against background where the two jets with highest $b$-tagging weights originate from a gluon splitting. Table~\ref{tab:SignalRegionAB} summarizes the selection criteria that are used in these two signal regions. The categories are statistically combined within SRA and SRB to maximize the sensitivity to signal.

\begin{table}[htb]
  \caption{Selection criteria for SRA and SRB, in addition to the common preselection requirements described in the text. The signal regions are separated into topological categories based on reconstructed top-candidate masses.}
  \begin{center}
  \def\arraystretch{1.4}
   \begin{tabular}{c||l|c|c|c} \hline\hline
      {\bf Signal Region}      &                    & {\bf TT}    & {\bf TW}     & {\bf T0}     \\ \hline \hline
                               & \mantikttwelvezero & \multicolumn{3}{c}{$>120\gev$}            \\ \cline{2-5}
                               & \mantikttwelveone  & $>120\gev$  & $[60,120]\gev$ & $<60\gev$    \\ \cline{2-5}
                               & \mtbmin            & \multicolumn{3}{c}{$>200\gev$}            \\ \cline{2-5}
                               & \nBJet    & \multicolumn{3}{c}{$\ge2$}                \\ \cline{2-5}
                               & $\tau$-veto        & \multicolumn{3}{c}{yes}                   \\ \cline{2-5} 
                               & \dphijetthreemet   & \multicolumn{3}{c}{$>0.4$}                \\ \cline{2-5}\hline \hline
      \multirow{3}{*}{{\bf A}} & \mantikteightzero  & \multicolumn{3}{c}{$>60\gev$}             \\ \cline{2-5}
                               & \drbjetbjet        & $>1$        & \multicolumn{2}{c}{-}       \\ \cline{2-5}
                               & \mttwo             & $>400$ GeV  & $>400$ GeV   & $>500$ GeV   \\ \cline{2-5}
                               & \met               & $>400 \gev$ & $> 500 \gev$ & $> 550 \gev$ \\ \hline \hline
      \multirow{2}{*}{{\bf B}} & \mtbmax            & \multicolumn{3}{c}{$>200\gev$}            \\ \cline{2-5}
                               & \drbjetbjet        & \multicolumn{3}{c}{$>1.2$}                \\ \cline{2-5}              
\hline\hline
    \end{tabular}
\end{center}
\label{tab:SignalRegionAB}
\end{table}

\subsubsection*{Signal Regions C}

SRC is optimized for direct top-squark pair production where $\Delta m(\stop,\ninoone)\approx m_t$, a regime in which the signal topology is similar to SM \ttbar\ production. In the presence of high-momentum ISR, which can be reconstructed as multiple jets forming an ISR system, the di-top-squark system is boosted in the transverse plane. The ratio of the \met\ to the \pt\ of the ISR system in the centre-of-mass (CM) frame of the entire (ISR plus di-top-squark) system (\pTISR), defined as \rISR, is proportional to the ratio of the $\ninoone$ and $\stop$ masses~\cite{An,Macaluso}:
\begin{equation*}
\rISR \equiv \frac{\met}{\pTISR} \sim \frac{m_{\ninoone}}{m_{\stop}}. 
\end{equation*}
A ``recursive jigsaw reconstruction technique'', as described
in Ref.~\cite{RJR_ISR}, is used to divide each event into an ISR hemisphere
and a sparticle hemisphere, where the latter consists of the pair of
candidate top squarks, each of which decays via a top quark and a $\ninoone$. Objects are grouped together based on their proximity in
the lab frame's transverse plane by minimizing the reconstructed
transverse masses of the ISR system and sparticle system simultaneously over all choices of object assignment. Kinematic variables are then defined based on this assignment of objects to either the ISR system or the sparticle system.
This method is equivalent to grouping the event objects according to the axis of maximum back-to-back \pt\ in the event's CM frame where the \pt\ of all accepted objects sums vectorially to zero. In events with a high-\pT\ ISR gluon, the axis of maximum back-to-back \pt, also known as the thrust axis, approximates the direction of the ISR and sparticles' back-to-back recoil.

The selection criteria for this signal region are summarized in
Table~\ref{tab:SignalRegionC}. The events are divided into five
windows (SRC1--5) defined by non-overlapping ranges of the reconstructed $\RISR$, which target
different top-squark and $\ninoone$ masses: e.g., SRC2 is optimized
for $\mstop = 300\GeV$ and $\mLSP=127\GeV$, and SRC4 is optimized for $\mstop
= 500\GeV$ and $\mLSP=327\GeV$. At least five jets must be
assigned to the sparticle hemisphere of the event (\nJetS), and at least one of those jets (\nBJetS) must be $b$-tagged. Transverse-momentum requirements on \pTISR, the highest-\pt\ $b$-jet in the sparticle hemisphere (\pTSBZero), and the fourth-highest-\pt\ jet in the sparticle hemisphere (\pTSFour) are applied. The transverse mass formed by the sparticle system and the \met, defined as \mS, is required to be $> 300\GeV$. The ISR system is also required to be separated in azimuth from the \ptmiss\ in the CM frame; this variable is defined as $\dPhiISRMET$.
Similarly to the categories defined for SRA and SRB, the individual SRCs are statistically combined to improve signal sensitivity.

\begin{table}[htpb]
  \caption{Selection criteria for SRC, in addition to the common preselection requirements described in the text. The signal regions are separated into windows based on ranges of $\RISR$.}
  \begin{center}
    \def\arraystretch{1.4}
    \begin{tabular}{c||c|c|c|c|c} \hline\hline
      {\bf Variable} & SRC1 & SRC2 & SRC3 & SRC4 & SRC5 \\ \hline \hline
       \nBJet & \multicolumn{5}{c}{$\ge1$} \\ \hline
      \nBJetS & \multicolumn{5}{c}{$\ge1$} \\ \hline
      \nJetS & \multicolumn{5}{c}{$\ge5$}  \\ \hline
      \pTSBZero & \multicolumn{5}{c}{$>40\gev$}  \\ \hline
      \mS & \multicolumn{5}{c}{$>300\gev$}  \\ \hline
      \dPhiISRMET & \multicolumn{5}{c}{$>3.0$}  \\ \hline
      \pTISR & \multicolumn{5}{c}{$>400$ GeV}   \\ \hline
      \pTSFour & \multicolumn{5}{c}{$>50$ GeV}   \\ \hline
      \rISR & 0.30--0.40 & 0.40--0.50 & 0.50--0.60 & 0.60--0.70 & 0.70--0.80\\  \hline \hline
    \end{tabular}
  \end{center}
  \label{tab:SignalRegionC}
\end{table}

\subsubsection*{Signal Regions D}

SRD is optimized for direct top-squark pair production where both top squarks decay via $\stop\to b \chinoonepm$ where $m_{\chinoonepm}=2\mLSP$. In this signal region, at least five jets are required, two of which must be $b$-tagged. The scalar sum of the transverse momenta of the two jets with the highest $b$-tagging weights (\ptbzero+\ptbone) as well as the second (\ptone), fourth (\ptthree), and fifth (\ptfour) jet transverse momenta are used for additional background rejection. Subregions SRD-low and SRD-high are optimized for $\mstop = 400\GeV$ with $\mLSP=50\GeV$, and $\mstop = 700\GeV$ with $\mLSP=100\GeV$, respectively. 
Tighter leading and sub-leading jet $\pT$ requirements are made for SRD-high, as summarized in Table~\ref{tab:SignalRegionD}. 

\begin{table}[!htb]
  \caption{Selection criteria for SRD, in addition to the common preselection requirements described in the text.}
  \begin{center}
  \def\arraystretch{1.4}
  \begin{tabular}{c||c|c}
    \hline\hline
    {\bf Variable}       & {\bf SRD-low} & {\bf SRD-high} \\
    \hline \hline
    \dphijetthreemet     & \multicolumn{2}{c}{$>0.4$}     \\ \hline
    \nBJet      & \multicolumn{2}{c}{$\geq$2}    \\\hline
    \drbjetbjet     & \multicolumn{2}{c}{$>$ 0.8}    \\ \hline
    \ptbzero+\ptbone & $>300$ GeV    & $>400$ GeV     \\ \hline
    $\tau$-veto          & \multicolumn{2}{c}{yes}        \\ \hline
    \ptone\              & \multicolumn{2}{c}{$>150\GeV$} \\ \hline
    \ptthree\            & $>100\GeV$    & $>80\GeV$      \\ \hline
    \ptfour\             & \multicolumn{2}{c}{$>60\GeV$}  \\ \hline
    \mtbmin\             & $>250\GeV$    & $>350\GeV$     \\ \hline
    \mtbmax\             & $>300\GeV$    & $>450\GeV$     \\ 
    \hline\hline
  \end{tabular}
  \end{center}
  \label{tab:SignalRegionD}
\end{table}

\subsubsection*{Signal Region E}

SRE is designed for models which have highly boosted top quarks. Such signatures can arise from direct pair production of high-mass top partners, or from the gluino-mediated compressed \stop\ scenario with large $\Delta m(\gluino,\stop)$. In this regime, reclustered jets with $R=0.8$ are utilized to optimize the experimental sensitivity to these highly boosted top quarks. In this signal region, at least two jets out of the four or more required jets must be $b$-tagged. Additional discrimination is provided by the $\met$ significance: $\htsig$, where $\HT$ is the scalar sum of the $\pT$ of all reconstructed $R=0.4$ jets in an event. The selection criteria for SRE, optimized for $m_{\gluino} = 1700 \GeV, \mstop=400\GeV$, and $\mLSP=395\GeV$, are summarized in Table~\ref{tab:SignalRegionE}.

\begin{table}[!htb]
  \caption{Selection criteria for SRE in addition to the common preselection requirements described in the text.}
  \begin{center}
  \def\arraystretch{1.4}
  \begin{tabular}{c||c}
    \hline\hline
    {\bf Variable}    & {\bf SRE}          \\
    \hline \hline
    \dphijetthreemet  & $>0.4$             \\ \hline
    \nBJet   & $\geq$2            \\     \hline
    \mantikteightzero & $>120$ \gev        \\     \hline
    \mantikteightone  & $>80$ \gev         \\     \hline
    \mtbmin\          & $>200$ \gev        \\     \hline
    \met\             & $> 550 \gev$       \\     \hline
    \HT               & $>800 \gev$        \\     \hline
    \htsig            & $> 18 \sqrt{\GeV}$ \\     
\hline\hline
  \end{tabular}
  \end{center}
  \label{tab:SignalRegionE}
\end{table}



\section{Background estimation}
\label{sec:background}

The main SM background process in SRA, SRB, SRD, and SRE is $\Znunu$ production in association with heavy-flavour jets. The second most significant background is \ttbar\ production where one
\Wboson\ boson decays via a lepton and neutrino and the lepton (particularly a hadronically decaying $\tau$
lepton) is either not identified or is reconstructed as a jet. This
process gives the major background contribution in SRC and an important
background in SRB, SRD and SRE as well. Other important background
processes are $\Wboson\to\ell\nu$ plus heavy-flavour jets, single top quark, and the
irreducible background from $\ttZ$, where the $\Zboson$ boson decays into two neutrinos. 

The main background contributions are estimated primarily from
comparisons between data and simulation outside the signal regions. Control regions (CRs) are
designed to enhance a particular background process, and are orthogonal to the SRs while probing a similar event topology. The CRs are used to normalize the simulation to data, but extrapolation from the CR to the SR is taken from simulation. Sufficient data are needed to avoid large statistical uncertainties in the background estimates, and the CR definitions are chosen to be
kinematically as close as possible to all SRs, to minimize the
systematic uncertainties associated with extrapolating the background
yield from the CR to the SR. Where CR definitions are farther from the SR definition, validation regions are employed to cross-check the extrapolation. In addition, control-region selection criteria
are chosen to minimize potential contamination from signal that could shadow contributions in the signal regions. The signal contamination is below 8\% in all CRs for all signal points that have not been excluded by previous ATLAS searches. No significant difference in the background estimates was found between the case where only SM backgrounds were considered and when signal is included in the estimation. As the CRs are not $100\%$ pure in the process of
interest, the cross-contamination between CRs from other processes is
estimated. The normalization factors and the cross-contamination are determined simultaneously for all regions using a
fit described below. 

Detailed CR definitions are given in Tables~\ref{tab:selectionCRZs},~\ref{tab:selectionCRTs}, and~\ref{tab:selectionCRWSTTTGamma}.
They are used for the \Zboson\ (CRZs), \ttbar\ (CRTs), \Wboson\ (CRW), single top (CRST), and \ttZ\ (CRTTGamma) background estimation.
The \dphijetthreemet\ and
$\mT(\ell,\met)$ requirements are designed
to reduce contamination from SM multijet processes
. The number of leptons (from this point on, lepton is used to mean electron or muon) is
indicated by $N_{\ell}$ and the transverse momentum of the lepton is
indicated by $\pT^{\ell}$. In all one-lepton CRs, once the trigger and
minimum $\pT^{\ell}$ selection are applied, the lepton is treated as a
non-$b$-tagged jet (to emulate the hadronic $\tau$ decays in
the SRs) in the computation of all jet-related variables. In the
two-lepton CRZs, a lepton-\pt\ requirement of at least 28 GeV is made to ensure the trigger selection is fully efficient. 
The invariant mass of the two oppositely charged
leptons, denoted by $m_{\ell\ell}$, must be consistent
with the leptons having originated from a $\Zboson$ boson. The transverse
momenta of these leptons are
then vectorially added to the \ptmiss\ to mimic the $\Znunu$ decays in the SRs,
forming the quantity \metprime. Quantities that depend on the \met\ are recalculated in the CRZs using \metprime\ and identified by the addition of a prime (e.g. \mtbminprime\ and \mtbmaxprime). Requirements such as the maximum
$\mT(\ell,\met)$ and the minimum $\Delta R$ between the two
highest-weight $b$-tagged jets and the lepton, $\drblmin$, are used to
enforce orthogonality between CRT, CRW, and CRST. In CRST, the requirement on the $\Delta R$ between
 the two highest-weight $b$-tagged jets, $\drbjetbjet$, is used to reject
$\ttbar$ contamination from the control region enriched in single-top
events. Finally, the normalization of the \ttV\ background in the
signal region, which is completely dominated by $\ttbar+Z(\to\nu\nu)$,
is estimated with a $\ttbar+\gamma$ control region in a way similar to the method described in Ref.~\cite{GtcStop1L}. The
same lepton triggers and lepton-\pt\ requirements are used for the $\ttbar+\gamma$ control region as in the
CRZs. Additionally, the presence of an isolated photon with $\pt>150\gev$ is
required and it is used to model the \Zboson\ decay in the signal
regions because of the similarity between the diagrams for photon and \Zboson\ production. Similarly to the \Zboson\ control region, the photon is used in the estimation of \met-related variables. 

\begin{figure}[htpb]
  \begin{center}
    \subfloat[]{\includegraphics[width=0.44\textwidth]{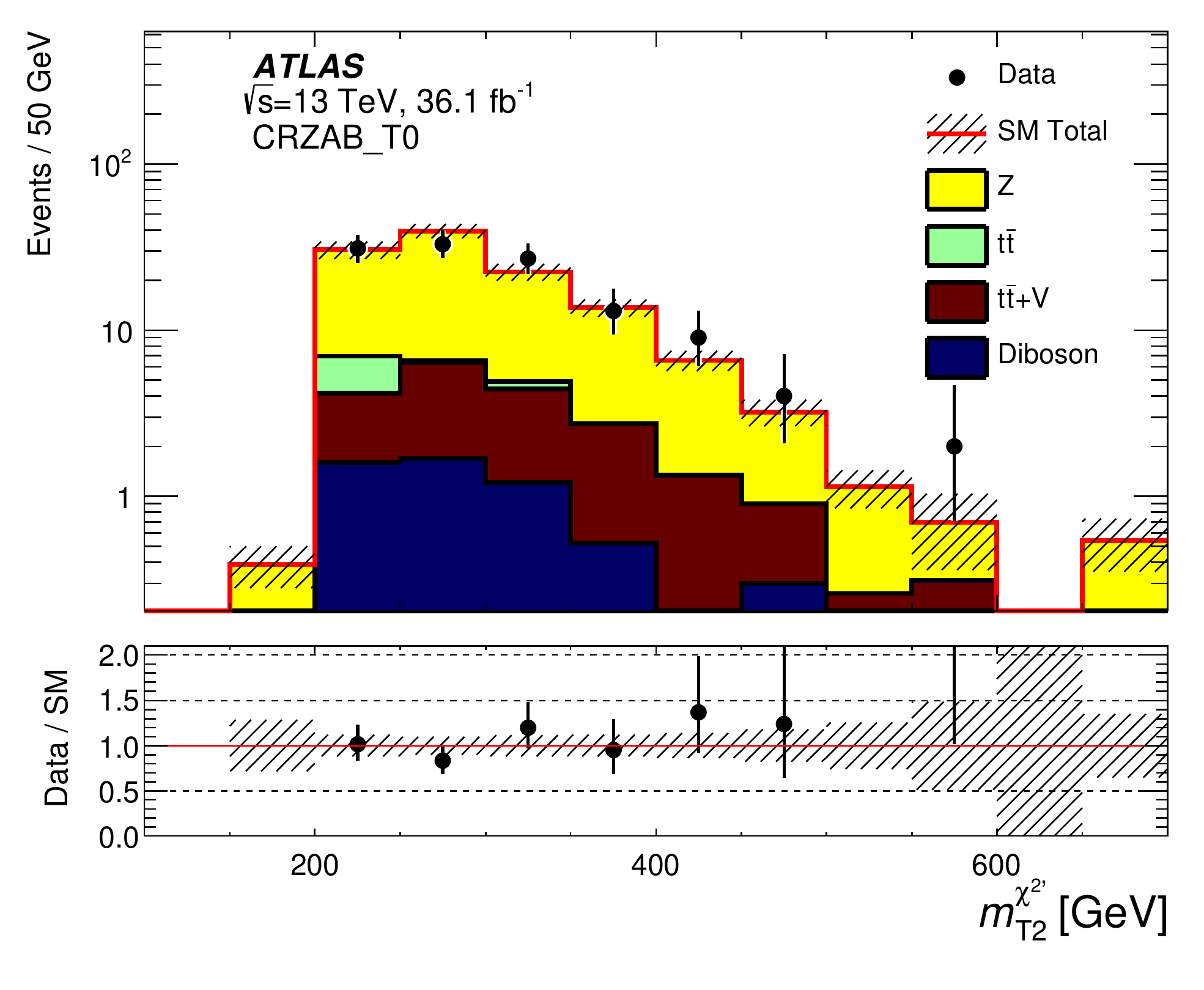}}
    \subfloat[]{\includegraphics[width=0.44\textwidth]{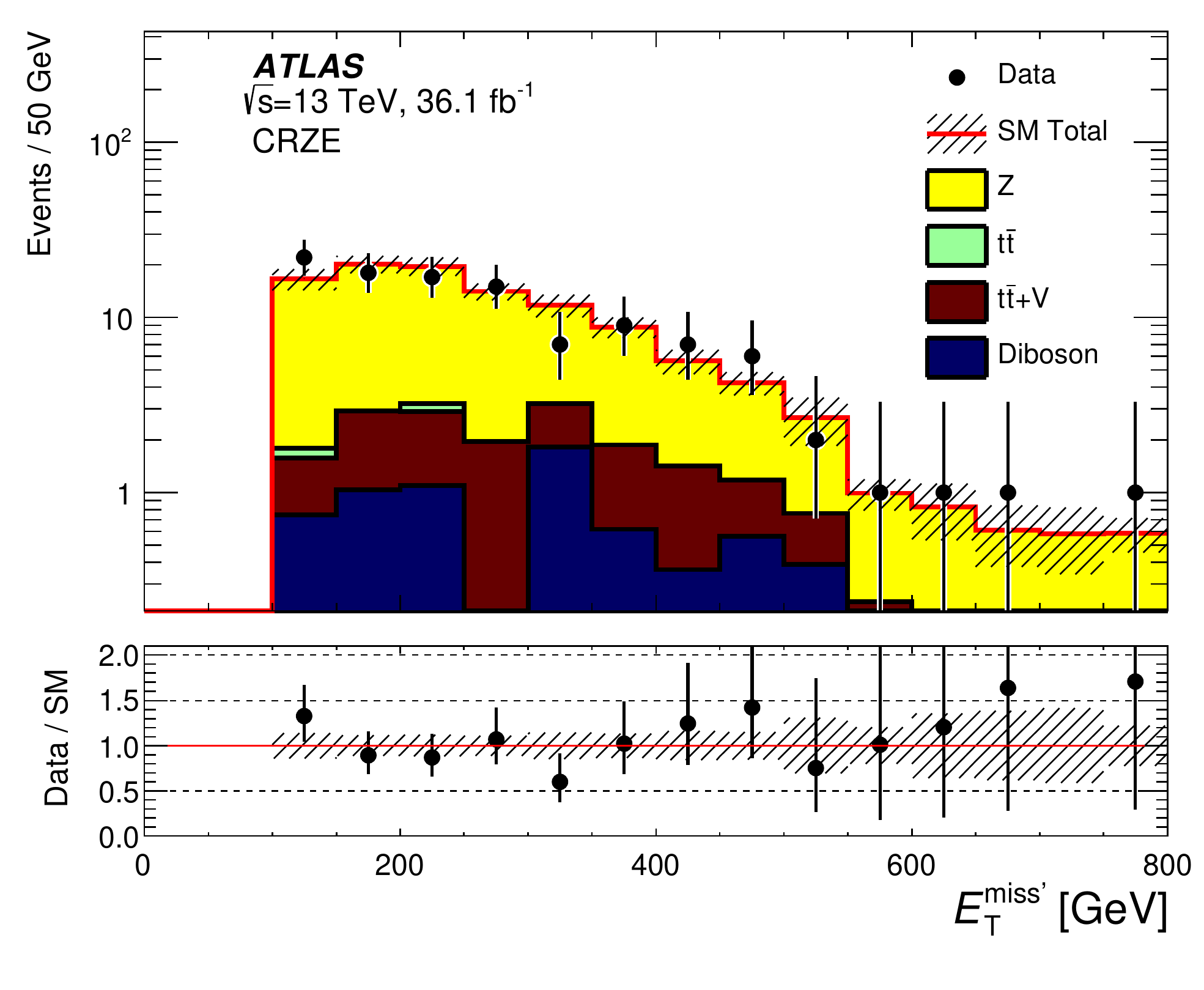}}\\
    \subfloat[]{\includegraphics[width=0.44\textwidth]{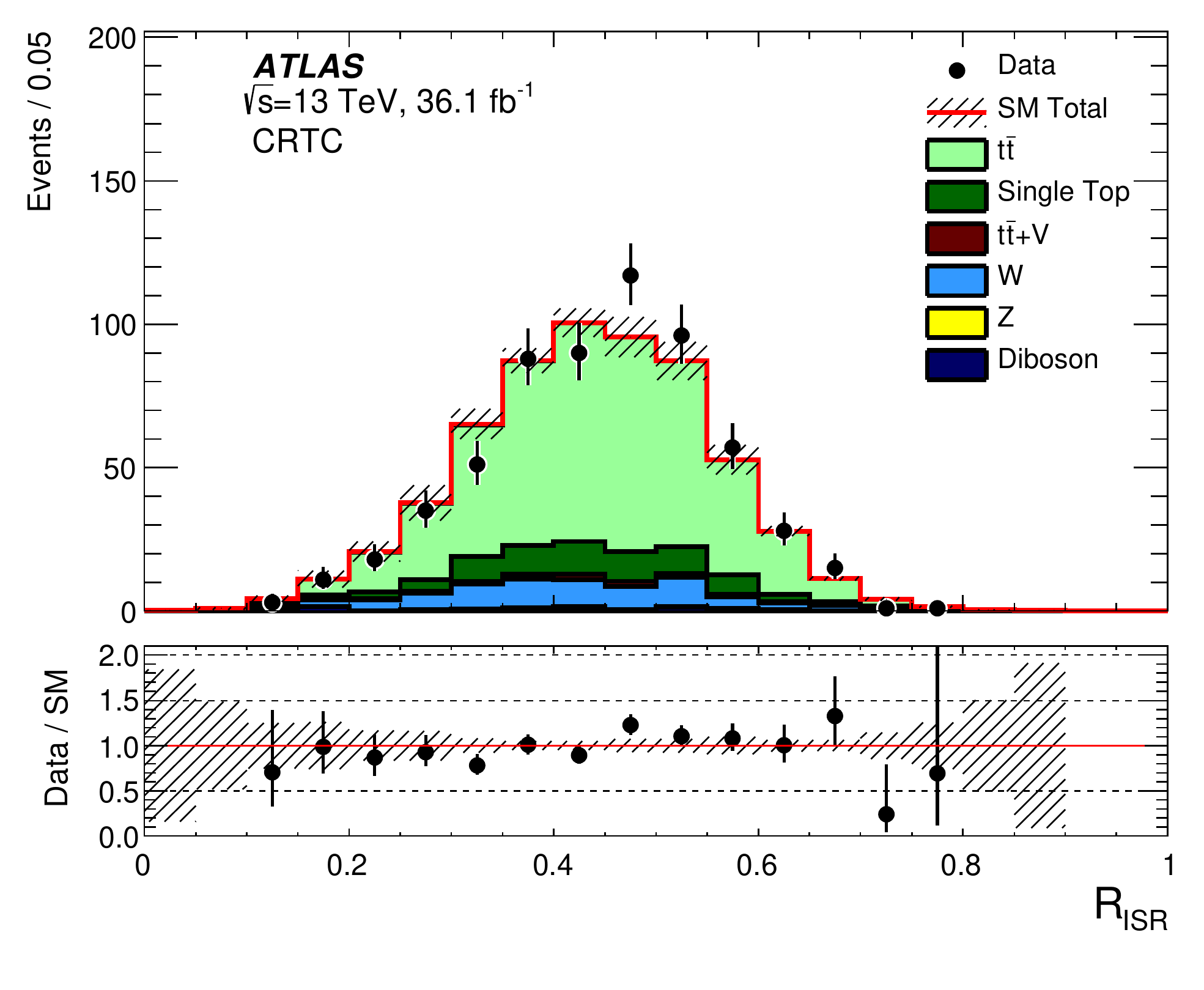}}
    \subfloat[]{\includegraphics[width=0.44\textwidth]{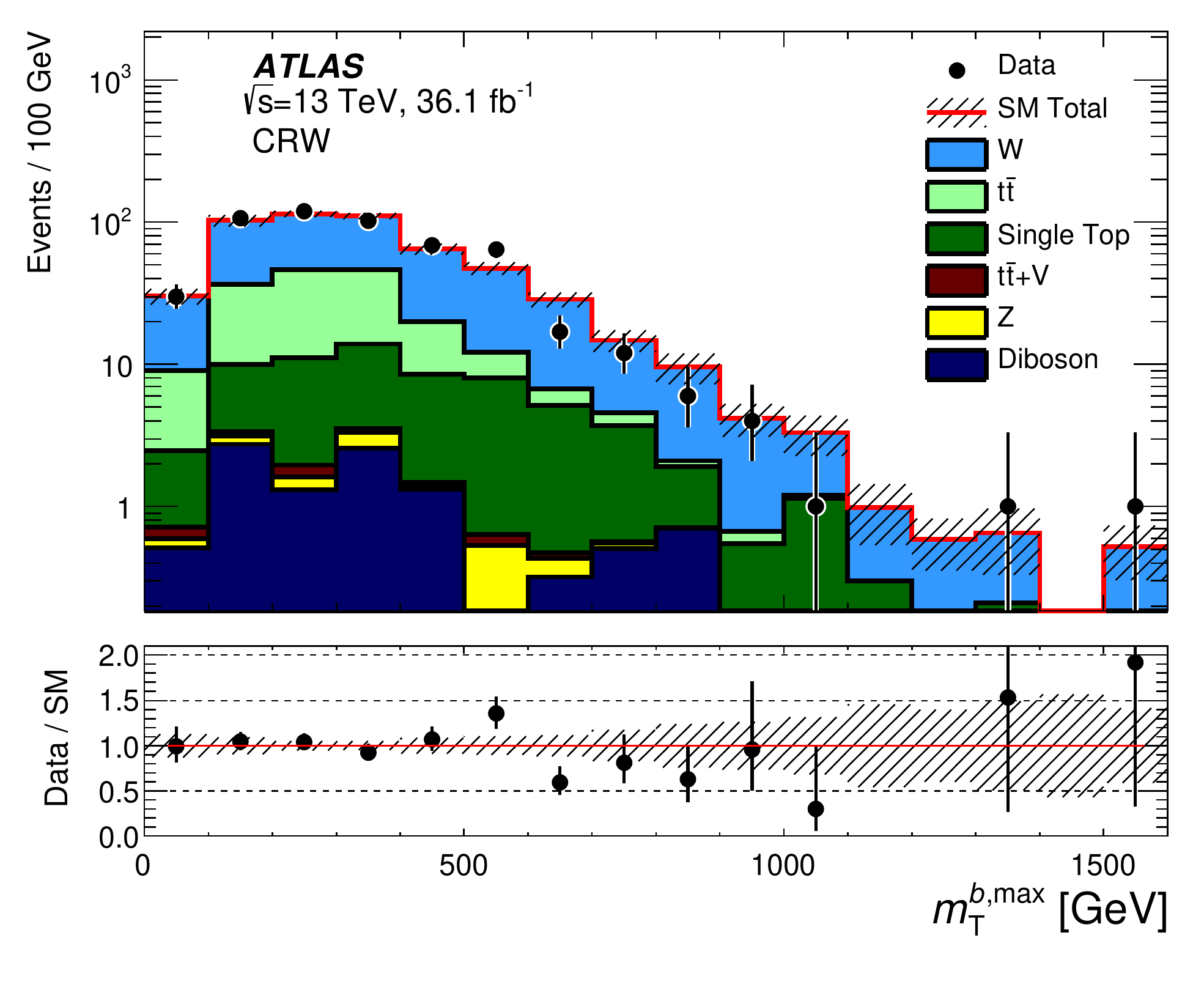}}\\
     \subfloat[]{\includegraphics[width=0.44\textwidth]{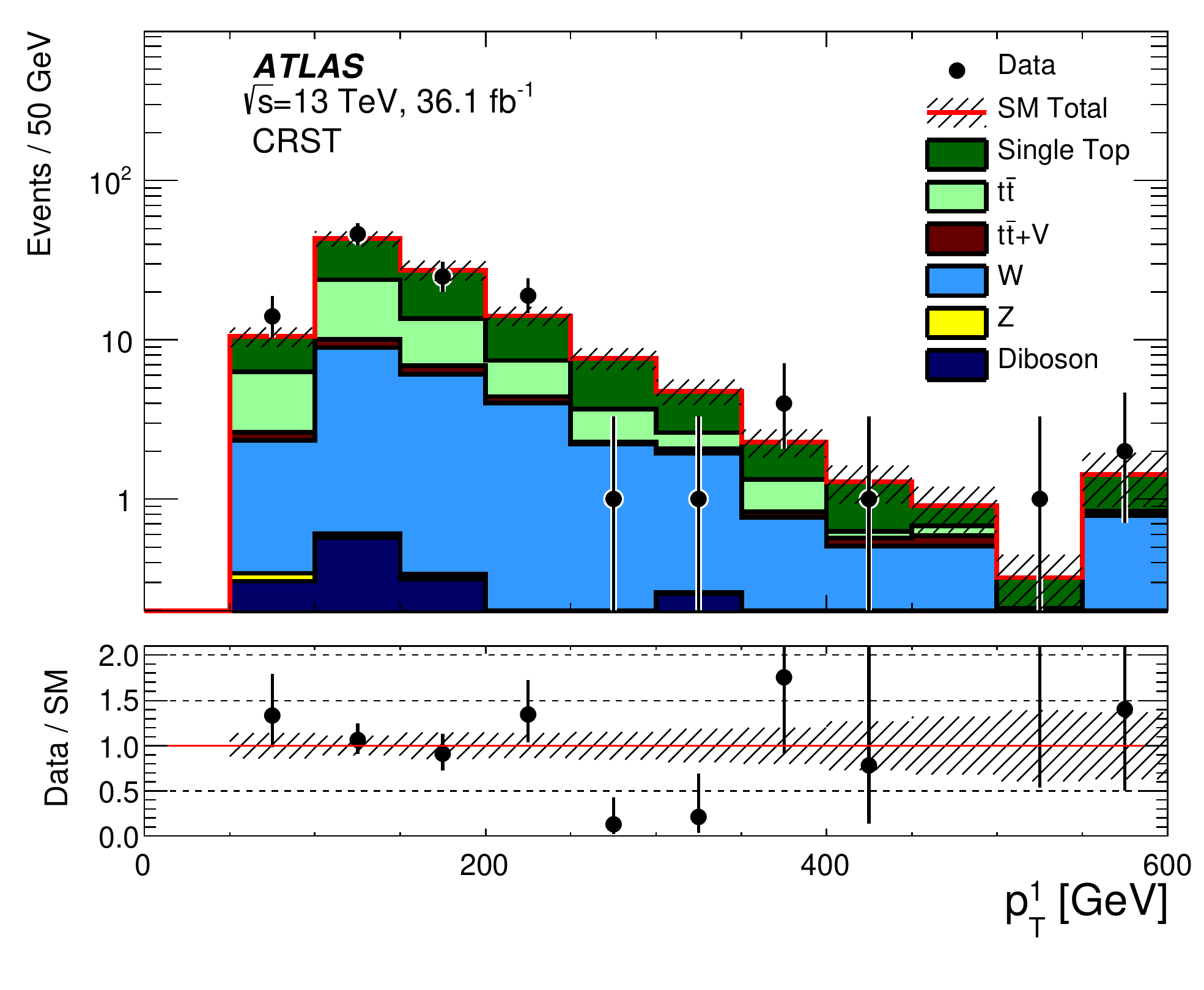}}
   \subfloat[]{\includegraphics[width=0.44\textwidth]{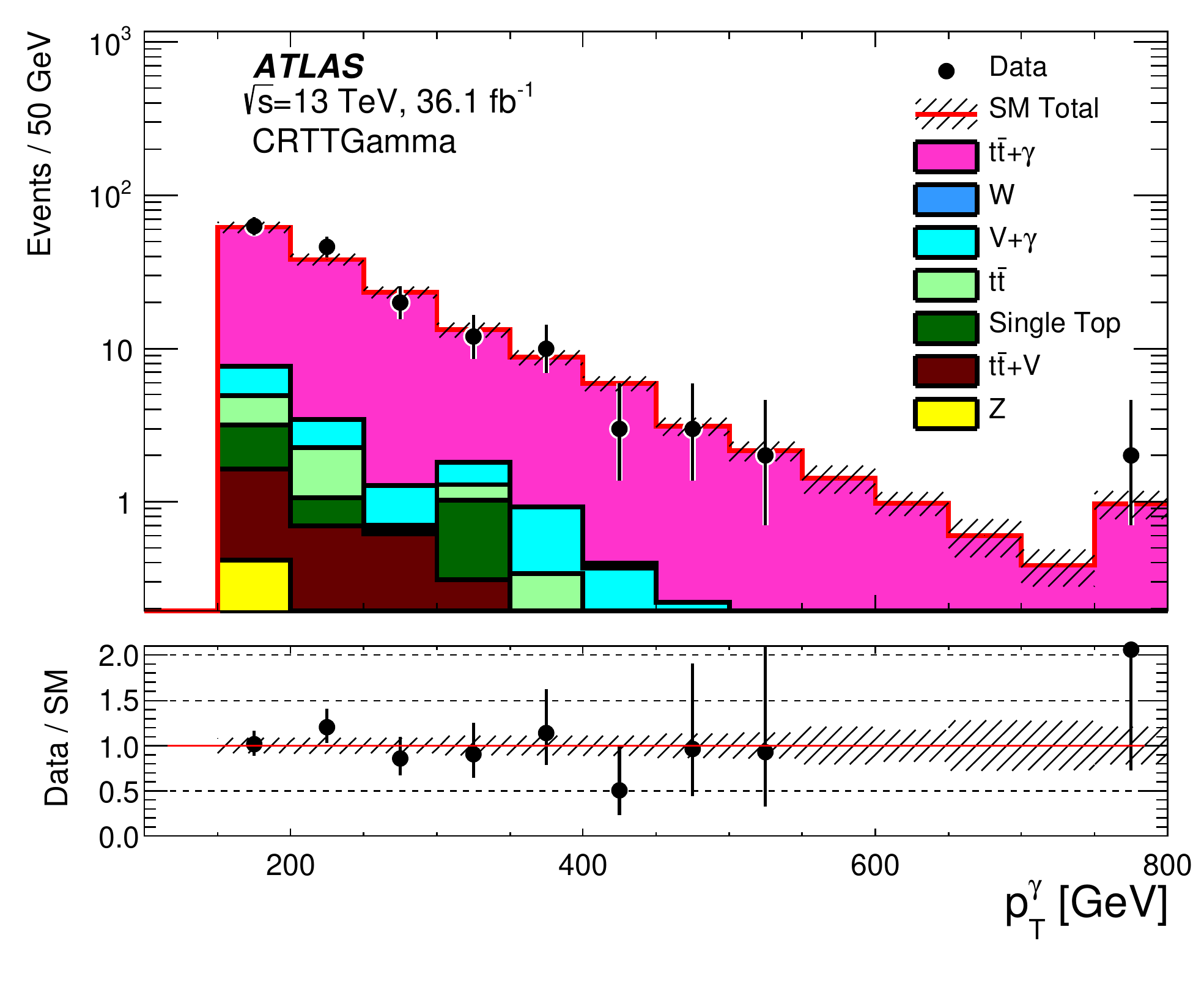}}

    \caption{Distributions of (a)~\mttwoprime\ in CRZAB-T0, (b)~\metprime\ in CRZE, (c)~\rISR\ in CRTC, 
      (d)~\mtbmax\ in CRW, (e)~the transverse momentum of the second-leading-\pT\ jet in CRST,  and
      (f)~the photon \pT\ in CRTTGamma. The stacked histograms show the SM prediction, normalized using scale factors derived from the simultaneous fit to all backgrounds. The ``Data/SM" plots show the ratio of data events to the total SM prediction. The hatched uncertainty band around the SM prediction and in the ratio plot illustrates the combination of MC statistical and detector-related systematic uncertainties. The rightmost bin includes overflow events.}
    \label{fig:CRs}
  \end{center}
\end{figure}

To estimate the \Zjets\ and \ttbar\ background in the different kinematic regions of the signal regions, individual control regions are designed for all signal regions where possible. Only if the statistical power of control regions is low, are they merged to form one control region for multiple signal regions. In the case of  CRST, CRW, and CRTTGamma, this results in the use of one common CR for all signal regions. Distributions from the $\Zjets$, $\ttbar$, $\Wjets$, single top, and $\ttbar\gamma$ control regions are shown in Figure~\ref{fig:CRs}.

\begin{table}[htpb]
  \caption{Selection criteria for the $\Zjets$ control regions used to estimate the $\Zjets$ background contributions in the signal regions.} 
  \begin{center}
    \def\arraystretch{1.4}
    \begin{tabular}{c||c|c|c|c}
      \hline \hline
      Selection           & CRZAB-TT-TW                    & CRZAB-T0     & CRZD & CRZE           \\ \hline \hline
     Trigger              & \multicolumn{4}{c}{electron or muon}                                   \\ 
     \hline
     $N_{\ell}$           & \multicolumn{4}{c}{2, opposite charge, same flavour}           \\ 
     \hline
     $\pT^{\ell}$         & \multicolumn{4}{c}{$>28 \gev$}                                        \\ 
     \hline
     $m_{\ell\ell}$       & \multicolumn{4}{c}{[86,96] \gev}                                      \\ 
     \hline
     $N_{\mathrm{jet}}$   & \multicolumn{4}{c}{$\ge 4$}                                           \\
      \hline     
      \ptzero, \ptone, \pttwo, \ptthree            & \multicolumn{4}{c}{$80,80,40,40\gev$}                              \\
      \hline
      $\met$              & \multicolumn{4}{c}{$<50 \gev$}                                        \\ 
      \hline
      $\metprime$         & \multicolumn{4}{c}{$ > 100$ \gev}                                     \\
      \hline
     \nBJet     & \multicolumn{4}{c}{$\ge 2 $}                                          \\
      \hline
       \mantikttwelvezero & \multicolumn{2}{c|}{$>120\gev$} & \multicolumn{2}{c}{-}                \\ 
      \hline
       \mantikttwelveone  & $>60\gev$                      & $<60\gev$    & \multicolumn{2}{c}{-} \\ 
       \hline
       \mtbminprime       & \multicolumn{2}{c|}{-}          & \multicolumn{2}{c}{$>200\,$\gev}     \\
      \hline  
      \mtbmaxprime        & \multicolumn{2}{c|}{-}          & $>200\,$\gev & -                     \\
      \hline
      \HT                 & \multicolumn{3}{c|}{-}          & $>500\,$\gev                         \\
       \hline\hline
    \end{tabular}
  \end{center}
  \label{tab:selectionCRZs}
\end{table}

\begin{landscape}
\begin{table}[htpb]
  \caption{Selection criteria for the \ttbar\ control regions used to estimate the \ttbar\ background contributions in the signal regions.} 
  \begin{center}
    \def\arraystretch{1.4}
\scriptsize
    \begin{tabular}{c||c|c|c|c|c|c|c|c|c}
      \hline \hline
      Selection  & CRTA-TT & CRTA-TW & CRTA-T0 & CRTB-TT & CRTB-TW & CRTB-T0 & CRTC & CRTD & CRTE \\ \hline \hline
      Trigger    & \multicolumn{9}{c}{\met}                                                                         \\ 
      \hline
      $N_{\ell}$ & \multicolumn{9}{c}{1}                                                                            \\ 
     \hline
     $\pT^{\ell}$ & \multicolumn{9}{c}{$>20 \gev$}   \\ 
     \hline
     $N_{\mathrm{jet}}$ & \multicolumn{9}{c}{$\ge 4$ (including
       electron or muon)} \\
      \hline     
      \ptzero, \ptone, \pttwo, \ptthree & \multicolumn{9}{c}{$80,80,40,40\gev$} \\
      \hline
      \nBJet & \multicolumn{9}{c}{$ \ge 2 $}\\
\hline   
      $\dphijettwomet$ & \multicolumn{9}{c}{$>0.4$} \\ 
      \hline
      $\dphijetthreemet$ & \multicolumn{6}{c|}{$>0.4$} & - & \multicolumn{2}{c}{$>0.4$}\\ 
      \hline
      $\mT(\ell,\met)$   &     \multicolumn{6}{c|}{$[30,100]\gev$} & $<100\gev$ & \multicolumn{2}{c}{$[30,100]\gev$}  \\ 
  \hline
       \mtbmin              & \multicolumn{6}{c|}{$>100\,$\gev} & - &  \multicolumn{2}{c}{$>100\,$\gev} \\ 
       \hline
       $\drblmin$ & \multicolumn{6}{c|}{$<1.5$} & $<2.0$ &  \multicolumn{2}{c}{$<1.5$}\\
      \hline
       \mantikttwelvezero      & \multicolumn{6}{c|}{$>120 \gev$} & \multicolumn{3}{c}{-} \\ 
      \hline
       \mantikttwelveone      & $>120 \gev$  & $[60, 120] \gev$  & $<60 \gev$  & $>120 \gev$  & $[60, 120] \gev$  & $<60 \gev$ & \multicolumn{3}{c}{-} \\ 
\hline
 \mantikteightzero      & \multicolumn{3}{c|}{$>60 \gev$} &  \multicolumn{5}{c|}{-} & $>120 \gev$ \\ 
      \hline
       \mantikteightone  &    \multicolumn{8}{c|}{-}  & $>80 \gev$  \\ 
\hline
 $\met$ & $ >250$ \gev & $ >300$ \gev & $ >350$ \gev & \multicolumn{6}{c}{$>250 \gev$} \\  
       \hline
       $\drbjetbjet$                       & $>1.0$ & \multicolumn{2}{c|}{-}& \multicolumn{3}{c|}{$>1.2$} & - & $>0.8$ & - \\
       \hline
       $\mtbmax$                        & \multicolumn{3}{c|}{-}& \multicolumn{3}{c|}{$>200\gev$} & - & $>100\gev$ & - \\
       \hline
       $\ptone$                        & \multicolumn{7}{c|}{-} & $>150\gev$ & - \\
       \hline
       $\ptthree$                        & \multicolumn{7}{c|}{-} & $>80\gev$ & - \\
       \hline
       $\ptbzero+\ptbone$                        & \multicolumn{7}{c|}{-} & $>300\gev$ & - \\
       \hline
       $\NjV$         & \multicolumn{6}{c|}{-}& $ \ge 5$ & \multicolumn{2}{c}{-} \\
      \hline
       $\NbV$         & \multicolumn{6}{c|}{-}& $ \ge 1$ & \multicolumn{2}{c}{-} \\
      \hline
       $\PTISR$         & \multicolumn{6}{c|}{-}& $>400\gev$ & \multicolumn{2}{c}{-} \\
\hline
       $\pTSFour$         & \multicolumn{6}{c|}{-}& $>40\gev$ & \multicolumn{2}{c}{-} \\
\hline
       $\HT$         & \multicolumn{8}{c|}{-}& $>500\gev$ \\

       \hline\hline
    \end{tabular}
  \end{center}
  \label{tab:selectionCRTs}
\end{table}
\end{landscape}

\begin{table}[htpb]
  \caption{Selection criteria for the common \Wjets, single-top, and $\ttbar+\gamma$ control-region definitions.} 
  \begin{center}
    \def\arraystretch{1.4}
    \begin{tabular}{c||c|c|c}
      \hline \hline
      Selection                       & CRW                                             & CRST   & CRTTGamma            \\ \hline \hline
      Trigger                         & \multicolumn{2}{c|}{\met}                       & electron or muon                        \\ 
      \hline
      $N_{\ell}$                      & \multicolumn{3}{c}{1}                                                           \\ 
     \hline
     $\pT^{\ell}$                     & \multicolumn{2}{c|}{$>20 \gev$}                 & $>28\gev$                     \\ 
     \hline
      $N_{\gamma}$                    & \multicolumn{2}{c|}{-}                          & 1                             \\ 
     \hline
     $\pT^{\gamma}$                   & \multicolumn{2}{c|}{-}                          & $>150\gev$                    \\ 
     \hline
     $N_{\mathrm{jet}}$               & \multicolumn{2}{c|}{$\ge 4$ (including electron or muon)} & $\ge4$                        \\
      \hline     
      \ptzero, \ptone,\pttwo,\ptthree & \multicolumn{3}{c}{$80,80,40,40\gev$}                                           \\
      \hline
      \nBJet                          & $1$                                             & \multicolumn{2}{c}{$ \ge 2 $} \\
\hline   
      $\dphijettwomet$                & \multicolumn{2}{c|}{$>0.4$}                     & -                             \\ 
      \hline
      $\mT(\ell,\met)$                & \multicolumn{2}{c|}{$[30,100]\gev$}             & -                             \\ 
\hline
       $\drblmin$                     & \multicolumn{2}{c|}{$>2.0$}                     & -                             \\ 
\hline
 $\met$                               & \multicolumn{2}{c|}{$>250 \gev$}                & -                             \\ 
\hline
 $\drbjetbjet$                        & -                                               & $>1.5$ & -                    \\ 
\hline
 $\mantikttwelvezero$                 & $<60\gev$                                               & $>120\gev$ & -                    \\ 
\hline
 $\mtbmin$                 & -                                               & $>200\gev$ & -                    \\ 
       \hline\hline
    \end{tabular}
  \end{center}
  \label{tab:selectionCRWSTTTGamma}
\end{table}

Contributions from all-hadronic \ttbar\ and multijet production are found to be negligible. These are 
estimated from data using a procedure described in
Ref.~\cite{jetSmearing}. The procedure determines the jet response from simulated dijet events, and then uses this response function to smear the jet response in low-\met\ events. The jet response is cross-checked with data where the \met\ can
be unambiguously attributed to the mismeasurement of one of the
jets. Diboson production, which is also subdominant, is estimated directly from simulation.

\subsubsection*{Simultaneous fit to determine SM background}
\label{sec:simultaneousfit}

The observed numbers of events in the various control regions are
included in a binned profile likelihood fit~\cite{likelihoodFit} to determine the SM background
estimates for \Zboson, \ttbar, \Wboson, single top, and \ttZ\ in each signal region. 
The normalizations of these backgrounds are determined simultaneously to best match the observed data in each control region, taking contributions from all backgrounds into account. A likelihood function is built as the
product of Poisson probability density functions, describing the observed and
expected numbers of events in the control regions~\cite{histFitter}. This procedure takes common
systematic uncertainties (discussed in
Section~\ref{sec:Systematics}) between the control and signal regions and
their correlations into account as they are treated as nuisance
parameters in the fit and are modelled by Gaussian probability density
functions. 
The contributions from all other background processes (dibosons and multijets) are fixed at the values expected from
the simulation, using the most accurate theoretical cross sections
available, as described in Section~\ref{sec:simulation}, while their uncertainties are used as nuisance parameters in the fit.

Zero-lepton VRs (VRZAB, VRZD, VRZE) are designed to validate the background estimate for
\Zjets\ in the signal regions. No VRZ is designed for SRC due to the negligible contribution of the \Zboson\ background in this region. The definitions of the VRZs, after the common zero-lepton preselection discussed in Section~\ref{sec:signalregions} is applied, are shown in Table~\ref{tab:selectionVRZs}. To provide orthogonality to the signal regions, the requirement on one or more of the following variables is inverted: \drbjetbjet, \mantikttwelvezero, \mantikteightzero.

To validate the \ttbar\ background, zero-lepton VRs sharing the same common preselection of the signal regions and which are close to the SRA and SRB definitions are designed for each of the categories (VRTA-TT, VRTA-TW, VRTA-T0, VRTB-TT, VRTB-TW, VRTB-T0). To avoid overlap with the signal regions the \mtbmin\ requirement is inverted in all validation regions. In VRTA, SRA requirements remain unchanged except for \mttwo\ not being applied, $100<\mtbmin<200\gev$, and the \met\ requirement being reduced by 100 GeV. For VRTB, all requirements in the VRs are the same as in the SRs except for \mtbmin, which is $100<\mtbmin<200\gev$ for VRTB-TT, $140<\mtbmin<200\gev$ for VRTB-TW, and $160<\mtbmin<200\gev$ for VRTB-T0. For SRC, the same requirements are used when defining the validation region (VRTC) except for the looser requirements of $\mS> 100\gev$, $\pTSFour>40\gev$ and $\nJetS > 4$. The \dPhiISRMET\ requirement is inverted and $\mV/\mS<0.6$, where \mV\ is the transverse mass defined by the visible objects of the sparticle system and the \met, is applied in addition to the existing selection. The validation region to validate the background estimates in SRD (VRTD) is formed by applying the following requirements: $100<\mtbmin<200\gev$, $\ptbzero+\ptbone>300\gev$, $\ptthree>80$ GeV, and $\mtbmax>300$ GeV. All other requirements are applied exactly as in SRD-low except for the requirement on \ptfour\ which is dropped. Finally, the validation region defined for SRE (VRTE) applies only the same requirements on the number of $b$-jets, \mantikteightzero, and \mantikteightone, and inverts the \mtbmin\
requirement to $100<\mtbmin<200\gev$. No other requirement is applied to VRTE. 

A one-lepton validation region for the \Wjets\ background (VRW) is used to test the \Wboson\ background estimates in all SRs. In this case the validation region is designed based on the definition of CRW. Compared to CRW,
the requirement that differs is \mindrblep, which is greater than 1.8 for the validation
region. Two additional requirements are included in the definition of VRW, namely $\mtbmin>150$ GeV and $\mantikttwelvezero<70$ GeV. 

\begin{figure}[htpb]
  \begin{center}
    \includegraphics[width=0.8\textwidth]{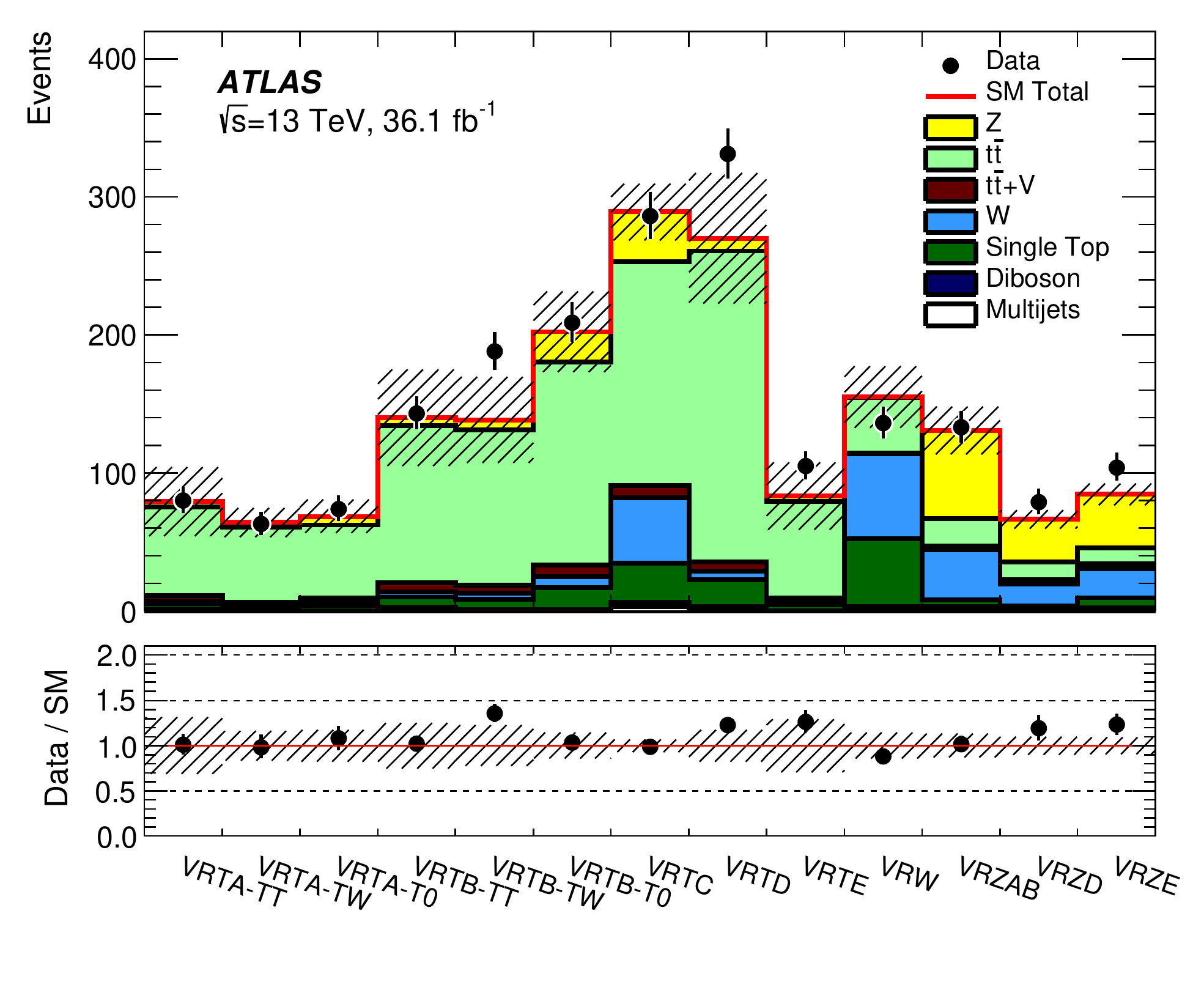}
    \caption{Yields for all validation regions after the likelihood fit. The stacked histograms show the SM prediction and the hatched uncertainty band around the SM prediction shows the total uncertainty, which consists of the MC statistical uncertainties, detector-related systematic uncertainties, and theoretical uncertainties in the extrapolation from CR to VR.} 
    \label{fig:VRs}
  \end{center}
\end{figure}

Signal contamination in all the validation regions for all considered signals that have not yet been excluded was also checked. The largest  contamination found is $\sim$25\% and occurs in the VRTs for top-squark masses below 350~\GeV\ and in VRZD and VRZE near top-squark masses of 700~\GeV.
The result of the simultaneous fit procedure, which is repeated with the VRs used as test signal regions, for each VR is shown in Figure~\ref{fig:VRs}, which displays agreement between data and MC predictions.

\begin{table}[htpb]
  \caption{Selection criteria for the \Zboson\ validation regions used
    to validate the \Zboson\ background estimates in the signal regions.} 
  \begin{center}
    \def\arraystretch{1.4}
    \begin{tabular}{c|c|c|c}
      \hline \hline
      Selection           & VRZAB                    & VRZD           & VRZE          \\
      \hline \hline
      Jet \ptzero, \ptone & $>80, >80 $ GeV            & $>150, >80 $ GeV & $>80, >80 $ GeV \\
      \hline
      $N_{\mathrm{jet}}$     & $\ge4$                   & $\ge5$         & $\ge4$        \\
      \hline
      \nBJet              & \multicolumn{3}{c}{$ \ge2 $}                              \\
      \hline
      $\tau$-veto         & \multicolumn{2}{c|}{yes} & no                             \\
      \hline
      \mtbmin             & \multicolumn{3}{c}{$>200$ GeV}                            \\
      \hline 
      \mantikttwelvezero  & $ <120 $ GeV             & \multicolumn{2}{c}{-}          \\
      \hline 
      \drbb               & $<1.0$                   & $<0.8$         & $<1.0$        \\
      \hline 
      \mtbmax             & -                        & $>200$ GeV     & -             \\
      \hline 
      \HT                 & \multicolumn{2}{c|}{-}   & $>500$ GeV                     \\
      \hline
      \htsig              & \multicolumn{2}{c|}{-}   & $>14$ $\sqrt{\gev}$            \\
      \hline 
      \mantikteightzero   & \multicolumn{2}{c|}{-}   & $ <120 $ GeV                   \\ 
      \hline\hline
    \end{tabular}
  \end{center}
  \label{tab:selectionVRZs}
\end{table}

\section{Systematic uncertainties}
\label{sec:Systematics}

Experimental and theoretical systematic
uncertainties in the SM predictions and signal
predictions are included in the profile likelihood fit described
in Section~\ref{sec:background}.

Statistical uncertainties dominate the total uncertainties of the background predictions in all SRs except SRB. 
The dominant systematic uncertainties for SRA and SRB are shown in Table~\ref{tab:srABSysts} while the systematic uncertainties for the remaining SRs are shown in Table~\ref{tab:srCDESysts}. The uncertainties are shown as a relative uncertainty to the total background estimate. The main sources of detector-related systematic uncertainty in the
SM background estimates are the jet energy scale
(JES) and jet energy resolution (JER), $b$-tagging efficiency, \met\ soft term, and pile-up. The effect of the JES and JER uncertainties on the
background estimates in the signal regions can reach 17\%. The uncertainty in the $b$-tagging
efficiency is nowhere more than 9\%. All jet- and lepton-related uncertainties are propagated
to the calculation of the \met, and additional uncertainties in the
energy and resolution of the soft term are also included~\cite{met}. The
uncertainty in the soft term of the \met\ is most significant in
SRC5 at 15\%. An uncertainty due to the pile-up modelling is also considered, with a contribution
up to 14\%. Lepton reconstruction and identification uncertainties are also considered but have a small impact.

The uncertainty in the combined 2015+2016 integrated luminosity is 3.2\%. It is derived, following a methodology similar to that detailed in Ref.~\cite{lumi}, from a preliminary calibration of the luminosity scale using $x$--$y$ beam-separation scans performed in August 2015 and May 2016.

Theoretical uncertainties in the modelling of the SM background are
estimated. For the \Vjets\ background processes, the modelling uncertainties
are estimated using \sherpa\ samples by varying the renormalization and
factorization scales, and the merging 
and resummation
scales (each varied up and down by a
factor of two). PDF uncertainties were found to have a negligible impact. The resulting impact on the total background yields from the
\Zjets\ theoretical uncertainties is up to 3\% while the uncertainties from the \Wjets\ sample variations are less than 3\%.  

For the \ttbar\ background, uncertainties are estimated from the
comparison of different matrix-element calculations, the choice of parton-showering model and
the emission of additional partons in the initial and final
states (comparing \textsc{Powheg-Box}+\pythia\ vs \herwig++ and \sherpa). More details are given in Ref.~\cite{top}. 
The largest impact of the \ttbar\ theory
systematic uncertainties on the total background yields arises for SRC and it
varies from 11\% to 71\% by tightening the \RISR requirement. For the \ttbar+$W/Z$ background, the theoretical uncertainty is
estimated through variations, in both \ttbar+$W/Z$ and $\ttbar\gamma$ MC simulation, including the choice of
renormalization and factorization scales (each varied up and down by a
factor of two), the choice of PDF, as well as a comparison between \mcatnlo\ and OpenLoops+\sherpa\ generators, resulting in a maximum uncertainty of 2\% in SRA-TT.
The single-top background is dominated by the $Wt$ subprocess. Uncertainties are estimated for
the choice of parton-showering model (\pythia\ vs \herwig++) and
for the emission of additional partons in the initial- and final-state
radiation. A 30\% uncertainty is assigned to the single-top background estimate to account for the effect of
interference between single-top-quark and \ttbar\ production. This uncertainty is estimated by comparing yields in the signal and control 
regions for a sample that includes resonant and non-resonant $WW$+$bb$ production with the sum of the yields of 
resonant \ttbar\ and single-top+$b$ production. The
final single-top uncertainty relative to the total background estimate
is up to 12\%. The detector systematic uncertainties are also applied to the signal
samples used for interpretation. Theoretical uncertainties in the
signal cross section as described in Section~\ref{sec:simulation} are
treated separately and limits on top-squark and neutralino masses are given for the $\pm 1\sigma$
values as well as the central cross section. 

Signal systematic uncertainties due to detector and acceptance effects are taken into account. 
The main sources of these uncertainties are the JER, ranging from 3\% to
6\%, the JES, ranging from 2\% to 5.7\%, pile-up, ranging from 0.5\% to
5.5\% and from $b$-tagging efficiency, ranging from 3\% to
5.5\%. Uncertainties in the acceptance due to theoretical variations
are taken into consideration. Those originate from variations of the QCD
coupling constant $\alpha_\mathrm{s}$, the variations of the renormalization
and factorization scales, the CKKW matching scale at which the parton-shower description and the matrix-element description are separate and
the parton-shower tune variations (each varied up and down by a
factor of two). These uncertainties range across
the SRs between 10\% and 25\% for the $\stop\to t^{(*)} \ninoone$ grid,
the mixed grid, the non-asymptotic higgsino grid, and the $\gluino\to
t\stop\to t\ninoone+$soft grid. For the wino-NLSP model, they range from 15\% to
20\%, and for the well-tempered neutralino pMSSM model they range from 10\% to 35\%. 
Finally, the uncertainty in the estimated number of signal events which arises from the cross-section uncertainties for the various
processes is taken into account by calculating two additional limits considering a $\pm1\sigma$ change in cross section. 
The cross-section uncertainty is $\sim$15--20\% for direct top-squark production and $\sim$15--30\% for gluino production~\cite{Beenakker:1997ut,Beenakker:2010nq,Beenakker:2011fu,Borschensky:2014cia} depending on the top-squark and gluino masses.

\begin{table}[htpb]
  \caption{Dominant systematic uncertainties (greater than 1\% for at
    least one SR) for SRA and SRB in percent relative to the total
    background estimates. The uncertainties due to the normalization
    from a control region for a given signal region and background are
    indicated by $\mu_{\ttZ}$, $\mu_{\ttbar}$, $\mu_{Z}$, $\mu_{W}$,
    and $\mu_{\mathrm{single~top}}$. The theory uncertainties are the
    total uncertainties for a given background. Additionally, the
    uncertainty due to the number of MC events in the background samples is shown as ``MC statistical''. }
  \begin{center}
{\renewcommand{\arraystretch}{1.2}
\begin{tabular}{
S[table-alignment=left]
S[table-number-alignment=right,table-alignment=right]
S[table-number-alignment=right,table-alignment=right]
S[table-number-alignment=right,table-alignment=right]
S[table-number-alignment=right,table-alignment=right]
S[table-number-alignment=right,table-alignment=right]
S[table-number-alignment=right,table-alignment=right]
}
\hline\hline
 & {SRA-TT} & {SRA-TW} & {SRA-T0} & {SRB-TT} & {SRB-TW} & {SRB-T0}\\ \hline 
{Total syst. unc.} & 24 & 23 & 15 & 19 & 14 & 15\\ \hline  
{\ttbar\ theory} & 10 & 6 & 3 & 10 & 11 & 12\\  
{\ttbar+V\ theory} & 2 & {$<$1\phantom{15}} & {$<$1\phantom{15}} & 1 & {$<$1\phantom{15}} & {$<$1\phantom{15}}\\  
{\Zboson\ theory} & 1 & 3 & 2 & {$<$1\phantom{15}} & 1 & {$<$1\phantom{15}}\\  
{Single top theory} & 6 & 3 & 5 & 3 & 4 & 5\\  
{Diboson theory} & {$<$1\phantom{15}} & 2 & {$<$1\phantom{15}} & {$<$1\phantom{15}} & {$<$1\phantom{15}} & {$<$1\phantom{15}}\\  
{$\mu_{\ttbar}$} & {$<$1\phantom{15}} & {$<$1\phantom{15}} & {$<$1\phantom{15}} & 2 & 2 & 1\\  
{$\mu_{\ttZ}$} & 6 & 3 & 2 & 4 & 3 & 2\\  
{$\mu_{Z}$} & 6 & 10 & 7 & 5 & 6 & 4\\  
{$\mu_{W}$} & 1 & 1 & 1 & 2 & 1 & 2\\  
{$\mu_{\mathrm{single~top}}$} & 5 & 3 & 5 & 4 & 4 & 5\\  
{JER} & 10 & 12 & 4 & 3 & 4 & 3\\  
{JES} & 4 & 7 & 1 & 7 & 4 & {$<$1\phantom{15}}\\  
{b-tagging} & 1 & 3 & 2 & 5 & 4 & 4\\  
{\met\ soft term} & 2 & 2 & {$<$1\phantom{15}} & 1 & {$<$1\phantom{15}} & {$<$1\phantom{15}}\\  
{Multijet estimate} & 1 & {$<$1\phantom{15}} & {$<$1\phantom{15}} & 2 & 2 & {$<$1\phantom{15}}\\  
{Pileup} & 10 & 5 & 5 & 8 & 1 & 3\\  
\hline\hline
\end{tabular}

}
\end{center}
\label{tab:srABSysts}
\end{table}

\begin{table}[htpb]
  \caption{Dominant systematic uncertainties (greater than 1\% for at
    least one SR) for SRC, SRD, and SRE in percent relative to the
    total background estimates. The uncertainty due to the
    normalization from a control region for a given signal region and
    background are indicated by $\mu_{\ttZ}$, $\mu_{\ttbar}$,
    $\mu_{Z}$, $\mu_{W}$, and $\mu_{\mathrm{single~top}}$. The theory
    uncertainties are the total uncertainties for a given
    background. Additionally, the uncertainty due to the number of MC events in the background samples is shown as ``MC statistical''.}
  \begin{center}
{\renewcommand{\arraystretch}{1.2}
\begin{tabular}{
S[table-alignment=left]
S[table-number-alignment=right,table-alignment=right]
S[table-number-alignment=right,table-alignment=right]
S[table-number-alignment=right,table-alignment=right]
S[table-number-alignment=right,table-alignment=right]
S[table-number-alignment=right,table-alignment=right]
S[table-number-alignment=right,table-alignment=right]
S[table-number-alignment=right,table-alignment=right]
S[table-number-alignment=right,table-alignment=right]
}
\hline\hline
 & {SRC1} & {SRC2} & {SRC3} & {SRC4} & {SRC5} & {SRD-low} & {SRD-high} & {SRE}\\ \hline 
{Total syst. unc.} & 31 & 18 & 18 & 16 & 80 & 25 & 18 & 22\\ \hline  
{\ttbar\ theory} & 27 & 11 & 14 & 11 & 71 & 12 & 10 & 11\\  
{\ttbar+V\ theory} & {$<$1\phantom{15}} & {$<$1\phantom{15}} & {$<$1\phantom{15}} & {$<$1\phantom{15}} & {$<$1\phantom{15}} & {$<$1\phantom{15}} & {$<$1\phantom{15}} & 1\\  
{\Zboson\ theory} & {$<$1\phantom{15}} & {$<$1\phantom{15}} & {$<$1\phantom{15}} & {$<$1\phantom{15}} & {$<$1\phantom{15}} & {$<$1\phantom{15}} & {$<$1\phantom{15}} & 2\\  
{\Wboson\ theory} & {$<$1\phantom{15}} & {$<$1\phantom{15}} & 1 & 3 & 2 & {$<$1\phantom{15}} & {$<$1\phantom{15}} & 1\\  
{Single top theory} & 3 & 2 & 2 & 3 & {$<$1\phantom{15}} & 5 & 6 & 12\\  
{$\mu_{\ttbar}$} & 4 & 6 & 6 & 5 & 5 & 1 & 1 & {$<$1\phantom{15}}\\  
{$\mu_{\ttZ}$} & {$<$1\phantom{15}} & {$<$1\phantom{15}} & {$<$1\phantom{15}} & {$<$1\phantom{15}} & {$<$1\phantom{15}} & 2 & 2 & 4\\  
{$\mu_{Z}$} & {$<$1\phantom{15}} & {$<$1\phantom{15}} & {$<$1\phantom{15}} & {$<$1\phantom{15}} & {$<$1\phantom{15}} & 4 & 5 & 5\\  
{$\mu_{W}$} & {$<$1\phantom{15}} & {$<$1\phantom{15}} & 1 & 3 & 3 & 3 & 1 & 2\\  
{$\mu_{\mathrm{single~top}}$} & 3 & 2 & 2 & 3 & {$<$1\phantom{15}} & 5 & 6 & 6\\  
{JER} & 4 & 10 & 6 & 5 & 10 & 3 & 6 & 4\\  
{JES} & 4 & 5 & 2 & 2 & 17 & 8 & 4 & 5\\  
{b-tagging} & 2 & 2 & {$<$1\phantom{15}} & 2 & 4 & 9 & 7 & {$<$1\phantom{15}}\\  
{\met\ soft term} & 1 & 3 & 2 & 3 & 15 & 4 & 3 & 2\\  
{Multijet estimate} & 12 & 3 & {$<$1\phantom{15}} & {$<$1\phantom{15}} & {$<$1\phantom{15}} & 2 & 2 & {$<$1\phantom{15}}\\  
{Pileup} & {$<$1\phantom{15}} & 1 & {$<$1\phantom{15}} & 2 & 14 & 9 & {$<$1\phantom{15}} & 2\\  
\hline\hline
\end{tabular}

}
\end{center}
\label{tab:srCDESysts}
\end{table}



\section{Results and interpretation}
\label{sec:result}

\begin{figure}[htpb]
  \begin{center}
    \includegraphics[width=0.8\textwidth]{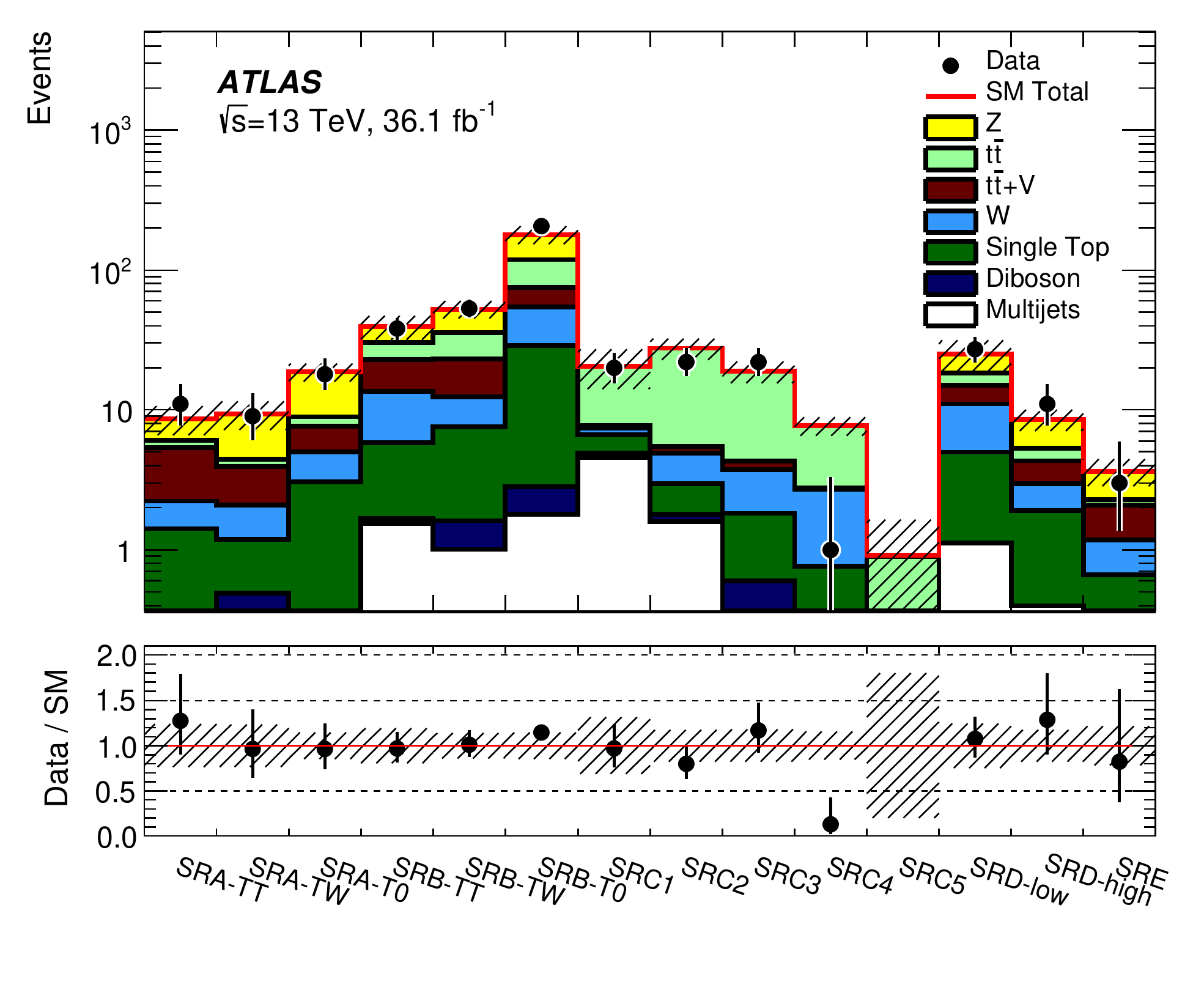}
    \caption{Yields for all signal regions after the likelihood fit. The stacked histograms show the SM prediction and the hatched uncertainty band around the SM predicttion shows total uncertainty, which consists of the MC statistical uncertainties, detector-related systematic uncertainties, and theoretical uncertainties in the extrapolation from CR to SR.} 
    \label{fig:srSummary}
  \end{center}
\end{figure}

\begin{figure}[htpb]
  \begin{center}
    \includegraphics[width=0.49\textwidth]{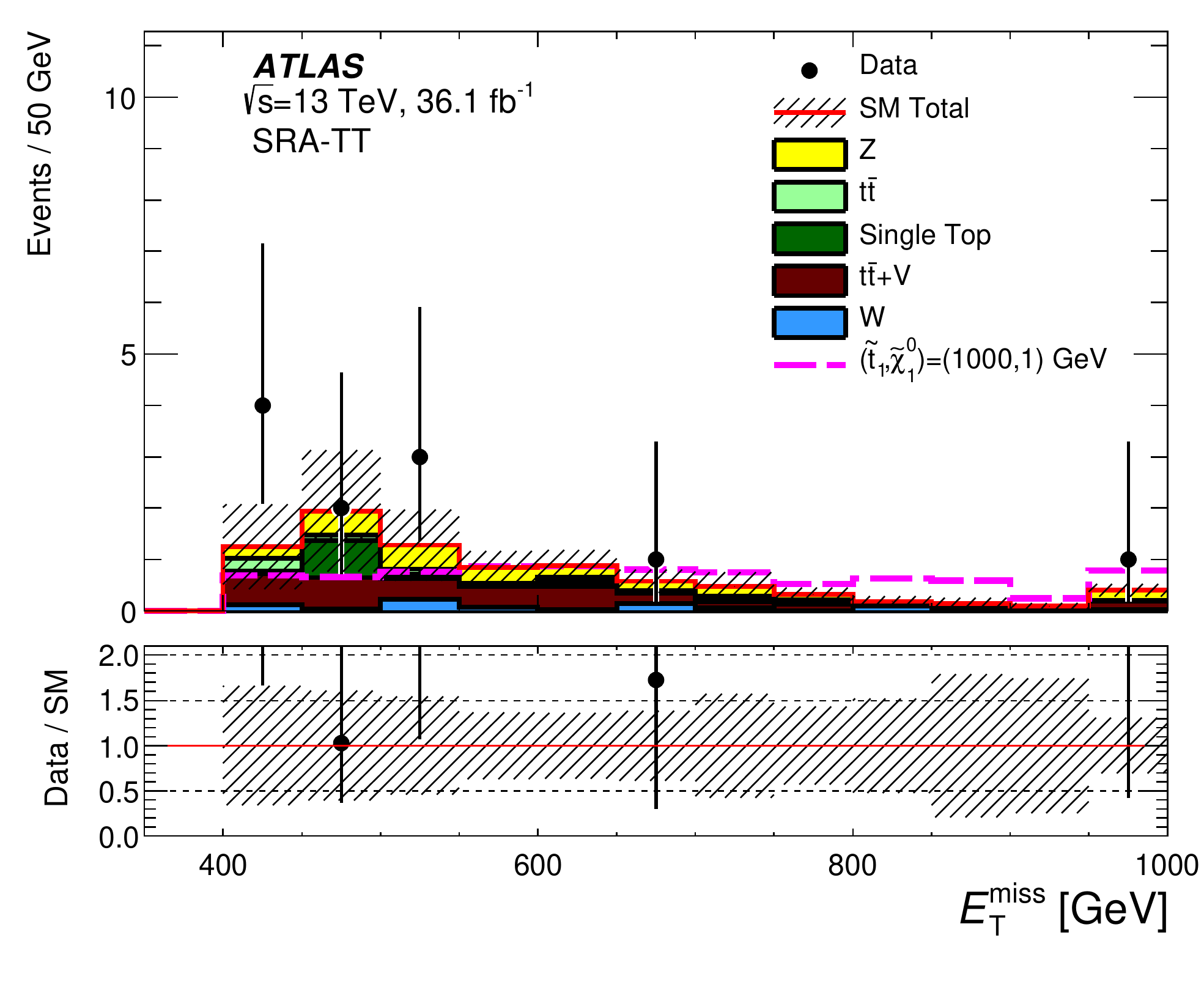}
    \includegraphics[width=0.49\textwidth]{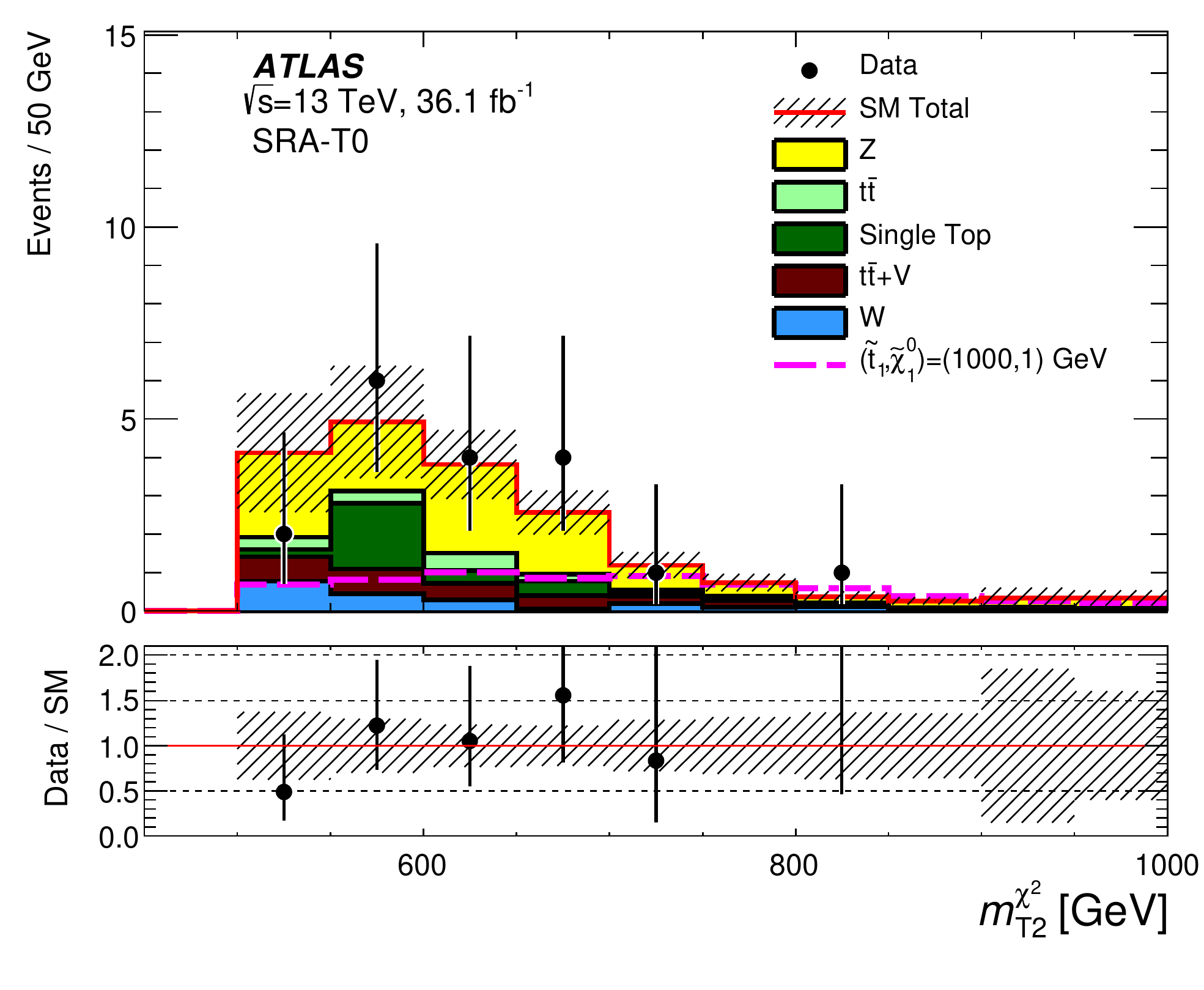}\\
    \includegraphics[width=0.49\textwidth]{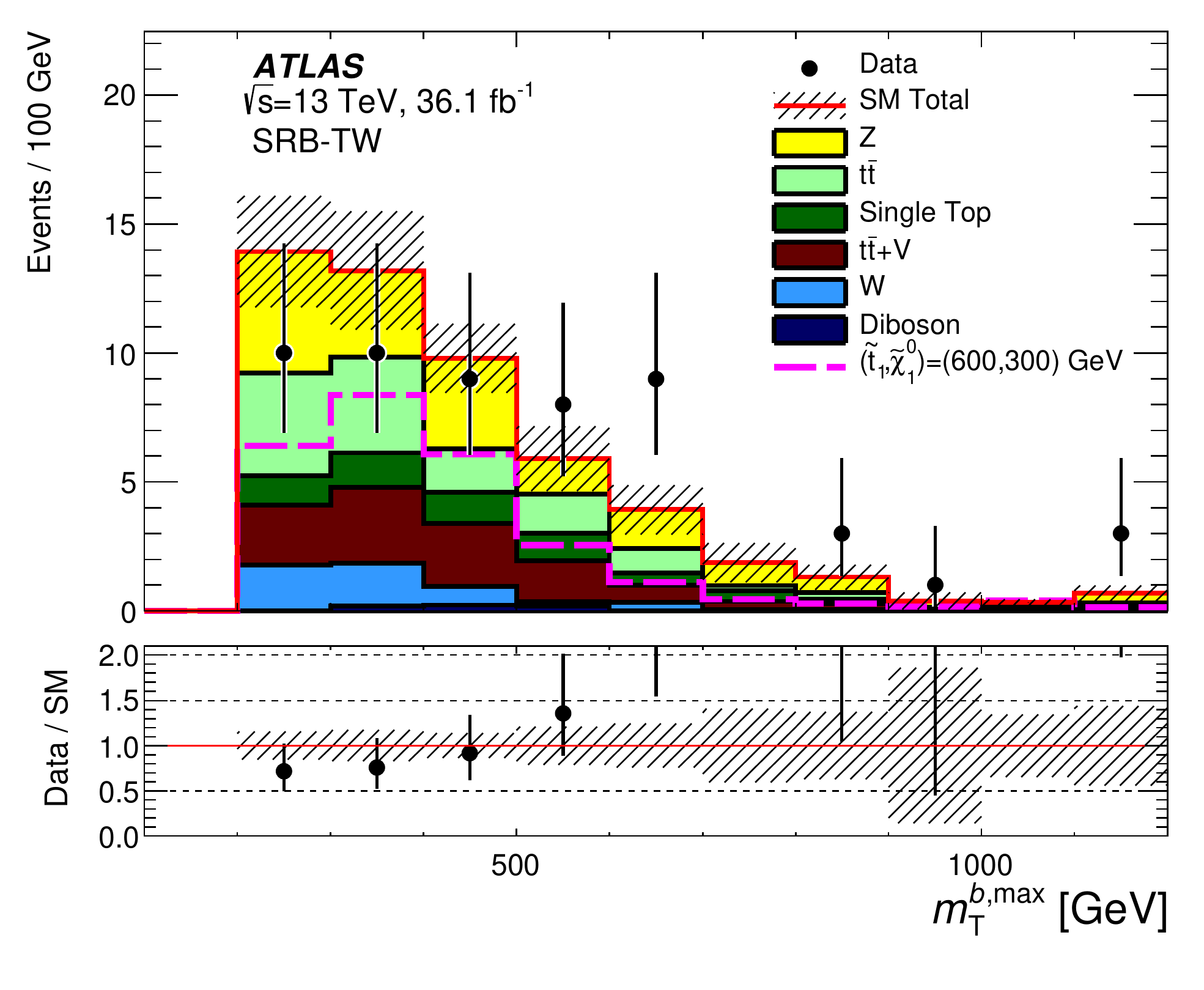}
    \includegraphics[width=0.49\textwidth]{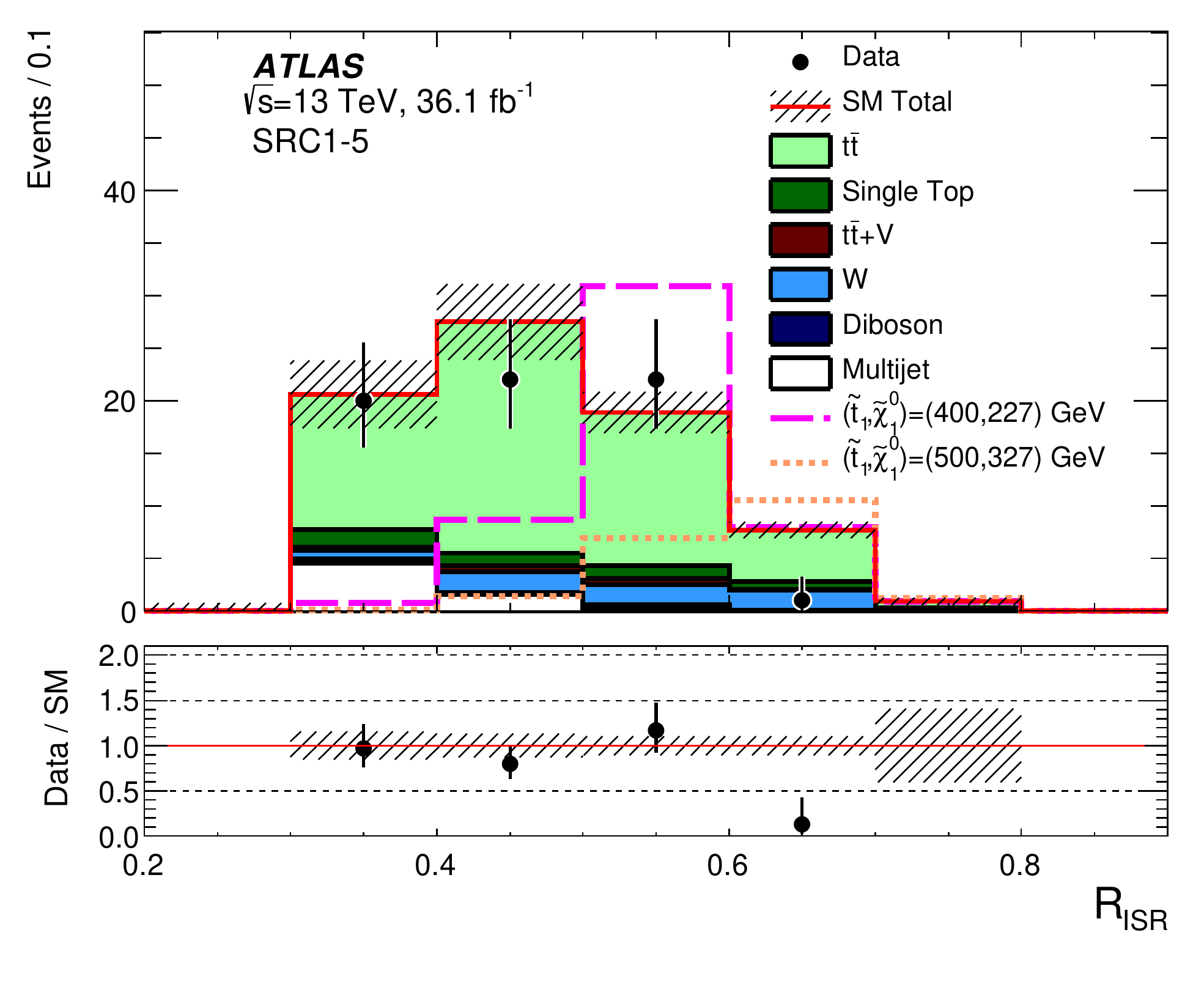}\\
    \includegraphics[width=0.49\textwidth]{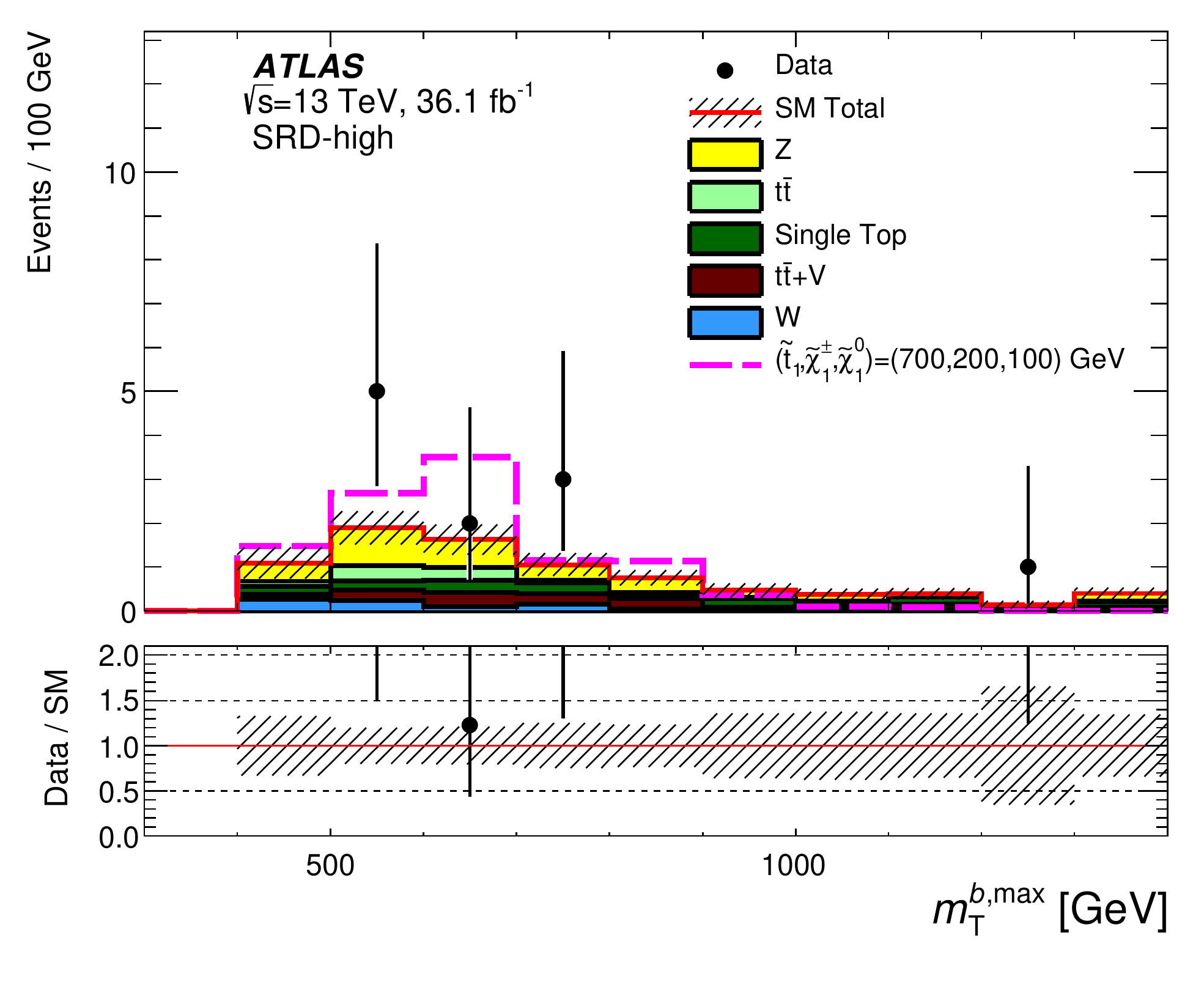}
    \includegraphics[width=0.49\textwidth]{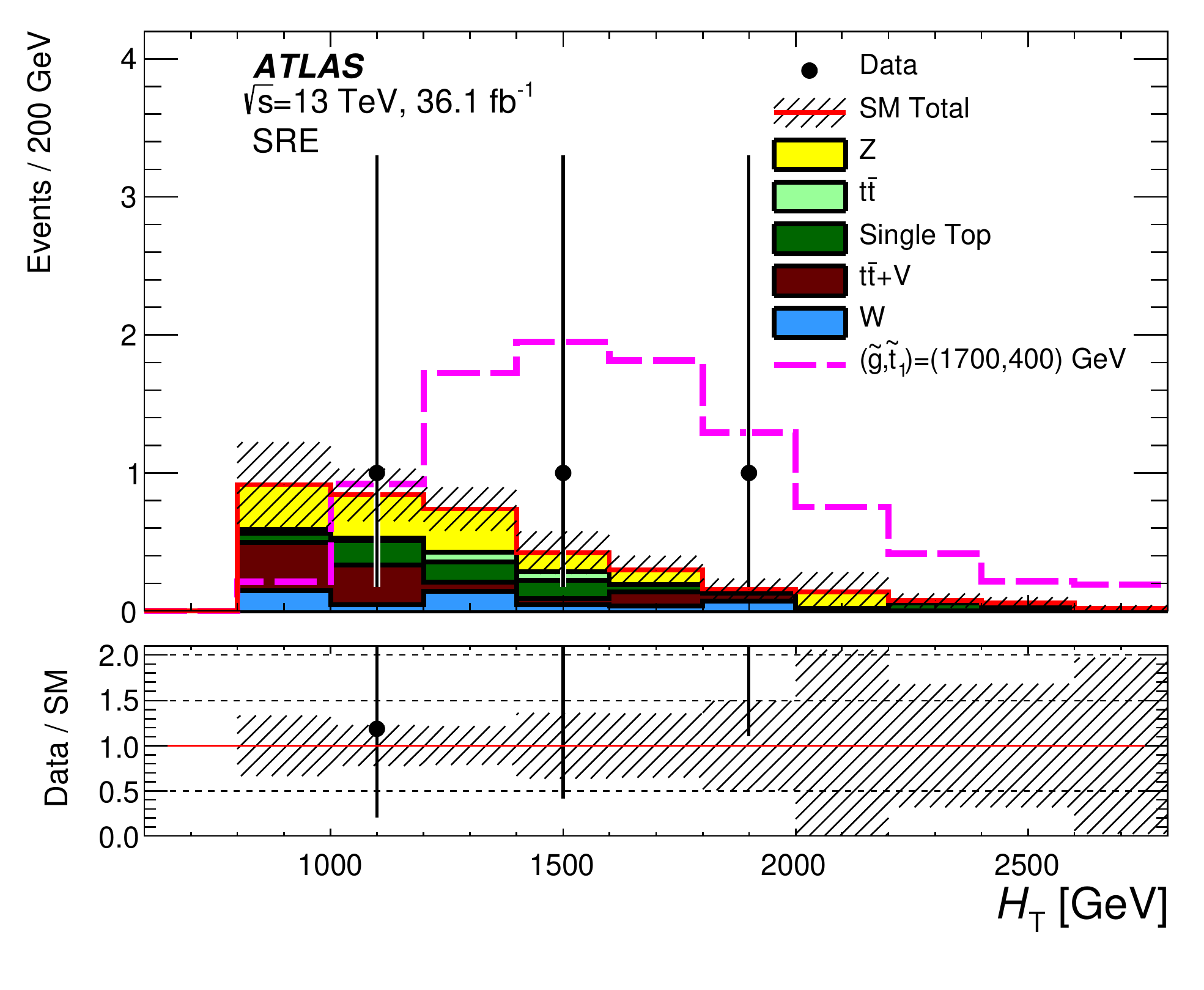}
    \caption{Distributions of \met\ for SRA-TT, \mttwo\ for SRA-T0, \mtbmax\ for SRB-TW, \rISR\ for SRC1--5, \mtbmax\ for SRD-high and \HT\ for SRE after the likelihood fit. The stacked histograms show the SM prediction and the hatched uncertainty band around the SM prediction shows the MC statistical and detector-related systematic uncertainties. For each variable, the distribution for a representative signal point is shown.}
    \label{fig:SRs}
  \end{center}
\end{figure}

The observed event yields are compared to the expected total number of
background events in
Tables~\ref{tab:srABYields},~\ref{tab:srCYields}, ~\ref{tab:srDEYields},
and Figure~\ref{fig:srSummary}. The total background estimate is
determined from a simultaneous fit to all control regions, based on a procedure described in Section~\ref{sec:simultaneousfit} but including the corresponding signal regions as well as control regions. Figure~\ref{fig:SRs}
shows the distribution of \met, \mttwo, \mtbmax, \mt, \rISR, and \HT\ for the various signal regions, with \rISR\ being shown combining SRC1--5. In these distributions, the background predictions are scaled to the values determined from the simultaneous fit.

\begin{table}[htpb]
  \caption{Observed and expected yields, before and after the fit, for SRA and SRB.
The uncertainties include MC statistical uncertainties, detector-related systematic uncertainties, and theoretical uncertainties in the extrapolation from CR to SR.}
  \begin{center}
{\renewcommand{\arraystretch}{1.2}
\begin{tabular}{
S[table-alignment=left]
S[table-number-alignment=left]
S[table-number-alignment=left]
S[table-number-alignment=left]
S[table-number-alignment=left]
S[table-number-alignment=left]
S[table-number-alignment=left]
S[table-number-alignment=left]
S[table-number-alignment=left]
S[table-number-alignment=left]
S[table-number-alignment=left]
S[table-number-alignment=left]
S[table-number-alignment=left]
}
\hline\hline
 & \multicolumn{2}{c}{SRA-TT} & \multicolumn{2}{c}{SRA-TW} & \multicolumn{2}{c}{SRA-T0} & \multicolumn{2}{c}{SRB-TT} & \multicolumn{2}{c}{SRB-TW} & \multicolumn{2}{c}{SRB-T0}\\ \hline 
{Observed} & \multicolumn{1}{l}{$11$} &  & \multicolumn{1}{l}{$9$} &  & \multicolumn{1}{l}{$18$} &  & \multicolumn{1}{l}{$38$} &  & \multicolumn{1}{l}{$53$} &  & \multicolumn{1}{l}{$206$} & \\ \hline \hline 
 \multicolumn{3}{r}{Fitted background events} & \multicolumn{10}{r}{}\\ \hline 
{Total SM} & \multicolumn{2}{l}{$8.6\phantom{0} \pm 2.1\phantom{0}$} & \multicolumn{2}{l}{$9.3\phantom{0} \pm 2.2\phantom{0}$} & \multicolumn{2}{l}{$18.7\phantom{0} \pm 2.7\phantom{0}$} & \multicolumn{2}{l}{$39.3\phantom{0} \pm 7.6\phantom{0}$} & \multicolumn{2}{l}{$52.4\phantom{0} \pm 7.4\phantom{0}$} & \multicolumn{2}{l}{$179\phantom{000} \pm 26\phantom{000}$}\\ \hline 
{\ttbar} & \multicolumn{2}{l}{$0.71\;_{-\;0.71}^{+\;0.91}$} & \multicolumn{2}{l}{$0.51\;_{-\;0.51}^{+\;0.55}$} & \multicolumn{2}{l}{$\phantom{1}1.31 \pm 0.64$} & \multicolumn{2}{l}{$\phantom{3}7.3\phantom{0} \pm 4.3\phantom{0}$} & \multicolumn{2}{l}{$12.4\phantom{0} \pm 5.9\phantom{0}$} & \multicolumn{2}{l}{$\phantom{1}43\phantom{000} \pm 22\phantom{000}$}\\ 
{\Wjets} & \multicolumn{2}{l}{$0.82 \pm 0.15$} & \multicolumn{2}{l}{$0.89 \pm 0.56$} & \multicolumn{2}{l}{$\phantom{1}2.00 \pm 0.83$} & \multicolumn{2}{l}{$\phantom{3}7.8\phantom{0} \pm 2.8\phantom{0}$} & \multicolumn{2}{l}{$\phantom{5}4.8\phantom{0} \pm 1.2\phantom{0}$} & \multicolumn{2}{l}{$\phantom{1}25.8\phantom{0} \pm \phantom{2}8.8\phantom{0}$}\\ 
{\Zjets} & \multicolumn{2}{l}{$2.5\phantom{0} \pm 1.3\phantom{0}$} & \multicolumn{2}{l}{$4.9\phantom{0} \pm 1.9\phantom{0}$} & \multicolumn{2}{l}{$\phantom{1}9.8\phantom{0} \pm 1.6\phantom{0}$} & \multicolumn{2}{l}{$\phantom{3}9.0\phantom{0} \pm 2.8\phantom{0}$} & \multicolumn{2}{l}{$16.8\phantom{0} \pm 4.1\phantom{0}$} & \multicolumn{2}{l}{$\phantom{1}60.7\phantom{0} \pm \phantom{2}9.6\phantom{0}$}\\ 
{\ttV} & \multicolumn{2}{l}{$3.16 \pm 0.66$} & \multicolumn{2}{l}{$1.84 \pm 0.39$} & \multicolumn{2}{l}{$\phantom{1}2.60 \pm 0.53$} & \multicolumn{2}{l}{$\phantom{3}9.3\phantom{0} \pm 1.7\phantom{0}$} & \multicolumn{2}{l}{$10.8\phantom{0} \pm 1.6\phantom{0}$} & \multicolumn{2}{l}{$\phantom{1}20.5\phantom{0} \pm \phantom{2}3.2\phantom{0}$}\\ 
{Single top} & \multicolumn{2}{l}{$1.20 \pm 0.81$} & \multicolumn{2}{l}{$0.70 \pm 0.42$} & \multicolumn{2}{l}{$\phantom{1}2.9\phantom{0} \pm 1.5\phantom{0}$} & \multicolumn{2}{l}{$\phantom{3}4.2\phantom{0} \pm 2.2\phantom{0}$} & \multicolumn{2}{l}{$\phantom{5}5.9\phantom{0} \pm 2.8\phantom{0}$} & \multicolumn{2}{l}{$\phantom{1}26\phantom{000} \pm 13\phantom{000}$}\\ 
{Dibosons} & \multicolumn{2}{c}{${-} {-}$} & \multicolumn{2}{l}{$0.35 \pm 0.26$} & \multicolumn{2}{c}{${-} {-}$} & \multicolumn{2}{l}{$\phantom{3}0.13 \pm 0.07$} & \multicolumn{2}{l}{$\phantom{5}0.60 \pm 0.43$} & \multicolumn{2}{l}{$\phantom{17}1.04 \pm \phantom{2}0.73$}\\ 
{Multijets} & \multicolumn{2}{l}{$0.21 \pm 0.10$} & \multicolumn{2}{l}{$0.14 \pm 0.09$} & \multicolumn{2}{l}{$\phantom{1}0.12 \pm 0.07$} & \multicolumn{2}{l}{$\phantom{3}1.54 \pm 0.64$} & \multicolumn{2}{l}{$\phantom{5}1.01 \pm 0.88$} & \multicolumn{2}{l}{$\phantom{17}1.8\phantom{0} \pm \phantom{2}1.5\phantom{0}$}\\ \hline \hline 
 \multicolumn{3}{r}{Expected events before fit} & \multicolumn{10}{r}{}\\ \hline 
{Total SM} & \multicolumn{2}{l}{$7.1\phantom{3}\phantom{3}$}  & \multicolumn{2}{l}{$7.9\phantom{3}\phantom{3}$}  & \multicolumn{2}{l}{$16.3\phantom{3}\phantom{3}$}  & \multicolumn{2}{l}{$32.4\phantom{3}\phantom{3}$}  & \multicolumn{2}{l}{$46.1\phantom{3}\phantom{3}$}  & \multicolumn{2}{l}{$162\phantom{3}\phantom{2}$} \\ \hline 
{\ttbar} & \multicolumn{2}{l}{$0.60\phantom{3}\phantom{4}$}  & \multicolumn{2}{l}{$0.45\phantom{3}\phantom{4}$}  & \multicolumn{2}{l}{$\phantom{1}1.45\phantom{3}\phantom{4}$}  & \multicolumn{2}{l}{$\phantom{3}6.1\phantom{3}\phantom{3}$}  & \multicolumn{2}{l}{$12.8\phantom{3}\phantom{3}$}  & \multicolumn{2}{l}{$\phantom{1}47\phantom{3}\phantom{2}$} \\ 
{\Wjets} & \multicolumn{2}{l}{$0.65\phantom{3}\phantom{4}$}  & \multicolumn{2}{l}{$0.70\phantom{3}\phantom{4}$}  & \multicolumn{2}{l}{$\phantom{1}1.58\phantom{3}\phantom{4}$}  & \multicolumn{2}{l}{$\phantom{3}6.1\phantom{3}\phantom{3}$}  & \multicolumn{2}{l}{$\phantom{5}3.83\phantom{3}\phantom{3}$}  & \multicolumn{2}{l}{$\phantom{1}20.4\phantom{3}\phantom{3}$} \\ 
{\Zjets} & \multicolumn{2}{l}{$2.15\phantom{3}\phantom{3}$}  & \multicolumn{2}{l}{$4.2\phantom{3}\phantom{3}$}  & \multicolumn{2}{l}{$\phantom{1}8.63\phantom{3}\phantom{3}$}  & \multicolumn{2}{l}{$\phantom{3}7.7\phantom{3}\phantom{3}$}  & \multicolumn{2}{l}{$14.4\phantom{3}\phantom{3}$}  & \multicolumn{2}{l}{$\phantom{1}53.6\phantom{3}\phantom{3}$} \\ 
{\ttV} & \multicolumn{2}{l}{$2.46\phantom{3}\phantom{4}$}  & \multicolumn{2}{l}{$1.43\phantom{3}\phantom{4}$}  & \multicolumn{2}{l}{$\phantom{1}2.02\phantom{3}\phantom{4}$}  & \multicolumn{2}{l}{$\phantom{3}7.3\phantom{3}\phantom{3}$}  & \multicolumn{2}{l}{$\phantom{5}8.4\phantom{3}\phantom{3}$}  & \multicolumn{2}{l}{$\phantom{1}15.9\phantom{3}\phantom{3}$} \\ 
{Single top} & \multicolumn{2}{l}{$1.03\phantom{3}\phantom{4}$}  & \multicolumn{2}{l}{$0.60\phantom{3}\phantom{4}$}  & \multicolumn{2}{l}{$\phantom{1}2.5\phantom{3}\phantom{3}$}  & \multicolumn{2}{l}{$\phantom{3}3.6\phantom{3}\phantom{3}$}  & \multicolumn{2}{l}{$\phantom{5}5.1\phantom{3}\phantom{3}$}  & \multicolumn{2}{l}{$\phantom{1}22.4\phantom{3}\phantom{2}$} \\ 
{Dibosons} & \multicolumn{2}{l}{${-} {-}$} & \multicolumn{2}{l}{$0.35\phantom{3}\phantom{4}$}  & \multicolumn{2}{l}{${-} {-}$} & \multicolumn{2}{l}{$\phantom{3}0.13\phantom{3}\phantom{4}$}  & \multicolumn{2}{l}{$\phantom{5}0.60\phantom{3}\phantom{4}$}  & \multicolumn{2}{l}{$\phantom{17}1.03\phantom{3}\phantom{4}$} \\ 
{Multijets} & \multicolumn{2}{l}{$0.21\phantom{3}\phantom{4}$}  & \multicolumn{2}{l}{$0.14\phantom{3}\phantom{4}$}  & \multicolumn{2}{l}{$\phantom{1}0.12\phantom{3}\phantom{4}$}  & \multicolumn{2}{l}{$\phantom{3}1.54\phantom{3}\phantom{4}$}  & \multicolumn{2}{l}{$\phantom{5}1.01\phantom{3}\phantom{4}$}  & \multicolumn{2}{l}{$\phantom{17}1.8\phantom{3}\phantom{3}$} \\ 
\hline\hline
\end{tabular}

}
\end{center}
\label{tab:srABYields}
\end{table}

\begin{table}[htpb]
  \caption{Observed and expected yields, before and after the fit.
The uncertainties include MC statistical uncertainties, detector-related systematic uncertainties, and theoretical uncertainties in the extrapolation from CR to SR.}
  \begin{center}
{\renewcommand{\arraystretch}{1.2}
\begin{tabular}{
S[table-alignment=left]
S[table-number-alignment=left]
S[table-number-alignment=left]
S[table-number-alignment=left]
S[table-number-alignment=left]
S[table-number-alignment=left]
S[table-number-alignment=left]
S[table-number-alignment=left]
S[table-number-alignment=left]
S[table-number-alignment=left]
S[table-number-alignment=left]
}
\hline\hline
 & \multicolumn{2}{c}{SRC1} & \multicolumn{2}{c}{SRC2} & \multicolumn{2}{c}{SRC3} & \multicolumn{2}{c}{SRC4} & \multicolumn{2}{c}{SRC5}\\ \hline 
{Observed} & \multicolumn{1}{l}{$20$} &  & \multicolumn{1}{l}{$22$} &  & \multicolumn{1}{l}{$22$} &  & \multicolumn{1}{l}{$1$} &  & \multicolumn{1}{l}{$0$} & \\ \hline \hline 
 \multicolumn{3}{r}{Fitted background events} & \multicolumn{8}{r}{}\\ \hline 
{Total SM} & \multicolumn{2}{l}{$20.6\phantom{0} \pm 6.5\phantom{0}$} & \multicolumn{2}{l}{$27.6\phantom{0} \pm 4.9\phantom{0}$} & \multicolumn{2}{l}{$18.9\phantom{0} \pm 3.4\phantom{0}$} & \multicolumn{2}{l}{$7.7\phantom{0} \pm 1.2\phantom{0}$} & \multicolumn{2}{l}{$0.91 \pm 0.73$}\\ \hline 
{\ttbar} & \multicolumn{2}{l}{$12.9\phantom{0} \pm 5.9\phantom{0}$} & \multicolumn{2}{l}{$22.1\phantom{0} \pm 4.3\phantom{0}$} & \multicolumn{2}{l}{$14.6\phantom{0} \pm 3.2\phantom{0}$} & \multicolumn{2}{l}{$4.91 \pm 0.97$} & \multicolumn{2}{l}{$0.63\;_{-\;0.63}^{+\;0.70}$}\\ 
{\Wjets} & \multicolumn{2}{l}{$\phantom{2}0.80 \pm 0.37$} & \multicolumn{2}{l}{$\phantom{2}1.93 \pm 0.49$} & \multicolumn{2}{l}{$\phantom{1}1.91 \pm 0.62$} & \multicolumn{2}{l}{$1.93 \pm 0.46$} & \multicolumn{2}{l}{$0.21 \pm 0.12$}\\ 
{\Zjets} & \multicolumn{2}{c}{${-} {-}$} & \multicolumn{2}{c}{${-} {-}$} & \multicolumn{2}{c}{${-} {-}$} & \multicolumn{2}{c}{${-} {-}$} & \multicolumn{2}{c}{${-} {-}$}\\ 
{\ttV} & \multicolumn{2}{l}{$\phantom{2}0.29 \pm 0.16$} & \multicolumn{2}{l}{$\phantom{2}0.59 \pm 0.38$} & \multicolumn{2}{l}{$\phantom{1}0.56 \pm 0.31$} & \multicolumn{2}{l}{$0.08 \pm 0.08$} & \multicolumn{2}{l}{$0.06 \pm 0.02$}\\ 
{Single top} & \multicolumn{2}{l}{$\phantom{2}1.7\phantom{0} \pm 1.3\phantom{0}$} & \multicolumn{2}{l}{$\phantom{2}1.2\phantom{0}\;_{-\;1.2}^{+\;1.4\phantom{0}}$} & \multicolumn{2}{l}{$\phantom{1}1.22 \pm 0.69$} & \multicolumn{2}{l}{$0.72 \pm 0.37$} & \multicolumn{2}{c}{${-} {-}$}\\ 
{Dibosons} & \multicolumn{2}{l}{$\phantom{2}0.39 \pm 0.33$} & \multicolumn{2}{l}{$\phantom{2}0.21\;_{-\;0.21}^{+\;0.23}$} & \multicolumn{2}{l}{$\phantom{1}0.28 \pm 0.18$} & \multicolumn{2}{c}{${-} {-}$} & \multicolumn{2}{c}{${-} {-}$}\\ 
{Multijets} & \multicolumn{2}{l}{$\phantom{2}4.6\phantom{0} \pm 2.4\phantom{0}$} & \multicolumn{2}{l}{$\phantom{2}1.58 \pm 0.77$} & \multicolumn{2}{l}{$\phantom{1}0.32 \pm 0.17$} & \multicolumn{2}{l}{$0.04 \pm 0.02$} & \multicolumn{2}{c}{${-} {-}$}\\ \hline \hline 
 \multicolumn{3}{r}{Expected events before fit} & \multicolumn{8}{r}{}\\ \hline 
{Total SM} & \multicolumn{2}{l}{$25.4\phantom{3}\phantom{3}$}  & \multicolumn{2}{l}{$36.0\phantom{3}\phantom{3}$}  & \multicolumn{2}{l}{$24.2\phantom{3}\phantom{3}$}  & \multicolumn{2}{l}{$9.2\phantom{3}\phantom{3}$}  & \multicolumn{2}{l}{$1.1\phantom{3}\phantom{4}$} \\ \hline 
{\ttbar} & \multicolumn{2}{l}{$18.2\phantom{3}\phantom{3}$}  & \multicolumn{2}{l}{$31.2\phantom{3}\phantom{3}$}  & \multicolumn{2}{l}{$20.6\phantom{3}\phantom{3}$}  & \multicolumn{2}{l}{$7.0\phantom{3}\phantom{4}$}  & \multicolumn{2}{l}{$0.89\phantom{3}\phantom{4}$} \\ 
{\Wjets} & \multicolumn{2}{l}{$\phantom{2}0.64\phantom{3}\phantom{4}$}  & \multicolumn{2}{l}{$\phantom{2}1.53\phantom{3}\phantom{4}$}  & \multicolumn{2}{l}{$\phantom{1}1.51\phantom{3}\phantom{4}$}  & \multicolumn{2}{l}{$1.53\phantom{3}\phantom{4}$}  & \multicolumn{2}{l}{$0.17\phantom{3}\phantom{4}$} \\ 
{\Zjets} & \multicolumn{2}{l}{${-} {-}$} & \multicolumn{2}{l}{${-} {-}$} & \multicolumn{2}{l}{${-} {-}$} & \multicolumn{2}{l}{${-} {-}$} & \multicolumn{2}{l}{${-} {-}$}\\ 
{\ttV} & \multicolumn{2}{l}{$\phantom{2}0.22\phantom{3}\phantom{4}$}  & \multicolumn{2}{l}{$\phantom{2}0.46\phantom{3}\phantom{4}$}  & \multicolumn{2}{l}{$\phantom{1}0.44\phantom{3}\phantom{4}$}  & \multicolumn{2}{l}{$0.07\phantom{3}\phantom{4}$}  & \multicolumn{2}{l}{$0.05\phantom{3}\phantom{4}$} \\ 
{Single top} & \multicolumn{2}{l}{$\phantom{2}1.44\phantom{3}\phantom{3}$}  & \multicolumn{2}{l}{$\phantom{2}1.0\phantom{3}\phantom{3}$}  & \multicolumn{2}{l}{$\phantom{1}1.04\phantom{3}\phantom{4}$}  & \multicolumn{2}{l}{$0.62\phantom{3}\phantom{4}$}  & \multicolumn{2}{l}{${-} {-}$}\\ 
{Dibosons} & \multicolumn{2}{l}{$\phantom{2}0.39\phantom{3}\phantom{4}$}  & \multicolumn{2}{l}{$\phantom{2}0.21\phantom{3}\phantom{4}$}  & \multicolumn{2}{l}{$\phantom{1}0.28\phantom{3}\phantom{4}$}  & \multicolumn{2}{l}{${-} {-}$} & \multicolumn{2}{l}{${-} {-}$}\\ 
{Multijets} & \multicolumn{2}{l}{$\phantom{2}4.6\phantom{3}\phantom{3}$}  & \multicolumn{2}{l}{$\phantom{2}1.58\phantom{3}\phantom{4}$}  & \multicolumn{2}{l}{$\phantom{1}0.32\phantom{3}\phantom{4}$}  & \multicolumn{2}{l}{$0.04\phantom{3}\phantom{4}$}  & \multicolumn{2}{l}{${-} {-}$}\\ 
\hline\hline
\end{tabular}

}
\end{center}
\label{tab:srCYields}
\end{table}

\begin{table}[htpb]
  \caption{Observed and expected yields, before and after the fit, for SRD and SRE.
The uncertainties include MC statistical uncertainties, detector-related systematic uncetainties, and theoretical uncertainties in the extrapolation from CR to SR.}
  \begin{center}
{\renewcommand{\arraystretch}{1.2}
\begin{tabular}{
S[table-alignment=left]
S[table-number-alignment=left]
S[table-number-alignment=left]
S[table-number-alignment=left]
S[table-number-alignment=left]
S[table-number-alignment=left]
S[table-number-alignment=left]
}
\hline\hline
 & \multicolumn{2}{c}{SRD-low} & \multicolumn{2}{c}{SRD-high} & \multicolumn{2}{c}{SRE}\\ \hline 
{Observed} & \multicolumn{1}{l}{$27$} &  & \multicolumn{1}{l}{$11$} &  & \multicolumn{1}{l}{$3$} & \\ \hline \hline 
 \multicolumn{3}{r}{Fitted background events} & \multicolumn{4}{r}{}\\ \hline 
{Total SM} & \multicolumn{2}{l}{$25.1\phantom{0} \pm 6.2\phantom{0}$} & \multicolumn{2}{l}{$8.5\phantom{0} \pm 1.5\phantom{0}$} & \multicolumn{2}{l}{$3.64 \pm 0.79$}\\ \hline 
{\ttbar} & \multicolumn{2}{l}{$\phantom{2}3.3\phantom{0} \pm 3.3\phantom{0}$} & \multicolumn{2}{l}{$0.98 \pm 0.88$} & \multicolumn{2}{l}{$0.21\;_{-\;0.21}^{+\;0.39}$}\\ 
{\Wjets} & \multicolumn{2}{l}{$\phantom{2}6.1\phantom{0} \pm 2.9\phantom{0}$} & \multicolumn{2}{l}{$1.06 \pm 0.34$} & \multicolumn{2}{l}{$0.52 \pm 0.27$}\\ 
{\Zjets} & \multicolumn{2}{l}{$\phantom{2}6.9\phantom{0} \pm 1.5\phantom{0}$} & \multicolumn{2}{l}{$3.21 \pm 0.62$} & \multicolumn{2}{l}{$1.36 \pm 0.25$}\\ 
{\ttV} & \multicolumn{2}{l}{$\phantom{2}3.94 \pm 0.85$} & \multicolumn{2}{l}{$1.37 \pm 0.32$} & \multicolumn{2}{l}{$0.89 \pm 0.19$}\\ 
{Single top} & \multicolumn{2}{l}{$\phantom{2}3.8\phantom{0} \pm 2.1\phantom{0}$} & \multicolumn{2}{l}{$1.51 \pm 0.74$} & \multicolumn{2}{l}{$0.66 \pm 0.49$}\\ 
{Dibosons} & \multicolumn{2}{c}{${-} {-}$} & \multicolumn{2}{c}{${-} {-}$} & \multicolumn{2}{c}{${-} {-}$}\\ 
{Multijets} & \multicolumn{2}{l}{$\phantom{2}1.12 \pm 0.37$} & \multicolumn{2}{l}{$0.40 \pm 0.15$} & \multicolumn{2}{c}{${-} {-}$}\\ \hline \hline 
 \multicolumn{3}{r}{Expected events before fit} & \multicolumn{4}{r}{}\\ \hline 
{Total SM} & \multicolumn{2}{l}{$22.4\phantom{3}\phantom{3}$}  & \multicolumn{2}{l}{$7.7\phantom{3}\phantom{3}$}  & \multicolumn{2}{l}{$3.02\phantom{3}\phantom{4}$} \\ \hline 
{\ttbar} & \multicolumn{2}{l}{$\phantom{2}3.4\phantom{3}\phantom{3}$}  & \multicolumn{2}{l}{$1.04\phantom{3}\phantom{4}$}  & \multicolumn{2}{l}{$0.21\phantom{3}\phantom{4}$} \\ 
{\Wjets} & \multicolumn{2}{l}{$\phantom{2}4.8\phantom{3}\phantom{3}$}  & \multicolumn{2}{l}{$0.84\phantom{3}\phantom{4}$}  & \multicolumn{2}{l}{$0.42\phantom{3}\phantom{4}$} \\ 
{\Zjets} & \multicolumn{2}{l}{$\phantom{2}6.7\phantom{3}\phantom{3}$}  & \multicolumn{2}{l}{$3.10\phantom{3}\phantom{4}$}  & \multicolumn{2}{l}{$1.15\phantom{3}\phantom{4}$} \\ 
{\ttV} & \multicolumn{2}{l}{$\phantom{2}3.06\phantom{3}\phantom{4}$}  & \multicolumn{2}{l}{$1.07\phantom{3}\phantom{4}$}  & \multicolumn{2}{l}{$0.69\phantom{3}\phantom{4}$} \\ 
{Single top} & \multicolumn{2}{l}{$\phantom{2}3.3\phantom{3}\phantom{3}$}  & \multicolumn{2}{l}{$1.30\phantom{3}\phantom{4}$}  & \multicolumn{2}{l}{$0.56\phantom{3}\phantom{4}$} \\ 
{Dibosons} & \multicolumn{2}{l}{${-} {-}$} & \multicolumn{2}{l}{${-} {-}$} & \multicolumn{2}{l}{${-} {-}$}\\ 
{Multijets} & \multicolumn{2}{l}{$\phantom{2}1.12\phantom{3}\phantom{4}$}  & \multicolumn{2}{l}{$0.40\phantom{3}\phantom{4}$}  & \multicolumn{2}{l}{${-} {-}$}\\ 
\hline\hline
\end{tabular}

}
\end{center}
\label{tab:srDEYields}
\end{table}

No significant excess above the SM prediction is observed in any of
the signal regions. 
The smallest $p$-values, which express the probability that the background fluctuates to the data or above,
are 27\%, 27\%, and 29\% for SRB-T0, SRD-high, and SRA-TT,
respectively. The largest deficit in the data can be found in SRC4 where one event is observed while 7.7 background events were expected. 
The $95\%$ confidence level (CL) upper limits on the number of beyond-the-SM (BSM) events in each signal region are derived using the CL$_\mathrm{s}$ prescription~\cite{CLs1,CLs2} and calculated from asymptotic formulae~\cite{likelihoodFit}. Model-independent limits on the visible BSM cross sections, defined as $\sigma_{\mathrm{vis}} = S^{95}_{\textnormal{obs}}/\int\!\!{\cal L}\,dt$, 
where $S^{95}_{\textnormal{obs}}$ is the 95\% CL upper limit on the number of signal events,
are reported in Table~\ref{tab:upLimits}. 

\begin{table}[htpb]
  \caption{Left to right: 95\% CL upper limits on the average visible cross section
($\langle\sigma A \epsilon\rangle_{\rm obs}^{95}$) where the average comes from possibly multiple production channels and on the number of
signal events ($S_{\rm obs}^{95}$ ).  The third column
($S_{\rm exp}^{95}$) shows the 95\% CL upper limit on the number of
signal events, given the expected number (and $\pm 1\sigma$
excursions of the expected number) of background events.
The two last columns indicate the CL$_\mathrm{B}$ value, i.e. the confidence level observed for the background-only hypothesis, and the discovery $p$-value ($p$) and the corresponding significance ($z$).
}
\label{tab:upLimits}

\begin{center}
    
{\renewcommand{\arraystretch}{1.3}
\begin{tabular*}{\textwidth}{@{\extracolsep{\fill}}lccccc}
\noalign{\smallskip}\hline\noalign{\smallskip}
{\bf Signal channel}                        & $\langle{\rm \sigma} A \epsilon\rangle_{\rm obs}^{95}$ [fb]  &  $S_{\rm obs}^{95}$  & $S_{\rm exp}^{95}$ & CL$_\mathrm{B}$ & $p$ ($z$)  \\
\noalign{\smallskip}\hline\noalign{\smallskip}

SRA-TT    & $0.30$ &  $11.0$ & $ { 8.7 }^{ +3.0 }_{ -1.4 }$ & $0.78$ & $ 0.23$~$(0.74)$ \\%
SRA-TW    & $0.27$ &  $9.6$ & $ { 9.6 }^{ +2.8 }_{ -2.1 }$ & $0.50$ & $ 0.50$~$(0.00)$ \\%
SRA-T0    & $0.31$ &  $11.2$ & $ { 11.5 }^{ +3.8 }_{ -2.0 }$ & $0.46$ & $ 0.50$~$(0.00)$ \\%
SRB-TT    & $0.54$ &  $19.6$ & $ { 20.0 }^{ +6.5 }_{ -4.9 }$ & $0.46$ & $ 0.50$~$(0.00)$ \\%
SRB-TW    & $0.60$ &  $21.7$ & $ { 21.0 }^{ +7.3 }_{ -4.3 }$ & $0.54$ & $ 0.50$~$(0.00)$ \\%
SRB-T0    & $2.19$ &  $80$ & $ { 58 }^{ +23 }_{ -17 }$ & $0.83$ & $ 0.13$~$(1.15)$ \\%
SRC1    & $0.42$ &  $15.1$ & $ { 15.8 }^{ +4.8 }_{ -3.5 }$ & $0.48$ & $ 0.50$~$(0.00)$ \\%
SRC2    & $0.31$ &  $11.2$ & $ { 13.9 }^{ +5.9 }_{ -3.6 }$ & $0.24$ & $ 0.50$~$(0.00)$ \\%
SRC3    & $0.42$ &  $15.3$ & $ { 12.3 }^{ +4.7 }_{ -3.4 }$ & $0.73$ & $ 0.27$~$(0.62)$ \\%
SRC4    & $0.10$ &  $3.5$ & $ { 6.7 }^{ +2.8 }_{ -1.8 }$ & $0.00$ & $ 0.50$~$(0.00)$ \\%
SRC5    & $0.09$ &  $3.2$ & $ { 3.0 }^{ +1.1 }_{ -0.1 }$ & $0.23$ & $ 0.23$~$(0.74)$ \\%
SRD-low    & $0.50$ &  $17.9$ & $ { 16.4 }^{ +6.3 }_{ -4.0 }$ & $0.62$ & $ 0.36$~$(0.35)$ \\%
SRD-high    & $0.30$ &  $10.9$ & $ { 8.0 }^{ +3.4 }_{ -1.3 }$ & $0.79$ & $ 0.21$~$(0.79)$ \\%
SRE    & $0.17$ &  $6.1$ & $ { 6.4 }^{ +1.4 }_{ -2.4 }$ & $0.42$ & $ 0.50$~$(0.00)$ \\%

\noalign{\smallskip}\hline\noalign{\smallskip}
\end{tabular*}
}
%

\end{center}
\end{table}

The detector acceptance multiplied by the efficiency ($A\cdot\epsilon$) is calculated for several signal regions and their
benchmark points. The $A\cdot\epsilon$ values for signal regions aimed at high-energy final states, SRA and SRE, are 9\% and 6\% for their respective signal benchmark points of $\mstop=1000\gev,\mLSP=1\gev$, and $m_{\gluino} = 1700\GeV, \mstop=400\GeV, \mLSP=395\GeV$. SRB, SRD-low, and SRD-high have $A\cdot\epsilon$ of 1.4\%, 0.05\%, and 0.5\% for $\mstop=600\gev,\mLSP=300\gev$; $\mstop =400\GeV, m_{\chinoonepm}=100\GeV, \mLSP=50\GeV$; and $\mstop =700\GeV, m_{\chinoonepm}=200\GeV, \mLSP=100\GeV$ where the branching ratio, $B$($\stop\to b \chinoonepm$) = 100\% is assumed for the SRD samples, respectively. Finally, SRC1--5 (combining the \rISR\ windows) has an $A\cdot\epsilon$ of 0.08\% for $\mstop= 400\GeV, \mLSP=227\GeV$.

The profile-likelihood-ratio test statistic is used to set limits on direct pair production of top squarks. The signal strength parameter is allowed to float in the fit~\cite{histFitter}, and any signal contamination in the CRs is taken into account. Again, limits are derived using the CL$_\mathrm{s}$ prescription and calculated from asymptotic formulae. Orthogonal signal subregions, such as SRA-TT, SRA-TW, and SRA-T0, are statistically combined by multiplying their likelihood functions. A similar procedure is performed for the signal subregions in SRB and SRC. For the overlapping signal regions defined for SRD (SRD-low and SRD-high), the signal region with the smallest expected CL$_\mathrm{s}$ value is chosen for each signal model. Once the signal subregions are combined or chosen, the signal region with the smallest expected CL$_\mathrm{s}$ is chosen for each signal model in the $\stop$--$\ninoone$ signal grid. The nominal event yield in each SR is set to the mean background expectation to determine the expected limits; contours that correspond to $\pm1\sigma$ uncertainties in the background estimates ($\sigma_{\mathrm{exp}}$) are also evaluated. The observed event yields determine the observed limits for each SR; these are evaluated for the nominal signal cross sections as well as for $\pm1\sigma$ theory uncertainties in those cross sections, denoted by $\sigma^{\mathrm{SUSY}}_{\mathrm{theory}}$. 

\begin{figure}[htpb]
  \begin{center} \includegraphics[width=0.7\textwidth]{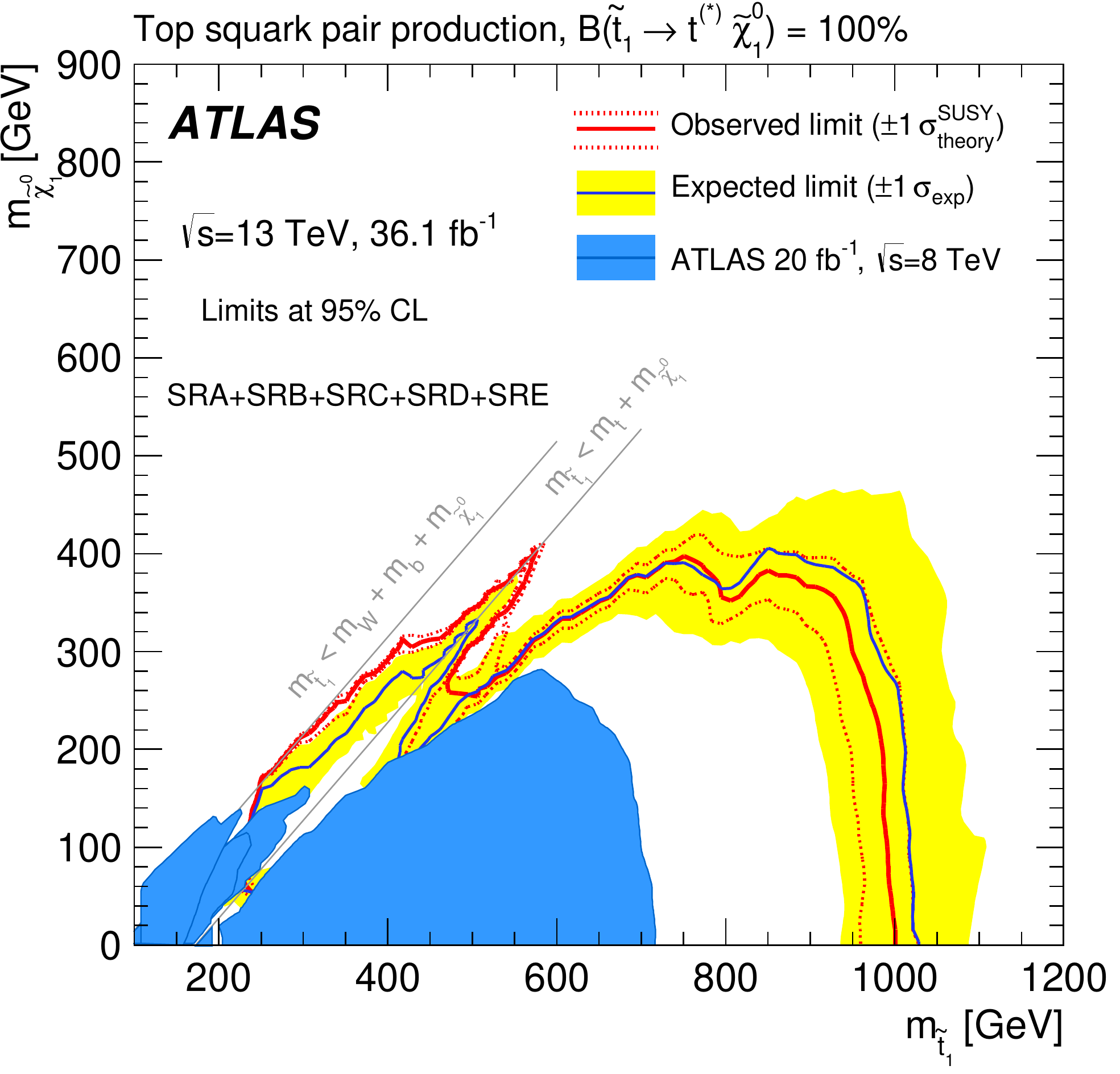}
    \caption{Observed (red solid line) and expected (blue solid line)
      exclusion contours at 95\% CL as a function of $\stop$ and
      $\ninoone$ masses in the scenario where both top squarks decay
      via $\stop\to t^{(*)} \ninoone$. Masses that are within the contours are excluded. Uncertainty bands corresponding to the $\pm 1
      \sigma$ variation of the expected limit (yellow band) and the
      sensitivity of the observed limit to $\pm 1\sigma$ variations of
      the signal theoretical uncertainties (red dotted lines) are also
      indicated. Observed limits from all third-generation Run-1 searches~\cite{Atlas8TeVSummary} at $\sqrt{s}=8$ TeV overlaid for comparison in blue.}
    \label{fig:SRABC_exclusion}
  \end{center}
\end{figure}

Figure~\ref{fig:SRABC_exclusion} shows the observed (solid red line) and
expected (solid blue line) exclusion contours at 95\% CL in the \stop--\ninoone\ mass plane for \lumi. 
The data excludes top-squark masses between
\stopLimLowLSPLow\ and \stopLimLowLSPHigh\ GeV for $\ninoone$ masses below $160\GeV$, extending Run-1 limits from the combination of zero- and one-lepton channels by 260~\GeV. Additional constraints are set in the case where $\mstop\approx m_t+\mLSP$, for which top-squark masses in the range \stopLimDiag\ GeV are excluded. The limits in this region of the exclusion are new compared to the 8~\TeV\ results and come from the inclusion of SRC, which takes advantage of an ISR system to discriminate between signal and the dominant \ttbar\ background.

\begin{figure}[htpb]
  \begin{center}
    \includegraphics[width=0.7\textwidth]{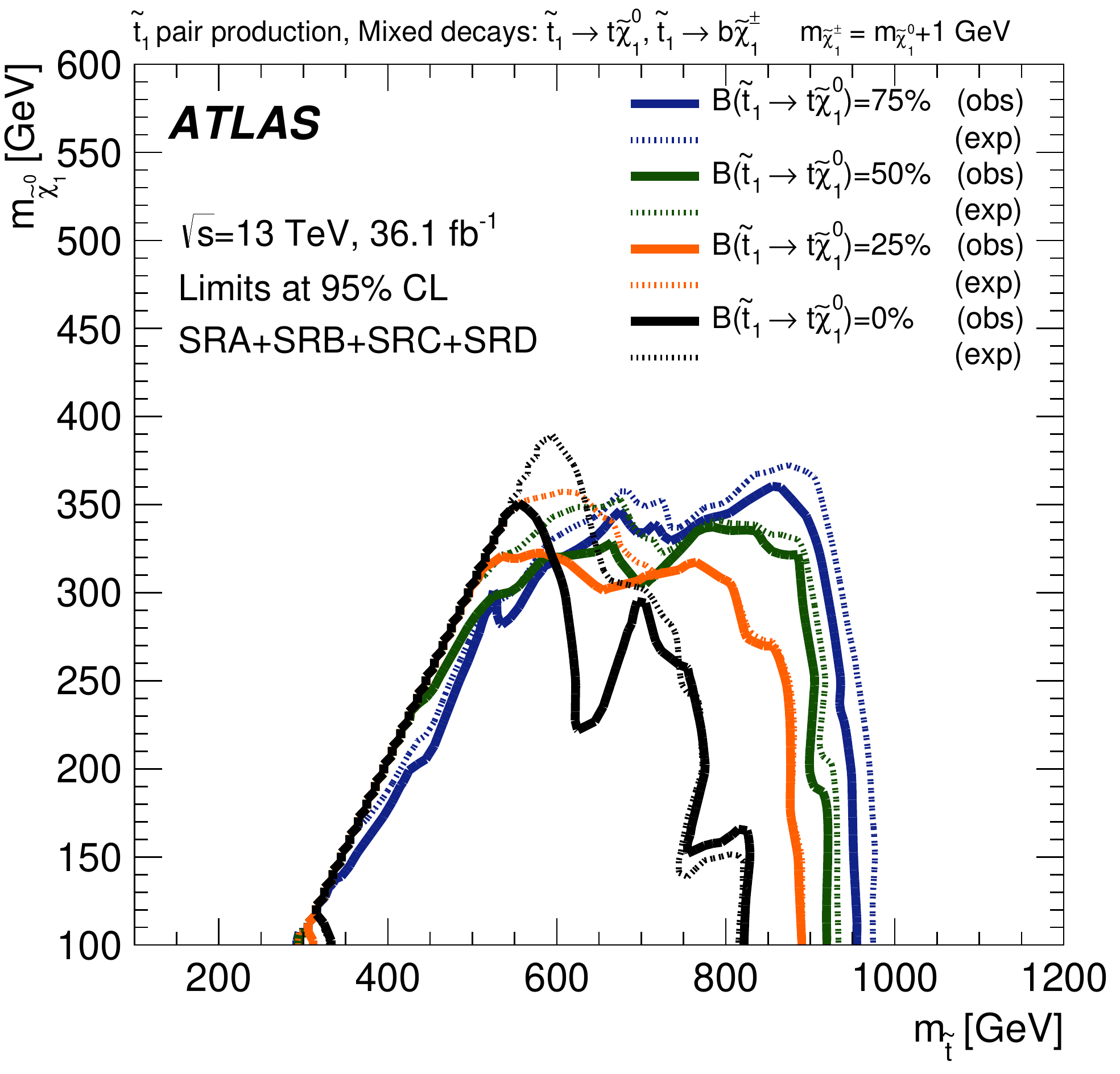}
    \caption{Observed (solid line) and expected (dashed line) exclusion contours at 95\% CL as a function of $\stop$ and $\ninoone$ masses and branching ratio to $\stop\to t\LSP$ in the natural SUSY-inspired mixed grid scenario where $m_{\chinoonepm}=\mLSP+1\gev$. 
}
    \label{fig:tbMet_exclusion}
  \end{center}
\end{figure}

\begin{figure}[htpb]
  \begin{center}
   \includegraphics[width=0.7\textwidth]{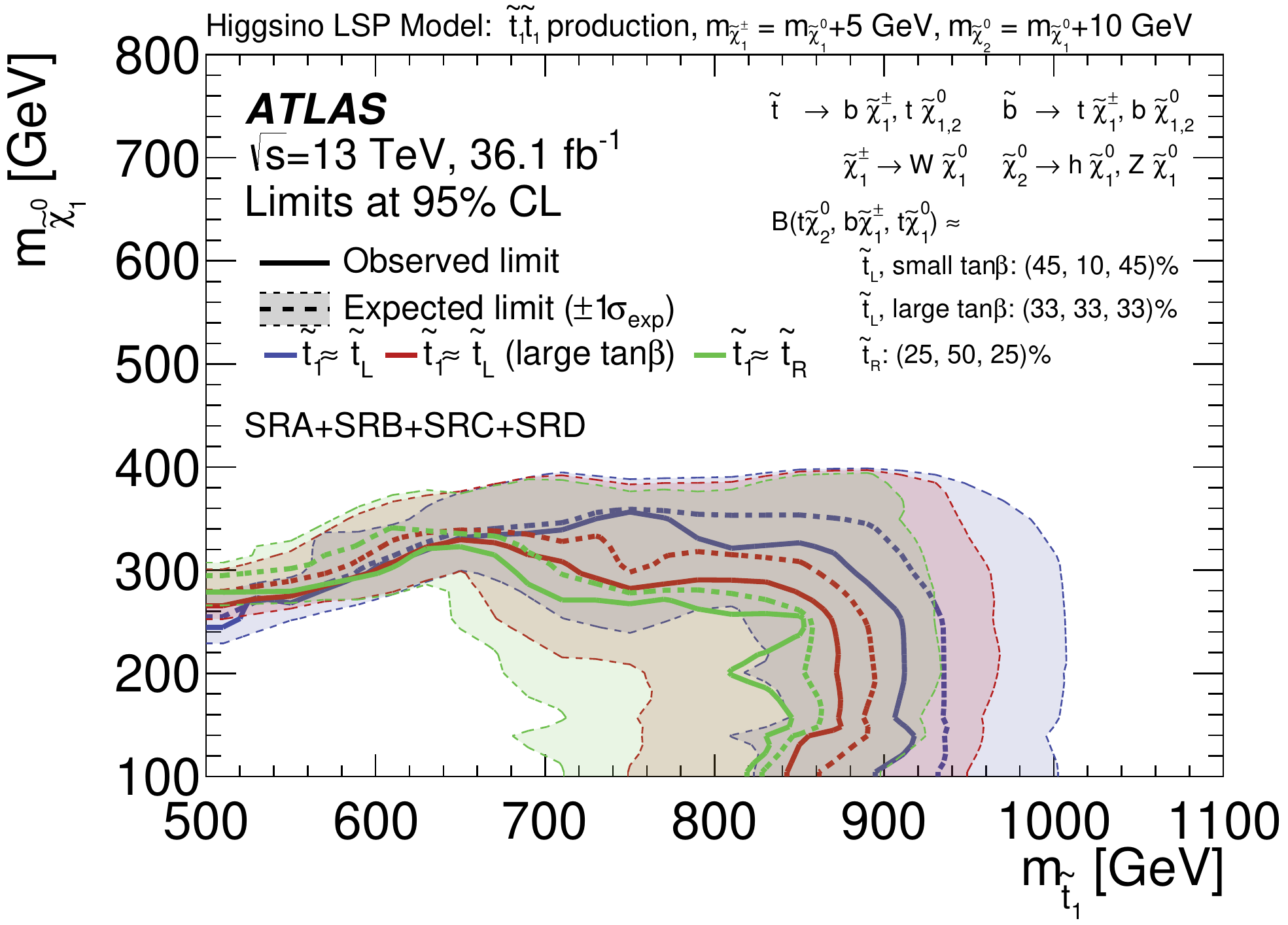}
    \caption{Observed (solid line) and expected (dashed line) exclusion contours at 95\% CL as a function of \mstop\ and \mLSP\ for the pMSSM-inspired non-asymptotic higgsino simplified model for a small tan$\beta$ with $B$($\stop\to t\ninotwo$, $\stop\to t\LSP$, $\stop\to b\chinoonepm$) = 45\%, 10\%, 45\% (blue), a large tan$\beta$ with $B$($\stop\to t\ninotwo$, $\stop\to t\LSP$, $\stop\to b\chinoonepm$) = 33\%, 33\%, 33\% (red), and a small $\tilde t_{R}$ with $B$($\stop\to t\ninotwo$, $\stop\to t\LSP$, $\stop\to b\chinoonepm$) = 25\%, 50\%, 25\% (green) assumption. Uncertainty bands correspond to the $\pm 1 \sigma$ variation of the expected limit.}
    \label{fig:nonAsymhiggsino_exclusion}
  \end{center}
\end{figure}

\begin{figure}[htpb]
  \begin{center}
    \includegraphics[width=0.7\textwidth]{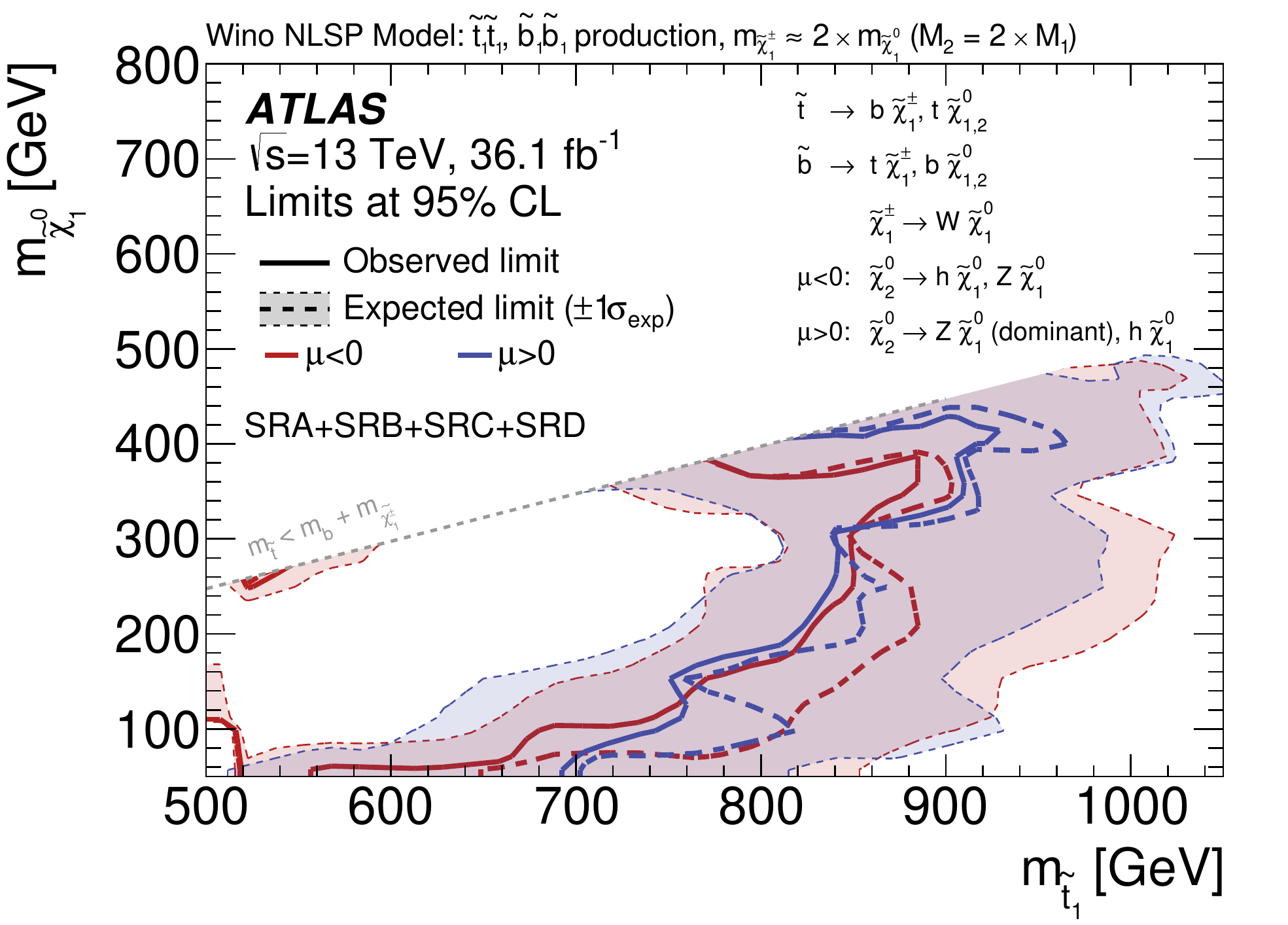}
    \caption{Observed (solid line) and expected (dashed line) exclusion contours at 95\% CL as a function of $\stop$ and $\ninoone$ masses for the Wino NLSP pMSSM model for both positive (blue) and negative (red) values of $\mu$. Uncertainty bands correspond to the $\pm 1 \sigma$ variation of the expected limit. 
    }
    \label{fig:winoNLSP_exclusion}
  \end{center}
\end{figure}

\begin{figure}[htpb]
  \begin{center}
    \includegraphics[width=0.7\textwidth]{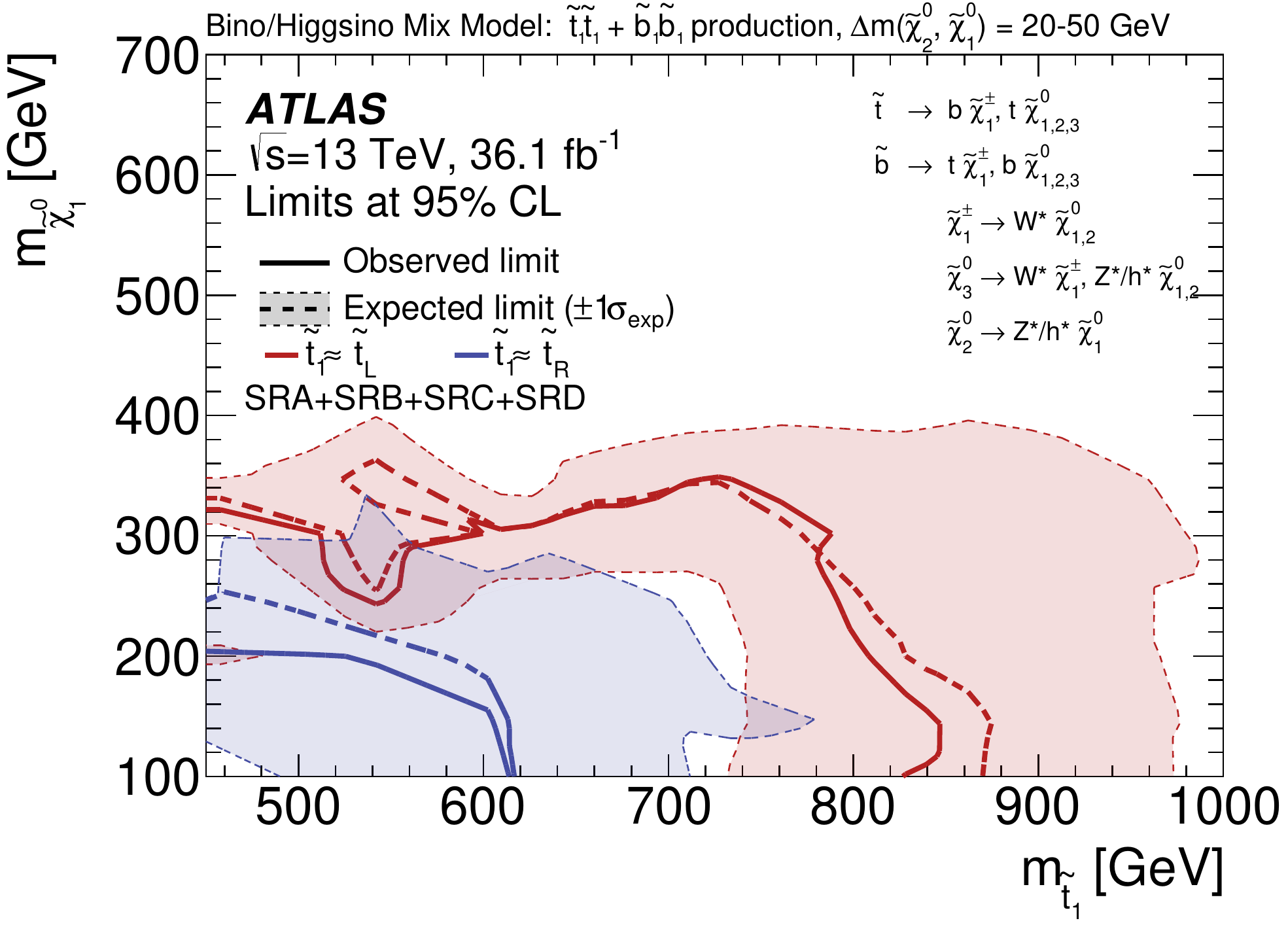}
    \caption{Observed (solid line) and expected (dashed line) exclusion contours at 95\% CL as a function of $\stop$ and $\ninoone$ masses for the $\tilde t_{L}$ scan (red) as well as for the $\tilde t_{R}$ scan (blue) in the well-tempered pMSSM model. Uncertainty bands correspond to the $\pm 1 \sigma$ variation of the expected limit.} 
    
    \label{fig:wellTemp_exclusion}
  \end{center}
\end{figure}

For signal models also considering top-squark decays into $b \chinoonepm$ or into additional massive neutralinos, four interpretations are considered:

\begin{description}
  \item[\boldmath Natural SUSY-inspired mixed grid:] A simplified model~\cite{naturalSUSY} where $m_{\chinoonepm}=\mLSP+1\gev$ with only two decay modes, $\stop\to b \chinoonepm$ and $\stop\to t\LSP$, and only on-shell top-quark decays are considered. The same 
maximal mixing between the partners of the left- and right-handed top quarks
and nature of the \LSP\ (pure bino) as for the $B$($\stop\to t\LSP$)=100\% case is assumed.  The branching ratio to $\stop\to t\LSP$ is set to 0\%, 25\%, 50\%, and 75\% and yield the limits shown in Figure~\ref{fig:tbMet_exclusion}.

\item[\boldmath Non-asymptotic higgsino:] A pMSSM-inspired simplified model with a higgsino LSP, ${m_{\chinoonepm}=\mLSP+5\gev}$, and ${m_{\ninotwo}=\mLSP+10\gev}$, assumes three sets of branching ratios for the considered decays of $\stop\to t\ninotwo$, $\stop\to t\LSP$, $\stop\to b\chinoonepm$~\cite{naturalSUSY}. A set of branching ratios with $B$($\stop\to t\ninotwo$, $\stop\to t\LSP$, $\stop\to b\chinoonepm$) = 33\%, 33\%, 33\% is considered, which is equivalent to a pMSSM model with the lightest top squark mostly consisting of the superpartner of left-handed top quark and $\tanb=60$ (ratio of vacuum expectation values of the two Higgs doublets). Additionally, $B$($\stop\to t\ninotwo$, $\stop\to t\LSP$, $\stop\to b\chinoonepm$) = 45\%, 10\%, 45\% and $B$($\stop\to t\ninotwo$, $\stop\to t\LSP$, $\stop\to b\chinoonepm$) = 25\%, 50\%, 25\% are assumed, which correspond to scenarios with $\mqlthree < \mtr$ (regardless of the choice of \tanb) and $\mtr<\mqlthree$ with $\tanb=20$, respectively. Here \mqlthree\ represents the left-handed third-generation mass parameter and \mtr\ is the mass parameter of the superpartner to the right-handed top-quark. Limits in the \mstop\ and \mLSP\ plane are shown in Figure~\ref{fig:nonAsymhiggsino_exclusion}.  

 \item[\boldmath Wino-NLSP pMSSM:] A pMSSM model where the LSP is bino-like and has mass \mone\ and where the NLSP is wino-like with mass \mtwo, while $\mtwo=2\mone$ and $\mstop>\mone$~\cite{naturalSUSY}. Limits are set for both positive and negative $\mu$ (the higgsino mass parameter) as a function of the \stop\ and \ninoone\ masses which can be translated to different \mone\ and \mqlthree, and are shown in Figure~\ref{fig:winoNLSP_exclusion}. Only bottom and top-squark production are considered in this interpretation. Allowed decays in the top-squark production scenario are $\stop\to t \ninotwo\to h/Z \LSP$, at a maximum branching ratio of 33\%, and $\stop \to b \chinoonepm$. Whether the $\ninotwo$ dominantly decays into a $h$ or $Z$ is determined by the sign of $\mu$. Along the diagonal region, the $\stop\to t\LSP$ decay with 100\% branching ratio is also considered. The equivalent decays in bottom-squark production are $\sbottom\to t\chinoonepm$ and $\sbottom\to b\ninotwo$. The remaining pMSSM parameters have the following values: $\mthree=2.2$ TeV (gluino mass parameter), $\ms=\sqrt{m_{\stopone} m_{\stoptwo}}=1.2$ TeV (geometric mean of top-squark masses), $\xtms=\sqrt{6}$ (mixing parameter between the superpartners of left- and right-handed states, where $X_{t}=\at-\mu/\tanb$ and $\at$ is the trilinear coupling parameter in the top-quark sector), and $\tanb=20$. All other pMSSM masses are set to $>$3 TeV. 

 \item[\boldmath Well-tempered neutralino pMSSM:] A pMSSM model in which three light neutralinos and a light chargino, which are mixtures of bino and higgsino states, are considered with masses within $50$~\GeV\ of the lightest state~\cite{atlasDM,wellTemp}. The model is designed to satisfy the SM Higgs boson mass and the dark-matter relic density ($0.10<\Omega h^{2}<0.12$, where $\Omega$ is energy density parameter and $h$ is the Planck constant~\cite{relic_density}) with pMSSM parameters: $\mone=-(\mu+\delta)$ where $\delta=20$--$50\gev$, $\mtwo=2.0$ TeV, $\mthree=1.8$ TeV, $\ms=0.8$--$1.2$~\TeV, $\xtms\sim\sqrt{6}$, and $\tanb=20$. For this model, limits are shown in Figure~\ref{fig:wellTemp_exclusion}. Only bottom- and top-squark production are considered in this interpretation. The signal grid points were produced in two planes, $\mu$ vs \mtr\ and $\mu$ vs \mqlthree, and then projected to the corresponding \stop\ and \ninoone\ masses. All other pMSSM masses are set to $>$3 TeV.

\end{description}

\begin{figure}[htpb]
  \begin{center}
    \includegraphics[width=0.7\textwidth]{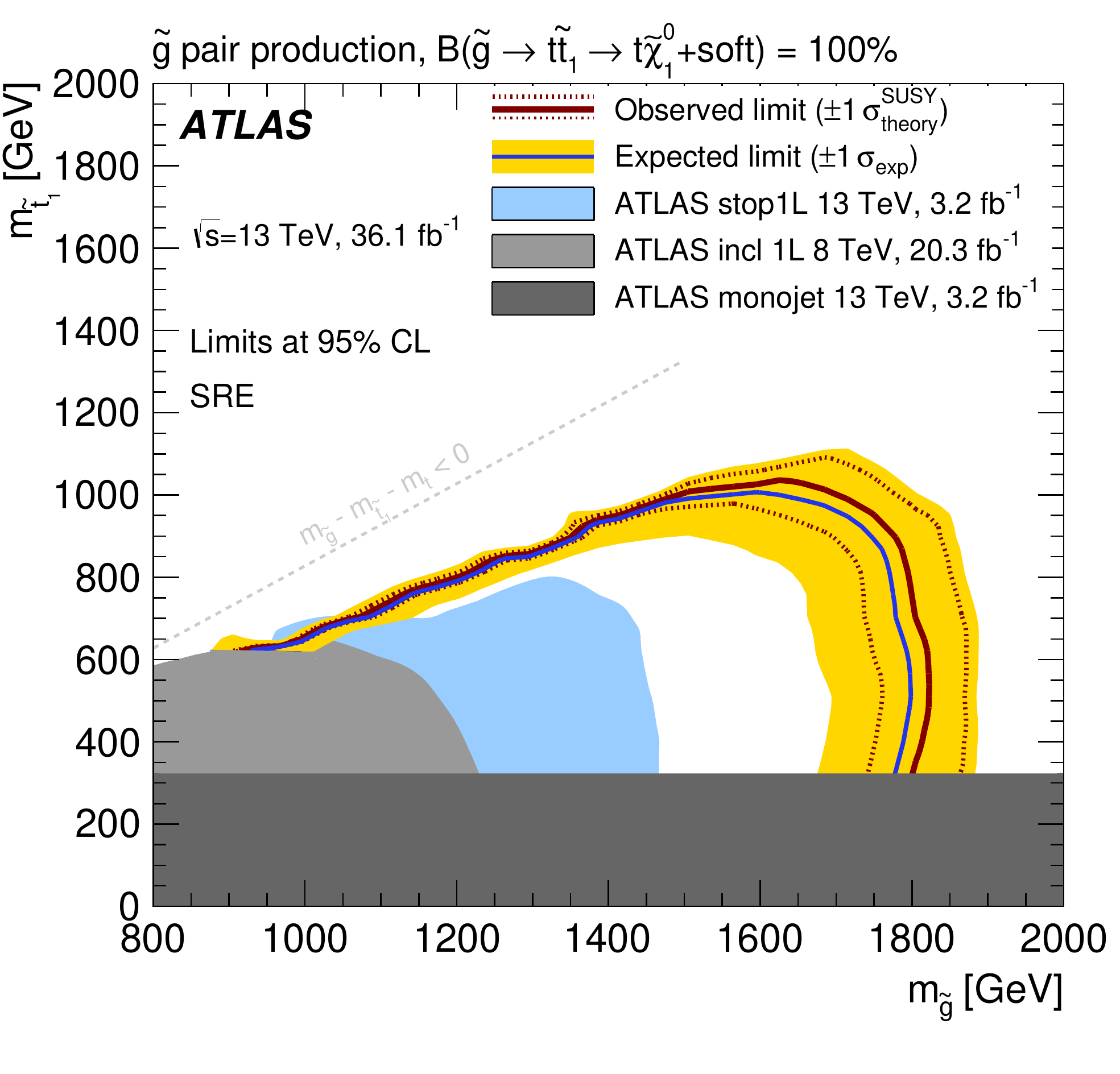}
    \caption{Observed (red solid line) and expected (blue solid line)
      exclusion contours at 95\% CL as a function
      of $\gluino$ and $\stop$ masses in the scenario where both
      gluinos decay via $\gluino\to t\stop\to t\ninoone+$soft
      and $\Delta m(\stop,\ninoone)=5\GeV$. Uncertainty bands corresponding to the $\pm 1
      \sigma$ variation of the expected limit (yellow band) and the
      sensitivity of the observed limit to $\pm 1\sigma$ variations of
      the signal theoretical uncertainties (red dotted lines) are also
      indicated. Observed limits from previous searches with the ATLAS detector at $\sqrt{s}=8$ and $\sqrt{s}=13$ TeV are overlaid in grey and blue~\cite{GtcStop1L,Gtc1L,GtcMonojet}.}
    \label{fig:SRE_exclusion}
  \end{center}
\end{figure}

The SRE results are interpreted for indirect top-squark production
through gluino decays in terms of the \stop\ vs $\gluino$ mass
plane with $\Delta m(\stop,\ninoone)=5\GeV$. 
Gluino masses up to $m_{\gluino}=1800\GeV$ with $\mstop<800\GeV$ are excluded as shown in Figure~\ref{fig:SRE_exclusion}.

\clearpage

%
%
%
\section{Conclusions}
\label{sec:conclusions}

Results from a search for top squark pair production based
on an integrated luminosity of \lumi\ of $\rts = 13 \tev ~pp$ 
collision data recorded by the ATLAS experiment at the LHC in 2015 and
2016 are presented. Top squarks are searched for in
final states with high-\pT\ jets and large missing transverse
momentum. In this paper, direct top squark production is studied assuming top squarks decay via $\stop
\rightarrow t^{(*)} \ninoone$ with large or small mass differences between the top squark and the neutralino $\Delta
m(\stop,\ninoone)$ and via $\stop\to b \chinoonepm$, where $m_{\chinoonepm}=\mLSP+1\gev$. 
Additionally, gluino-mediated \stop\ production is
studied, in which gluinos decay via $\gluino\rightarrow t\stop$, with a
small $\Delta m(\stop,\ninoone)$. 

No significant excess above the expected SM background is observed. Exclusion limits at 95\% confidence level in the plane of the top-squark and LSP masses are derived, resulting in the exclusion of top-squark masses in the range
\stopLimLowLSP\ GeV for $\ninoone$ masses below $160\GeV$. For the case where $\mstop\sim m_t+\mLSP$, top-squark masses in the range \stopLimDiag\ GeV are excluded. In addition, model-independent limits and $p$-values for each signal region are reported. Limits that take into account an additional decay of $\stop\to b \chinoonepm$ are also set with an exclusion of top-squark masses between 450 and 850 GeV for $\mLSP<240\gev$ and $B(\stop\to t \LSP)=50\%$ for $m_{\chinoonepm}=\mLSP+1\gev$. Limits are also derived in two pMSSM models, where one model assumes a wino-like NLSP and the other model is constrained by the dark-matter relic density. In addition to limits in pMSSM slices, limits are set in terms of one pMSSM-inspired simplified model where $m_{\chinoonepm}=\mLSP+5\gev$ and $m_{\ninotwo}=\mLSP+10\gev$. Finally, exclusion contours are reported for gluino production where $\mstop=\mLSP+5\gev$, resulting in gluino masses being constrained to be above 1800 GeV for \stop\ masses below 800 GeV.

%

We thank CERN for the very successful operation of the LHC, as well as the
support staff from our institutions without whom ATLAS could not be
operated efficiently.

We acknowledge the support of ANPCyT, Argentina; YerPhI, Armenia; ARC, Australia; BMWFW and FWF, Austria; ANAS, Azerbaijan; SSTC, Belarus; CNPq and FAPESP, Brazil; NSERC, NRC and CFI, Canada; CERN; CONICYT, Chile; CAS, MOST and NSFC, China; COLCIENCIAS, Colombia; MSMT CR, MPO CR and VSC CR, Czech Republic; DNRF and DNSRC, Denmark; IN2P3-CNRS, CEA-DRF/IRFU, France; SRNSF, Georgia; BMBF, HGF, and MPG, Germany; GSRT, Greece; RGC, Hong Kong SAR, China; ISF, I-CORE and Benoziyo Center, Israel; INFN, Italy; MEXT and JSPS, Japan; CNRST, Morocco; NWO, Netherlands; RCN, Norway; MNiSW and NCN, Poland; FCT, Portugal; MNE/IFA, Romania; MES of Russia and NRC KI, Russian Federation; JINR; MESTD, Serbia; MSSR, Slovakia; ARRS and MIZ\v{S}, Slovenia; DST/NRF, South Africa; MINECO, Spain; SRC and Wallenberg Foundation, Sweden; SERI, SNSF and Cantons of Bern and Geneva, Switzerland; MOST, Taiwan; TAEK, Turkey; STFC, United Kingdom; DOE and NSF, United States of America. In addition, individual groups and members have received support from BCKDF, the Canada Council, CANARIE, CRC, Compute Canada, FQRNT, and the Ontario Innovation Trust, Canada; EPLANET, ERC, ERDF, FP7, Horizon 2020 and Marie Sk{\l}odowska-Curie Actions, European Union; Investissements d'Avenir Labex and Idex, ANR, R{\'e}gion Auvergne and Fondation Partager le Savoir, France; DFG and AvH Foundation, Germany; Herakleitos, Thales and Aristeia programmes co-financed by EU-ESF and the Greek NSRF; BSF, GIF and Minerva, Israel; BRF, Norway; CERCA Programme Generalitat de Catalunya, Generalitat Valenciana, Spain; the Royal Society and Leverhulme Trust, United Kingdom.

The crucial computing support from all WLCG partners is acknowledged gratefully, in particular from CERN, the ATLAS Tier-1 facilities at TRIUMF (Canada), NDGF (Denmark, Norway, Sweden), CC-IN2P3 (France), KIT/GridKA (Germany), INFN-CNAF (Italy), NL-T1 (Netherlands), PIC (Spain), ASGC (Taiwan), RAL (UK) and BNL (USA), the Tier-2 facilities worldwide and large non-WLCG resource providers. Major contributors of computing resources are listed in Ref.~\cite{ATL-GEN-PUB-2016-002}.

%



%

\printbibliography



%
\newpage 

\begin{flushleft}
{\Large The ATLAS Collaboration}

\bigskip

M.~Aaboud$^\textrm{\scriptsize 137d}$,
G.~Aad$^\textrm{\scriptsize 88}$,
B.~Abbott$^\textrm{\scriptsize 115}$,
O.~Abdinov$^\textrm{\scriptsize 12}$$^{,*}$,
B.~Abeloos$^\textrm{\scriptsize 119}$,
S.H.~Abidi$^\textrm{\scriptsize 161}$,
O.S.~AbouZeid$^\textrm{\scriptsize 139}$,
N.L.~Abraham$^\textrm{\scriptsize 151}$,
H.~Abramowicz$^\textrm{\scriptsize 155}$,
H.~Abreu$^\textrm{\scriptsize 154}$
% ...
\bigskip
\\
% ...
$^{12}$ Institute of Physics, Azerbaijan Academy of Sciences, Baku, Azerbaijan\\
% ...
$^{88}$ CPPM, Aix-Marseille Universit{\'e} and CNRS/IN2P3, Marseille, France\\
% ...
$^{115}$ Homer L. Dodge Department of Physics and Astronomy, University of Oklahoma, Norman OK, United States of America\\
% ...
$^{119}$ LAL, Univ. Paris-Sud, CNRS/IN2P3, Universit{\'e} Paris-Saclay, Orsay, France\\
% ...
$^{137}$ $^{(a)}$ Facult{\'e} des Sciences Ain Chock, R{\'e}seau Universitaire de Physique des Hautes Energies - Universit{\'e} Hassan II, Casablanca; $^{(b)}$ Centre National de l'Energie des Sciences Techniques Nucleaires, Rabat; $^{(c)}$ Facult{\'e} des Sciences Semlalia, Universit{\'e} Cadi Ayyad, LPHEA-Marrakech; $^{(d)}$ Facult{\'e} des Sciences, Universit{\'e} Mohamed Premier and LPTPM, Oujda; $^{(e)}$ Facult{\'e} des sciences, Universit{\'e} Mohammed V, Rabat, Morocco\\
% ...
$^{139}$ Santa Cruz Institute for Particle Physics, University of California Santa Cruz, Santa Cruz CA, United States of America\\
% ...
$^{151}$ Department of Physics and Astronomy, University of Sussex, Brighton, United Kingdom\\
% ...
$^{154}$ Department of Physics, Technion: Israel Institute of Technology, Haifa, Israel\\
$^{155}$ Raymond and Beverly Sackler School of Physics and Astronomy, Tel Aviv University, Tel Aviv, Israel\\
% ...
$^{161}$ Department of Physics, University of Toronto, Toronto ON, Canada\\
% ...
$^{a}$ Also at Department of Physics, King's College London, London, United Kingdom\\
$^{b}$ Also at Institute of Physics, Azerbaijan Academy of Sciences, Baku, Azerbaijan\\
$^{c}$ Also at Novosibirsk State University, Novosibirsk, Russia\\
$^{d}$ Also at TRIUMF, Vancouver BC, Canada\\
$^{e}$ Also at Department of Physics {\&} Astronomy, University of Louisville, Louisville, KY, United States of America\\
% ...
$^{*}$ Deceased
\end{flushleft}


\end{document}